\documentclass[aps,prd]{revtex4} 
 
\input{epsf} 
 
\newcommand{\half}{{\frac{1}{2}}} 
\newcommand{\om}{\omega} 
\newcommand{\bom}{{\bar\omega}} 
\newcommand{\ep}{\epsilon} 
 
\newcommand{\sinY}{\sin\theta \partial_{\theta} Y_l^m}

\newcommand{\bi}{\begin{itemize}} 
\newcommand{\ei}{\end{itemize}} 
\newcommand{\be}{\begin{equation}} 
\newcommand{\ee}{\end{equation}} 
\newcommand{\beqa}{\begin{eqnarray}} 
\newcommand{\eeqa}{\end{eqnarray}} 
\newcommand{\ba}{\begin{array}} 
\newcommand{\ea}{\end{array}} 
\newcommand{\bea}{\begin{eqnarray}} 
\newcommand{\eea}{\end{eqnarray}} 
\newcommand{\bean}{\begin{eqnarray*}} 
\newcommand{\eean}{\end{eqnarray*}}

\newcommand{\ds}{\displaystyle}

\newcommand{\nn}{\nonumber} 
 
\newcommand{\aap}{Astron. and Astrophys.}

\newcommand{\mnras}{ Mon. Not. Roy. Astr. Soc.}

\begin{document} 
 
\title{The rotational modes of relativistic stars: Numerical results} 
 
\author{Keith H. Lockitch} 
\affiliation{Department of Physics, University of Illinois Urbana-Champaign, 
\\ 1110 E. Green St., Champaign-Urbana, IL 61801, USA} 
 
\author{John L. Friedman\thanks{Email address: friedman@uwm.edu}} 
\affiliation{Department of Physics, University of Wisconsin-Milwaukee, 
\\ P.O. Box 413, Milwaukee, WI 53201, USA} 
 
\author{Nils Andersson} 
\affiliation{Department of Mathematics, University of
Southampton, Southampton SO17 1BJ, UK}

\date{\today}

\begin{abstract} 
We study the inertial modes of slowly rotating, fully relativistic  
compact stars. The equations that govern perturbations of both  
barotropic and non-barotropic models are discussed, but we present 
numerical results only for the barotropic case. For barotropic stars 
all inertial modes are a hybrid mixture of axial and polar perturbations.  
We use a spectral method to solve for such modes of various polytropic 
models. Our main attention is on modes that can be driven unstable  
by the emission of gravitational waves. Hence, we calculate the  
gravitational-wave growth timescale for these unstable modes and 
compare the results to previous estimates obtained in Newtonian  
gravity (i.e. using post-Newtonian radiation formulas). We find that  
the inertial modes are slightly stabilized by relativistic effects,  
but that previous conclusions concerning eg. the unstable  
r-modes remain essentially unaltered when the problem  
is studied in full general relativity. 
\end{abstract} 
 
\maketitle 
 
\section{Introduction}

 
That gravitational waves can drive various modes of  
oscillation in a rotating neutron star unstable was first 
suggested by Chandrasekhar \cite{chandra}. The detailed 
mechanism behind this instability 
was explained by Friedman and Schutz \cite{fs78a,fs78b}, who 
showed that the instability sets in when an originally retrograde  
mode becomes prograde (according to an inertial observer) 
due to the rotation of the star.  They also showed that  
the instability  is generic, that, for arbitrarily slow rotation, 
any perfect-fluid stellar model has unstable modes with sufficiently 
large values of the azimuthal eigenvalue $m$  (the mode depends on 
$\varphi$ as $\exp(im \varphi)$). The existence of this radiation-driven  
instability is potentially important since it could limit 
the attainable spin rate of astrophysical neutron stars \cite{jlf83}.  
However, a detailed assessment of its astrophysical relevance 
is complicated by the fact that viscosity tends to  
counteract the growth of an unstable mode.  One must account  
not only for the familiar hydrodynamic shear and bulk viscosities  
\cite{il}, but also exotic mechanisms like the  
mutual friction that is relevant in superfluid neutron  
stars. Once a star has cooled 
below the superfluid transition temperature, mutual friction 
appears to suppress the instability of f-modes \cite{lm95}, 
and gravitational waves from f-modes appeared to set a limit on rotation 
only slightly more stringent that the maximum spin of an equilibrium 
model.  
 
Recent work has modified this conclusion in two ways.  First, 
numerical studies of the marginally stable  
``neutral'' f-modes of fully relativistic, rapidly rotating polytropes   
by Stergioulas and Friedman \cite{sf} showed that  
relativistic effects destabilize the modes of a rotating star 
considerably. Most importantly, one finds that in general relativity 
the $m=2$ f-mode may  have a neutral 
point for attainable rates of rotation. This is in contrast  
with the Newtonian result that the $m=2$ mode is unlikely to  
become unstable in uniformly rotating stars and it is important 
since the quadrupole mode is the most efficient emitter of gravitational 
radiation. Hence one would expect it to lead to the fastest  
growing instability. For realistic equations of state it has been shown 
\cite{msb98} that in a typical  
$1.4M_\odot$ star the $m=2$ f-mode has a neutral point 
near $\Omega \approx 0.85\Omega_K$ where $\Omega_K$ represents 
the mass-shedding limit. 
 
The second piece of evidence follows from the fact that  
the inertial r-modes are unstable at any rate of 
rotation in a perfect fluid star \cite{na98,fm98}.  
The great surprise of a few years ago was the discovery that,  
despite the fact that they radiate mainly through  
the current multipoles, the unstable r-modes  
could potentially limit  the spin of a  
rotating neutron star significantly \cite{lom98,aks99}.  
Since its original discovery the r-mode instability has been  
discussed in a number of papers, and we refer the interested 
reader to recent review articles for detailed discussions of 
the literature \cite{fl1,ak,lindblom,fredlamb,fl2}. 
At the present time key issues concern the nonlinear evolution of 
an unstable mode \cite{sfont,tohl1,tohl2,arras,lin}, the possible presence of 
hyperons in the neutron star core (since hyperons may lead to a very strong  
bulk viscosity) \cite{pbj,lo02,hae} and the role of superfluidity \cite{lm00,ac01}.

The present paper concerns the  
effect that general relativity has on the instability 
of the r-modes and other inertial modes. Intuitively, one 
would expect this to be a relevant issue since the 
relativistic framedragging is an order $\Omega$ effect 
which would affect inertial modes at leading order. 
In the last couple of years significant progress toward an  
understanding of the nature 
of inertial modes in general relativity has been made. 
The picture that is emerging is largely similar to that  
in Newtonian gravity. In the Newtonian case the r-modes are 
a purely axial parity subset of the large class of inertial  
modes which generally have velocity fields that are described by a 
hybrid mixture of axial and polar components to leading order in $\Omega$.  
Non-barotropic Newtonian stars 
have an infinite set of r-modes for each  
permissible $l$ and $m$ (corresponding to the spherical harmonic $Y_l^m$ used to
describe  
the velocity field), while barotropic models  
retain only a vestigial r-mode for each  
$l=m$, see \cite{lf,yl} for detailed discussions.  
In relativity one can prove that barotropic stars 
(where the perturbations are described by the same  
one-parameter equation of state as the background star~\footnote{In Paper~I  
such stars were refered to as ``isentropic''.})  
have no purely axial inertial modes \cite{laf1}. 
All inertial modes of such stars  
are hybrids.  
 
As in  Newtonian theory \cite{ak},  
the relativistic {\em non-barotropic} r-mode problem is 
significantly different. Like their Newtonian counterparts, relativistic 
non-barotropic stars have modes that are purely axial in the spherical 
limit.  
These modes are  
determined by solving a single ordinary differential equation 
for one of the perturbed metric components~\cite{koj}, cf. eq~(\ref{singeq2}).  
We have previously shown \cite{laf1} that discrete mode-solutions  
to this equation exist for uniform density stars.  
For more realistic equations of state, the  
problem is complicated by the fact that this equation 
corresponds to a singular eigenvalue problem which also  
admits a continuous spectrum \cite{koj,bk}. 
The dynamical role of this continuous spectrum,  
or indeed if it remains present when more physics 
is included in the model, is not clear at the present time.  
In order to determine a purely axial mode of a non-barotropic 
star one must typically  identify a discrete mode embedded in 
the continuous spectrum.  
This technical difficulty has led to suggestions that  
the r-modes may not even ``exist'' for certain relativistic 
stars \cite{yosh,kr01}. However, from eq.~(\ref{singeq2}),  
it is clear that the slow-rotation approximation is no longer 
consistent in regions where $\alpha - \tilde{\omega} \sim O(\Omega^2)$  
or smaller. This means that the problem likely requires a ``boundary  
layer'' approach \cite{regul} where either $\Omega^2$ terms,  
viscosity or the coupling to polar perturbations are included 
in the analysis of the region near the singular point.  
Recent results by Yoshida and Lee \cite{yl02} 
and Ruoff et al. \cite{rsk02} support this view.  
 
So far there have been two studies of the growth  
timescale for the unstable relativistic r-modes \cite{rk02,yf01}.  
Both concern the non-barotropic problem. 
The results suggest that post-Newtonian 
estimates of the instability growth time are surprisingly good,  
but also indicate a weakening of the instability  
once the star becomes very compact.  
The aim of the present paper is to extend these results  
by considering the effect of radiation reaction on various 
inertial modes of barotropic stellar models.  
We expect the results obtained from a barotropic model  
to be relevant even though real neutron stars will have internal  
stratification associated with composition gradients (eg. due to the  
varying proton fraction). As discussed by Reisenegger and Goldreich \cite{rg}  
this will lead to a Brunt-V\"ais\"al\"a frequency  
$N\sim 500 \mbox{ s}^{-1} (\rho /\rho_{\rm nuclear})^{1/2}$ 
and the presence of core g-modes with frequencies in the range  
100-200~Hz. In a very slowly rotating star the buoyancy force will dominate  
the Coriolis force and the g-modes will remain largely unchanged, but  
when $\Omega >> N$ the situation will be reversed and the  
g-modes will be almost entirely rotationally restored. 
In other words, one would expect all low frequency modes 
to be well described by the inertial modes of a barotropic  
model for rotation periods shorter than 2-3~ms. This means that  
the results obtained in this paper should be a reasonable approximation  
for neutron stars in Low-Mass X-ray binaries (expected to  
have rotation rates in the range 250-500~Hz), and an  
accurate representation of the inertial modes 
of a newly born neutron star spinning at or near the break-up limit.   
 
The layout of the paper is as follows: 
In section~II we summarize the equations that  
describe the oscillations of a slowly rotating  
relativistic star.  
Section~III provides a description of the numerical 
method we use to solve the eigenvalue problem, while 
the obtained results are discussed in Section~IV.  
Section~V is devoted to a discussion of our method  
for estimating the radiation reaction timescale for  
inertial modes. Our conclusions   
are given in Section~VI. 
 
Throughout the paper we use the following conventions: 
We refer to ref.~\cite{lf} as  
Paper I, ref.~\cite{laf1} as  Paper II , while Paper III is \cite{regul}. Our 
numbering 
convention for equations from these papers is such that e.g. (II,2.5) means Eqn.
(2.5)  
from Paper II. Unless otherwise stated we use geometrized units 
$c=G=1$.

 
\section{Perturbations of Slowly Rotating Stars} 
 
\subsection{The Equilibrium Star} 
 
We consider a perfect fluid star rotating slowly with uniform angular  
velocity $\Omega$.  By slow rotation we mean the assumption that  
$\Omega$ is small compared to the Kepler velocity,  
$\Omega_K\simeq 0.67\sqrt{\pi G\bar\ep}$, at which the star is unstable  
to mass shedding at its equator ($\bar\ep$ represents the average  
energy density in the star).  In particular, we neglect all  
quantities of order $\Omega^2$ or higher.  In this approximation the  
star retains its spherical shape, because the centrifugal deformation  
of its figure is an order $\Omega^2$ effect \cite{h67}. The only  
new order $\Omega$ effect that arises because of general relativity is the  
rotational framedragging, denoted by $\omega(r)$ below.  
 
To first order in $\Omega$ 
the equilibrium state is described  by a  
stationary, axisymmetric spacetime with metric, $g_{\alpha\beta}$, of  
the form \cite{h67,cm74}  
\be 
ds^2 = -e^{2\nu(r)} dt^2 + e^{2\lambda(r)} dr^2 + r^2 d \theta^2
+ r^2 sin^2 \theta d \varphi^2  - 2 \omega(r) r^2 sin^2\theta dt d\varphi
\label{equil_metric}
\ee 
with perfect fluid matter source 
\be 
T_{\alpha\beta} = (\epsilon+p)u_\alpha u_\beta + p g_{\alpha\beta}. 
\label{emom} 
\ee 
Here, $\ep$ and $p$ are, respectively, the total energy density and  
pressure of the fluid as measured by an observer moving with unit 
4-velocity, 
\be 
u^\alpha = e^{-\nu} ( t^\alpha + \Omega \varphi^\alpha); 
\label{equil_4v} 
\ee 
$t^\alpha=(\partial_t)^\alpha$ and 
$\varphi^\alpha=(\partial_\varphi)^\alpha$ being, respectively, the 
timelike and rotational Killing vectors of the spacetime. 
 
The metric and fluid variables are required to satisfy Einstein's  
equation, $G_{\alpha\beta}=8\pi T_{\alpha\beta}$.  This reduces to  
the well-known Tolman-Oppenheimer-Volkov (TOV) equations for the  
``spherical quantities'' (those of order $\Omega^0$) together with  
Hartle's equation \cite{h67} for the order  $\Omega$ quantity, 
\be 
\bom(r)\equiv\Omega-\om(r), 
\label{bom} 
\ee 
that governs the dragging of inertial frames induced by the rotation 
of the star.  For the exact form of these equations, we refer the reader  
to Paper II; Eqs. (II,3.4-3.7) and (II,4.3).

To complete our specification of the equilibrium star we must  
provide an equation of state (EOS) relating the density and pressure. 
In this paper, we will always require our equilibrium solution to  
satisfy a one-parameter EOS, $\ep = \ep(p)$. This is an accurate assumption 
for equilibrium neutron stars since their temperature is likely to  
be significantly below the Fermi temperature $T_F \sim 10^{12}$~K.  
For simplicity we use the polytropic EOS, 
\bea 
p &=& K \rho^{1+\frac{1}{n}} \nn\\ 
\ep &=& \rho + np, 
\eea 
where $\rho$ is the rest-mass density, $n$ is the polytropic index and  
$K$ is the polytropic constant.  We use a set of polytropic indices  
($n=0.0,0.5,1.0,1.5$) that span the range of compressibilities of  
proposed realistic neutron star equations of state \cite{glend,akmal}. 
 
Although our equilibrium model obeys a one-parameter EOS, we do not  
necessarily require the perturbed fluid to satisfy the {\em same} EOS.  For an 
adiabatic perturbation of an equilibrium star obeying a one-parameter  
EOS, the perturbed pressure and energy density are customarily 
related by 
\be 
\frac{\delta p}{\Gamma_1 p} = \frac{\delta \ep}{(\ep+p)}  
+ \xi^\alpha A_\alpha 
\label{ad_osc} 
\ee 
where $\Gamma_1(r)$ is the adiabatic index, $\xi^\alpha$ is the Lagrangian  
fluid displacement and where the Schwarzschild discriminant, 
\be 
A_\alpha \equiv \frac{1}{(\ep+p)}\nabla_\alpha\ep  
- \frac{1}{\Gamma_1 p}\nabla_\alpha p, 
\ee 
governs convective stability in the star. 
In general, the adiabatic index $\Gamma_1$ need not be equal to the constant 
\be 
\Gamma \equiv \frac{(\ep+p)}{p}\frac{dp}{d\ep} = 1 + \frac{1}{n} 
\label{differ} 
\ee 
associated with the equilibrium (polytropic) EOS.  
In terms of this constant we have 
\be 
A_\alpha = \left(\frac{1}{\Gamma} - \frac{1}{\Gamma_1}\right) 
\frac{1}{p}\nabla_\alpha p . 
\label{schwarz} 
\ee 
We will call a model barotropic if and only if 
the perturbed configuration satisfies the same one-parameter EOS as 
the unperturbed configuration. In this case $\Gamma_1\equiv\Gamma$  
and the Schwarzschild discriminant vanishes identically. Such stars  
are marginally stable to convection, and since they have no 
internal stratification they do not admit 
finite frequency modes restored by buoyancy (g-modes) \cite{rg}.  In this paper 
we will consider perturbations of both barotropic and nonbarotropic stars, 
but will focus mainly on the barotropic case in our numerical work. 
The non-barotropic case has already been discussed in detail in  
\cite{laf1,yosh,kr01,regul}. 
 
\subsection{The Perturbation Equations} 
\label{sect:pert} 
 
We now consider the rotationally restored (inertial) modes of  
a slowly rotating relativistic star.  The equations governing such  
perturbations were derived in detail in Paper II. Here we will  
simply quote the main results needed for the present analysis. 
 
Since the equilibrium spacetime is stationary and axisymmetric,  
we may decompose our perturbations into oscillation modes proportional to 
$e^{i(\sigma t + m \varphi)}$. For convenience, we will always choose  
$m \geq 0$, since the complex conjugate of an $m<0$ mode with real  
frequency $\sigma$ is an $m>0$ mode with frequency $-\sigma$.  Note  
that $\sigma$ is the mode frequency measured by an inertial observer at  
infinity. 
 
In the Lagrangian perturbation formalism \cite{f78,fi92}, the basic  
variables are the metric perturbation $h_{\alpha\beta}$ and  
the Lagrangian displacement $\xi^\alpha$.  We begin by expanding  
these variables in vector and tensor spherical harmonics. 
The Lagrangian displacement vector can be written 
\beqa 
\xi^{\alpha} \equiv \frac{1}{i\kappa\Omega} \sum_{l=m}^\infty  
\biggl\{  
\ds{\frac{1}{r} W_l(r) Y_l^m r^{\alpha}  
+ V_l(r) \nabla^{\alpha} Y_l^m } 
- \ds{i U_l(r) P^\alpha_{\ \mu} \epsilon^{\mu\beta\gamma\delta}  
\nabla_{\beta} Y_l^m \nabla_{\!\gamma} \, t \nabla_{\!\delta} \, r} 
\biggr\} e^{i\sigma t} \ , 
\label{xi_exp} 
\eeqa 
where we have defined, 
\be 
P^\alpha_{\ \mu} \equiv e^{(\nu+\lambda)}  
\left( \delta^\alpha_{\ \mu} 
- t_\mu \nabla^\alpha t 
\right) 
\ee 
and introduced the ``comoving'' frequency, 
\be 
\kappa\Omega \equiv \sigma+m\Omega. 
\ee 
The exact form of expression (\ref{xi_exp}) has been chosen for  
convenience. In particular, we have used the gauge freedom inherent in 
the Lagrangian formalism \cite{cq76,ss77} to set $\xi_t\equiv 0$.   
Note also the chosen relative phase between the terms in (\ref{xi_exp})  
with polar parity (those with coefficients $W_l$ and $V_l$) and the terms  
with axial parity (those with coefficients $U_l$). 
 
Working in the Regge-Wheeler gauge \cite{rw57}, we express our metric  
perturbation as 
\beqa 
h_{\mu\nu} = e^{i\sigma t}  \sum_{l=m}^\infty    
 \left[ \ba[c]{cccc} 
H_{0,l}(r)e^{2\nu}Y_l^m & H_{1,l}(r)Y_l^m &  
h_{0,l}(r) \,(\frac{m}{sin\theta})Y_l^m & i h_{0,l}(r)\,\sinY\\ 
H_{1,l}(r)Y_l^m & H_{2,l}(r)e^{2\lambda}Y_l^m &  
h_{1,l}(r) \,(\frac{m}{sin\theta})Y_l^m & i h_{1,l}(r)\,\sinY\\ 
\mbox{\scriptsize symm} & \mbox{\scriptsize symm} & 
r^2K_l(r)Y_l^m & 0\\ 
\mbox{\scriptsize symm} & \mbox{\scriptsize symm} & 
0 & r^2sin^2\theta K_l(r)Y_l^m 
\ea 
\right]  
\label{h_components} 
\eeqa 
Again, note the choice of phase between the polar-parity components 
(those with coefficients $H_{0,l}$, $H_{1,l}$, $H_{2,l}$ and $K_l$)  
and the axial-parity components (those with coefficients 
$h_{0,l}$ and $h_{1,l}$). 
 
In Paper II, we also found it convenient to make use of the Eulerian 
perturbation formalism, whose basic variables are the metric  
perturbation $h_{\alpha\beta}$ and the perturbed density, $\delta\ep$, 
pressure, $\delta p$, and fluid 4-velocity, $\delta u^\alpha$. 
Since they represent scalar quantities the Eulerian changes  
in the density and pressure may be written as 
\be 
\delta\ep = \sum_{l=m}^\infty \, \delta\ep_l(r) \, Y_l^m \, e^{i\sigma t} 
\label{GR:del_ep_exp} 
\ee 
and 
\be 
\delta p = \sum_{l=m}^\infty \, \delta p_l(r) \, Y_l^m \, e^{i\sigma t}, 
\label{GR:del_p_exp} 
\ee 
respectively, while the Eulerian change in the fluid velocity may be 
expressed in terms of $\xi^\alpha$ and $h_{\alpha\beta}$ as defined 
above, 
\be 
\delta u^\alpha = q^\alpha_{\ \beta}\pounds_u\xi^\beta  
+ \mbox{$\half$} u^\alpha u^\beta u^\gamma h_{\beta\gamma}, 
\label{delu} 
\ee 
where $q^{\alpha\beta} \equiv g^{\alpha\beta} + u^\alpha u^\beta$. 
 
We showed in Paper~II that the rotationally restored modes of a  
slowly rotating star have a fundamentally different character  
depending on whether the star is barotropic or nonbarotropic.   
The difference pertains to the character of the modes in the limit  
as the star's angular velocity, $\Omega$, goes to zero. This is yet  
another reason why it is appropriate to consider the problem  
within  the slow rotation approximation. 
In particular, we proved in Paper~II that a relativistic barotrope does not 
admit distinct classes of r-modes or g-modes (modes whose limit as  
$\Omega\rightarrow 0$ are purely axial or purely polar, respectively). 
Instead, the generic inertial mode of such a star is a hybrid mixture of  
axial and polar components to lowest order in $\Omega$.   
In contrast, non-barotropic stars allow distinct g-modes already 
in the non-rotating case and purely axial r-modes may exist at   
lowest order in $\Omega$.  
To reflect the fundamental difference between the two cases,  
we organized our slow-rotation  
expansion by requiring our perturbation variables to obey the following  
ordering in powers of $\Omega$, 
\begin{displaymath} 
U_l, h_{0,l}  \sim  \phantom{1} O(1) 
\end{displaymath} 
\begin{displaymath} 
W_l, V_l, H_{1,l}  \sim    
\Biggl\{  \ba{ll} 
O(1) & \mbox{barotropic stars} \\ 
O(\Omega^2) & \mbox{nonbarotropic stars} 
\ea  
\end{displaymath} 
\be 
H_{0,l}, H_{2,l}, K_l, h_{1,l}, \delta \ep_l, \delta p_l, \sigma   
\sim  \phantom{1} O(\Omega) \ . 
\label{ordering} 
\ee 
The perturbation equations may then be grouped in powers of $\Omega$ 
and solved order by order.  To compare our results with the Newtonian  
r-modes and hybrid modes we need only find the leading order mode  
solutions; that is, we need only find the mode frequency to order 
$\Omega$ and the eigenfunctions to $O(1)$.  As discussed in Paper II,  
these turn out to be determined by a subset of the perturbation 
equations up to first order in $\Omega$. 
 
It is relevant to point out that the equations derived by this procedure  
will be somewhat different in different regions of spacetime  
(see Table~\ref{tab1}).   
In the ``near zone,'' the region in which $\sigma r << 1$, we  
will be able to ignore second time derivatives, whereas we cannot do this 
in the ``wave zone,'' the region in which $\sigma r >> 1$.   
Because the inertial modes that we are primarily interested in 
are restored by the Coriolis force, their 
frequencies scale with the angular velocity of the star,  
$\sigma\sim\Omega$.  For slow rotation, this implies that the near  
zone extends far away from the star into the nonrelativistic region 
($M/r<< 1$) and that the wave zone will be located entirely within  
the nonrelativistic region \cite{ip71}.  
This will be important in Sect.~\ref{times_sect} when we calculate  
the energy radiated in gravitational waves and the timescales on which  
gravitational radiation reaction drives the unstable modes.  
For now, we will focus on the equations that are relevant in the near zone,
which 
were derived in Paper II.  These will allow us to find the eigenvalues  
and eigenfunctions of the modes that we are interested in. In  
Sect.~\ref{sect:dEdt} we will consider the equations more generally  
when we derive an expression for the radiated energy. 
The various spacetime regions are depicted schematically in Table~\ref{tab1}. 
 
\begin{table}[h] 
\begin{tabular}{c|c|c|c|cc} 
\hline\hline 
 & \multicolumn{2}{c}{} & \multicolumn{2}{|c}{} &  
\\ 
 & \multicolumn{2}{c} 
{ \hspace{0.5em} relativistic zone, \ \ $\frac{M}{r}\sim 1$ \hspace{0.5em} } 
 & \multicolumn{2}{|c}{ nonrelativistic zone, \ \  $\frac{M}{r}<< 1$ }  
 &  
\\ 
 & \multicolumn{2}{c|}{} & \multicolumn{2}{|c}{} &  
\\ 
\cline{2-6} 
 & \multicolumn{1}{||c||}{} & \multicolumn{3}{|c}{} &  
\\ 
 $r=0$ \ & \multicolumn{1}{||c||}{neutron star}  
 & \multicolumn{1}{|l}{\ $r=R$} &  
 \multicolumn{2}{c}{exterior spacetime} & 
 \multicolumn{1}{l}{$r\rightarrow\infty$}  
\\ 
 & \multicolumn{1}{||c||}{} & \multicolumn{3}{|c}{} &  
\\ 
\cline{2-6} 
 & \multicolumn{3}{c|}{} & &  
\\ 
 & \multicolumn{3}{c|} 
{\hspace{8em} near zone, \ \ $\sigma r \sim \Omega r << 1$ \hspace{8em} }  
 & \hspace{1em} wave zone, \ \ $\sigma r \sim \Omega r >> 1$ \hspace{1em} 
 &  
\\ 
 & \multicolumn{3}{c|}{} & &  
\\ 
\hline\hline 
\end{tabular} 
\caption{The spatial regions relevant to the relativistic inertial-mode  
problem.} 
\label{tab1}
\end{table} 

The ordering (\ref{ordering}) for the barotropic case is slightly more 
general than for the nonbarotropic case because the polar-parity  
coefficients $W_l$, $V_l$ and $H_{1,l}$ are not assumed to be negligible 
compared to the axial-parity coefficients $U_l$ and $h_{0,l}$.   
We will retain all of these variables in presenting the relevant 
equations and then specialize to each of the two cases.  (Because  
$h_{1,l}$ is an order $\Omega$ variable, we ignore it and drop the  
``0'' subscript on $h_{0,l}$, writing it as $h_l$. Only in Sect.  
\ref{sect:dEdt} will it be necessary to restore this distinction.) 
 
As presented in Paper II, the relevant $O(1)$ equations (those that apply  
to the perturbed spherical star) are: 
 
\noindent 
Eq. (II,3.20), 
\be 
H_{1,l} +\frac{16\pi(\epsilon+p)}{l(l+1)} e^{2\lambda} r W_l = 0. 
\label{GR:sph_H1}  
\ee 
 
\noindent 
Eq. (II,3.23), 
\be 
V_l = \frac{e^{-(\nu+\lambda)}}{l(l+1)(\epsilon+p)}  
\left[(\epsilon+p)e^{\nu+\lambda} r W_l \right]'. 
\label{GR:sph_V} 
\ee 
and 
\noindent 
Eq. (II,3.22), 
\beqa 
h_l^{''} - (\nu'+\lambda') h_l'  
+ \biggl[ \frac{(2-l^2-l)}{r^2}e^{2\lambda}  
- \frac{2}{r}(\nu'+\lambda') 
- \frac{2}{r^2} \biggr] h_l  
= \frac{4}{r}(\nu'+\lambda') U_l 
\label{GR:sph_h_0''} 
\eeqa 
where a prime denotes a derivative with respect to $r$.  Throughout this  
work we have used Eq. (\ref{GR:sph_H1}) [i.e., Eq. (II,3.20)] to  
eliminate the metric variable $H_{1,l}(r)$ in favor of $W_l(r)$. 
 
To close the system of equations we need to retain only  
two of the equations that arise at $O(\Omega)$.  
As discussed in Paper II, the relevant pair are the two  
independent components of Eq. (II,4.16), which enforces the conservation  
of vorticity in constant entropy surfaces [see Eq. (\ref{Del_om_NI}) below]. 
The $[\theta\varphi]$ component of Eq. (II,4.16) leads to Eq. (II,4.53), 
\be 
0 = \ba[t]{l} 
\left[ l(l+1)\kappa\Omega(h_l+U_l)-2m{\bar\omega}U_l\right] \\ 
 \\ 
+(l+1)Q_l \left[ 
\frac{e^{2\nu}}{r}\partial_r\left(r^2{\bar\omega}e^{-2\nu}\right)W_{l-1} 
-2(l-1){\bar\omega}V_{l-1} 
\right]  \\ 
 \\ 
-l Q_{l+1} \left[\frac{e^{2\nu}}{r}\partial_r 
\left(r^2{\bar\omega}e^{-2\nu}\right)W_{l+1} 
+2(l+2){\bar\omega}V_{l+1} \right] \ , 
\ea 
\label{om_th_ph} 
\ee 
 
\noindent 
while the $[r\theta]$ component of Eq. (II,4.16) gives Eq. (II,4.54), 
\bea 
 0 &=& 
\frac{A_r}{(\ep+p)} \left[ 
(l-1)Q_l \left(i\delta p_{l-1}  
+ \frac{\partial_r p}{\kappa\Omega r} W_{l-1}\right) 
- (l+2)Q_{l+1} \left(i\delta p_{l+1}  
+ \frac{\partial_r p}{\kappa\Omega r} W_{l+1}\right) 
\right] \nn \\ 
&&\nn \\ 
&&+ (l-2)Q_{l-1}Q_l \left[ 
-2\partial_r\left({\bar\omega}e^{-2\nu}U_{l-2}\right) 
+\frac{(l-1)}{r^2}\partial_r\left(r^2{\bar\omega}e^{-2\nu}\right)U_{l-2} 
\right] \nn \\ 
&&\nn \\ 
&&+ Q_l \biggl[ 
	\ba[t]{l} 
	(l-1)\kappa\Omega\partial_r\left(e^{-2\nu}V_{l-1}\right) 
	-2m\partial_r\left({\bar\omega}e^{-2\nu}V_{l-1}\right) \\ 
	 \\ 
	+\frac{m(l-1)}{r^2}\partial_r\left(r^2{\bar\omega}e^{-2\nu}\right)V_{l-1} 
	+(l-1)\kappa\Omega e^{-2\nu}\left(\frac{16\pi r(\epsilon+p)}{(l-1)l} 
					-\frac{1}{r}\right)e^{2\lambda}W_{l-1} 
	\biggr] 
	\ea \nn \\ 
&&\nn \\ 
&&+\biggl[ 
	\ba[t]{l} 
	m\kappa\Omega\partial_r\left[e^{-2\nu}(h_l+U_l)\right] 
	+2\partial_r\left({\bar\omega}e^{-2\nu}U_l\right) 
	\left((l+1)Q_l^2-l Q_{l+1}^2\right) \\ 
	 \\ 
	+\frac{1}{r^2}\partial_r\left(r^2{\bar\omega}e^{-2\nu}\right)U_l 
	\left[m^2+l(l+1)\left(Q_{l+1}^2+Q_l^2-1\right)\right] 
	\biggr] 
	\ea \nn \\ 
&&\nn \\ 
&&- Q_{l+1} \biggl[ 
	\ba[t]{l} 
	(l+2)\kappa\Omega\partial_r\left(e^{-2\nu}V_{l+1}\right) 
	+2m\partial_r\left({\bar\omega}e^{-2\nu}V_{l+1}\right) \\ 
	 \\ 
	+\frac{m(l+2)}{r^2}\partial_r\left(r^2{\bar\omega}e^{-2\nu}\right)V_{l+1} 
	+(l+2)\kappa\Omega e^{-2\nu}\left(\frac{16\pi r(\epsilon+p)}{(l+1)(l+2)} 
					-\frac{1}{r}\right)e^{2\lambda}W_{l+1} 
	\biggr] 
	\ea \nn \\ 
&&\nn \\ 
&&+(l+3)Q_{l+1}Q_{l+2} \left[ 
2\partial_r\left({\bar\omega}e^{-2\nu}U_{l+2}\right) 
+\frac{(l+2)}{r^2}\partial_r\left(r^2{\bar\omega}e^{-2\nu}\right)U_{l+2} 
\right]. 
\label{om_r_th}  
\eea 
The constants $Q_l$ were defined in Paper II to be 
\be 
Q_l \equiv \left[ 
\frac{(l+m)(l-m)}{(2l-1)(2l+1)} 
\right]^{1/2}. 
\ee 
 
Notice that in writing Eq. (\ref{om_r_th}) we have retained (in the  
first line) the term containing the Schwarzschild discriminant $A_r$  
from the right hand side of Eq. (II,4.16), 
\be 
i\kappa\Omega e^{-\nu} \Delta \omega_{\alpha\beta} =  
\frac{2}{n} A_r \nabla_{\left[\alpha\right.} r  
\nabla_{\!\left.\beta\right]} \Delta p. 
\label{Del_om_NI} 
\ee 
We have done this in order to emphasize the difference between the  
barotropic and nonbarotropic cases.  To find the rotational modes to  
leading order in $\Omega$ we need a complete set of equations involving  
{\it only} our $O(1)$ variables.  If $A_r$ is not identically zero, the  
retained term in (\ref{om_r_th}) brings about a coupling between these  
variables and the $O(\Omega)$ variable $\delta p_l$.   
In a barotropic star, $A_r$ is identically zero, so this coupling between  
$O(1)$ and $O(\Omega)$ variables vanishes.  In this case, the five equations 
(\ref{GR:sph_H1})-(\ref{om_r_th}) do, indeed, involve only the $O(1)$  
variables $h_l$, $U_l$, $W_l$, $V_l$ and $H_{1,l}$.  These equations, 
therefore, comprise a complete set and fully determine our normal mode 
eigenvalue problem. 
In a nonbarotropic star, however, $A_r\neq 0$, and the coupling between  
$O(1)$ and $O(\Omega)$ variables in Eq. (\ref{om_r_th}) does not vanish.  
In this case, the equations (\ref{GR:sph_H1})-(\ref{om_r_th}) do not  
involve {\it only} the variables $h_l$, $U_l$, $W_l$, $V_l$ and $H_{1,l}$. 
There are then two options: The first is reminiscent of the situation 
for r-modes of non-barotropic Newtonian stars. The problem would become well
posed  
if we extended the analysis to one order higher in $\Omega$ \cite{ak} and thus 
determined also $\delta p_l$ etcetera. The second possibility  
would be to obtain a well-defined eigenvalue problem 
by assuming the non-barotropic ordering (\ref{ordering}) in which only the axial
variables  
$U_l$ and $h_l$ are $O(1)$.  With this choice, one obtains an  
eigenvalue problem from Eqs. (\ref{GR:sph_h_0''}) and (\ref{om_th_ph})  
- dropping $W_l$ and $V_l$ from the latter as $O(\Omega^2)$ quantities.   
These equations can be reexpressed as Eq. (II,4.24): Kojima's \cite{koj} 
eigenvalue equation governing $h_l$, 
\be 
(\alpha -\tilde{\omega}) 
\left\{  e^{\nu-\lambda} {d\over dr} \left[ e^{-\nu-\lambda}  
{dh_l \over dr} \right] - \left[{l(l+1) \over r^2} - {4M\over r^3} 
+8\pi(\ep+p) \right]  h_l \right\}  + 
16\pi(\ep+p) \alpha h_l = 0 \ , 
\label{singeq2} 
\ee 
where $\tilde{\omega}= \bar{\omega}/\Omega$ and $\alpha=l(l+1)\kappa/2m$ 
and where $U_l$ is determined (once the eigenvalue, $\alpha$, and $h_l$ are  
known) by Eq. (\ref{om_th_ph}), which becomes, 
\be 
U_l = \frac{\alpha}{(\tilde\om-\alpha)} h_l. 
\label{Usingeq} 
\ee 
 
{\em To summarize}:  
The eigenvalue problem governing the rotational modes of a  
{\it barotropic} relativistic  
star to lowest order in $\Omega$ is determined by the set  
of equations (\ref{GR:sph_H1})-(\ref{om_r_th}) for the variables $h_l$,  
$U_l$, $W_l$, $V_l$ and $H_{1,l}$, which are all assumed to be zeroth 
order in $\Omega$. By contrast, the eigenvalue problem governing the  
rotational modes of a {\it non-barotropic} star to lowest order in  
$\Omega$ is determined by Eqs. (\ref{singeq2}) and (\ref{Usingeq})  
for the variables  $h_l$ and $U_l$, which alone are assumed to be  
zeroth order in $\Omega$. 
 
To complete the specification of our eigenvalue problem we must impose 
appropriate boundary conditions.  We require our variables to be regular 
everywhere in the spacetime, which implies that they vanish at the  
origin, $r=0$. The fluid variables $U_l(r)$ and $V_l(r)$ are otherwise 
unconstrained - apart from the fact that they vanish outside the star. 
In Paper II, we derived the boundary condition (II,4.64) on the fluid 
variable $W_l(r)$ at the surface of the star, $r=R$, 
\be 
W_l(R)=0 \ \ \ \ \mbox{(all $l$)}. 
\label{bc_on_W} 
\ee 
This condition and Eq. (\ref{GR:sph_H1}) imply that the metric variable 
$H_{1,l}(r)$ vanishes at the surface, and in the exterior, of the star. 
By inspection, it is clear that all but one of the perturbation equations 
vanish in the vacuum exterior to the star ($r>R$).  The one surviving  
equation is that governing the metric variable $h_l(r)$, i.e.,  
Eq. (\ref{GR:sph_h_0''}) for barotropic stars or Eq. (\ref{singeq2}) 
for nonbarotropic stars.  In the near zone, where these equations apply,  
they both reduce to the same equation outside the star: Eq.~(II,4.66), or,  
\be 
(1-\frac{2M}{r}) \frac{d^2 h_l}{dr^2} - \left[ \frac{l(l+1)}{r^2}  
	- \frac{4M}{r^3} \right] h_l  = 0, 
\label{h_l''_ext} 
\ee 
where $M$ is the total gravitational mass of the star.   
This equation has a regular singular point at $r=\infty$, so it has  
at least one regular series expansion about this point.  
The other, linearly independent, solution turns out to be singular at  
$r=\infty$, growing like $r^{l+1}$. As we will see in  
Sect.~\ref{sect:dEdt} we may ignore this second solution in the near  
zone because it is of order $(\sigma r)^{2l+1}$ relative to the  
nonsingular solution.  To find the mode eigenvalues and eigenfunctions  
to lowest order in $\Omega$ we therefore need only the regular solution.  
As discussed in Paper II the regular solution can be immediately written  
down as Eq. (II,4.67), or 
\be 
h_l(r) = \sum_{s=0}^\infty {\hat h}_{l,s}  
\left(\frac{R}{r}\right)^{l+s}, 
\label{h_ext} 
\ee 
where the coefficients ${\hat h}_{l,s}$ are given by the recursion  
relation (II,4.68), 
\be 
{\hat h}_{l,s} = \left(\frac{2M}{R}\right)  
\frac{(l+s-2)(l+s+1)}{s(2l+s+1)} {\hat h}_{l,s-1} 
\label{h_ext_soln} 
\ee 
or, equivalently, by, 
\be 
{\hat h}_{l,s} =  
\frac{(l+s-2)!(l+s+1)!(2l+1)!}{s!(l-2)!(l+1)!(2l+s+1)!}  
\left(\frac{2M}{R}\right)^s  {\hat h}_{l,0} 
\ee 
with ${\hat h}_{l,0}$ an arbitrary normalization constant.   
This known exterior solution must be matched   
to the interior solution for $h_l(r)$ at the surface of the star. 
This provides a boundary condition 
on the interior solution.  
We require that the solutions be continuous at the surface, 
\be 
\lim_{\varepsilon\rightarrow 0}  \left[ 
h_l(R-\varepsilon) - h_l(R+\varepsilon) \right] = 0, 
\label{cont_cond} 
\ee 
for all $l$ (which fixes the normalization constant ${\hat h}_{l,0}$),  
and that the Wronskian of the interior and exterior solutions vanish  
at $r=R$, i.e. that 
\be 
\lim_{\varepsilon\rightarrow 0} \left[ 
h_l(R-\varepsilon) h'_l(R+\varepsilon)  
- h'_l(R-\varepsilon) h_l(R+\varepsilon) 
\right] = 0, 
\label{match_cond} 
\ee 
for all $l$. 
 
Finally, we note that since we are working in linearized perturbation 
theory there is a scale invariance to the equations.  If  
$(\xi^\alpha, h_{\alpha\beta})$ is a solution to the perturbation 
equations then $(S\xi^\alpha, Sh_{\alpha\beta})$ is also a  
solution, for constant $S$.  We will sometimes find it convenient to  
impose the following normalization condition in addition to the  
boundary and matching conditions just discussed: 
\be 
\left\{ \begin{array}{ll} 
U_m(r=R) = 1 & \mbox{for axial-led inertial modes and r-modes,} \\ 
U_{m+1}(r=R) = 1 & \mbox{for polar-led inertial modes.}  
\end{array} \right. 
\label{norm_cond} 
\ee

 
\section{Mode Frequencies and Eigenfunctions} 
\label{Sect:freqsandfuncs} 
 
From the equations in the previous section it is clear that the  
task of determining the relativistic analogues of the Newtonian r-modes 
is different depending on whether the star is barotropic or non-barotropic.  
Each problem presents it's own computational challenge.  
For non-barotropic stars one must deal with the fact that Kojima's equation 
(\ref{singeq2}) represents a singular eigenvalue problem.  
The barotropic case is conceptually easier because one does not have to 
worry about singularities. On the other hand, all inertial modes of a barotropic
star  
will have a hybrid nature which complicates the numerical solution of the
eigenvalue problem 
considerably. In this paper we focus our attention on barotropic stars 
since: i) one would expect the  
inertial modes of a more complex stellar model (with internal stratification) 
to be similar to those of a barotropic star for 
sufficiently rapid rotation, ii)  the nonbarotropic case 
has already been discussed by several authors \cite{yosh,kr01,regul}.  
 
We have used the spectral method described in Appendix~\ref{appendix1} to  
solve numerically for a sample of rotationally  
restored modes of slowly rotating, fully relativistic barotropes.  
The relevant set of equations for this problem is  
Eqs. (\ref{GR:sph_H1})-(\ref{om_r_th}), which comprise a system of ordinary  
differential equations for the variables $U_l(r)$, $V_l(r)$, $W_l(r)$  
$H_{1,l}(r)$ and $h_l(r)$ (for all allowed $l$).  Together with the  
boundary and matching conditions at the surface of the star, these  
equations form a nonlinear eigenvalue problem for the parameter $\kappa$,  
the dimensionless mode frequency in the rotating frame.   
[In practice, because the metric variable $H_{1,l}$ is related  
algebraically to the fluid variable $W_l$ by Eq. (\ref{GR:sph_H1}), we 
eliminate $H_{1,l}$ from the other equations and solve the system 
(\ref{GR:sph_V})-(\ref{om_r_th}).] 
 
For simplicity, we have restricted our study to relativistic polytropes 
even though it would be straightforward to generalise our calculation  
to tabulated realistic equations of state. This is an important 
step that should eventually be taken, but we feel that we should  
first try to understand the overall effect that general relativity  
has on the inertial modes of a compact star.  
 
In Newtonian barotropic stars there remained a large set of modes that 
were purely axial to lowest order in $\Omega$: the r-modes with spherical  
harmonic indices $l=m$.  The $l=m=2$ r-mode is the one expected to  
dominate the gravitational wave-driven instability of sufficiently hot  
and rapidly rotating neutron stars \cite{lom98,aks99}.   
We showed in Paper II, however, that relativistic barotropes do not 
admit such modes. Pure r-modes with $l=m\geq 2$ are not allowed by  
the perturbation equations (\ref{GR:sph_h_0''})-(\ref{om_r_th}). The
corresponding modes 
are instead axial-led hybrid modes.  We have previously solved explicitly  
for these important hybrid modes to first post-Newtonian order 
in uniform density stars [cf. Eqs. (II,5.33)-(II,5.40)].  
We now report on a more general numerical study of these and other  
hybrid inertial modes in fully relativistic stars. 
 
Apart from the replacement of the Newtonian r-modes with hybrids,  
the structure of the inertial mode spectrum in relativistic stars 
appears to be identical to that in Newtonian stars.  That is, there  
is a one-to-one correspondence between the rotationally restored modes  
of a Newtonian star and those of the corresponding relativistic model. 
One can make a more formal comparison by constructing sequences of  
relativistic models of increasing compactness, $M/R$, a suitable measure 
of the importance of relativistic effects.  A relativistic star with small  
$M/R$ agrees well in all of its physical properties with the Newtonian  
model having the same EOS and central density.  In each star along the  
relativistic sequence we search for inertial modes using the method  
described in Appendix~\ref{appendix1}.  We find that for each of the Newtonian 
modes considered, there exists a family of relativistic modes parametrized  
by $M/R$ that approaches the Newtonian mode as $M/R\rightarrow 0$. 
 
\begin{figure}[h] 
\centerline{\epsfysize=6cm \epsfbox{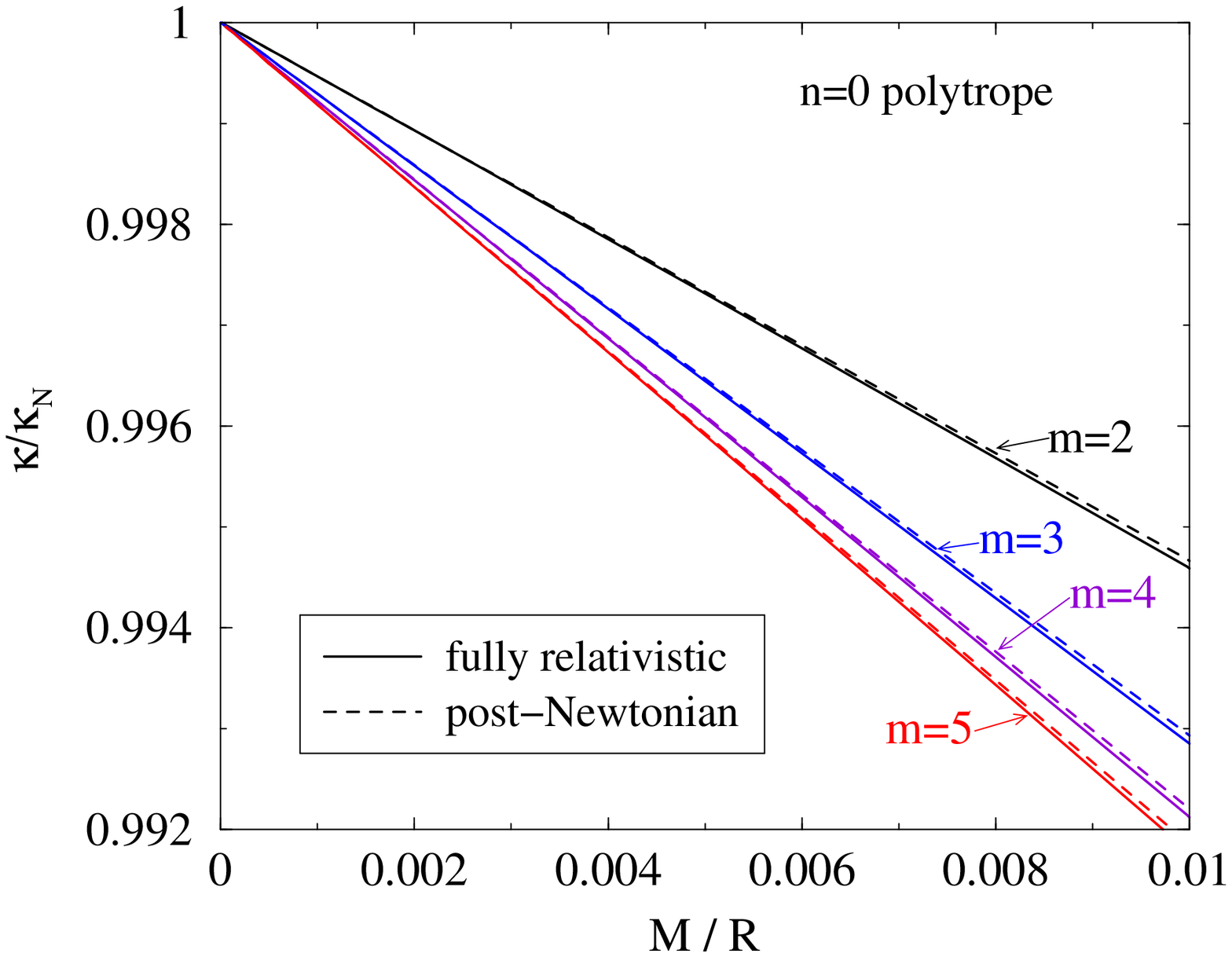} \epsfysize=6cm \epsfbox{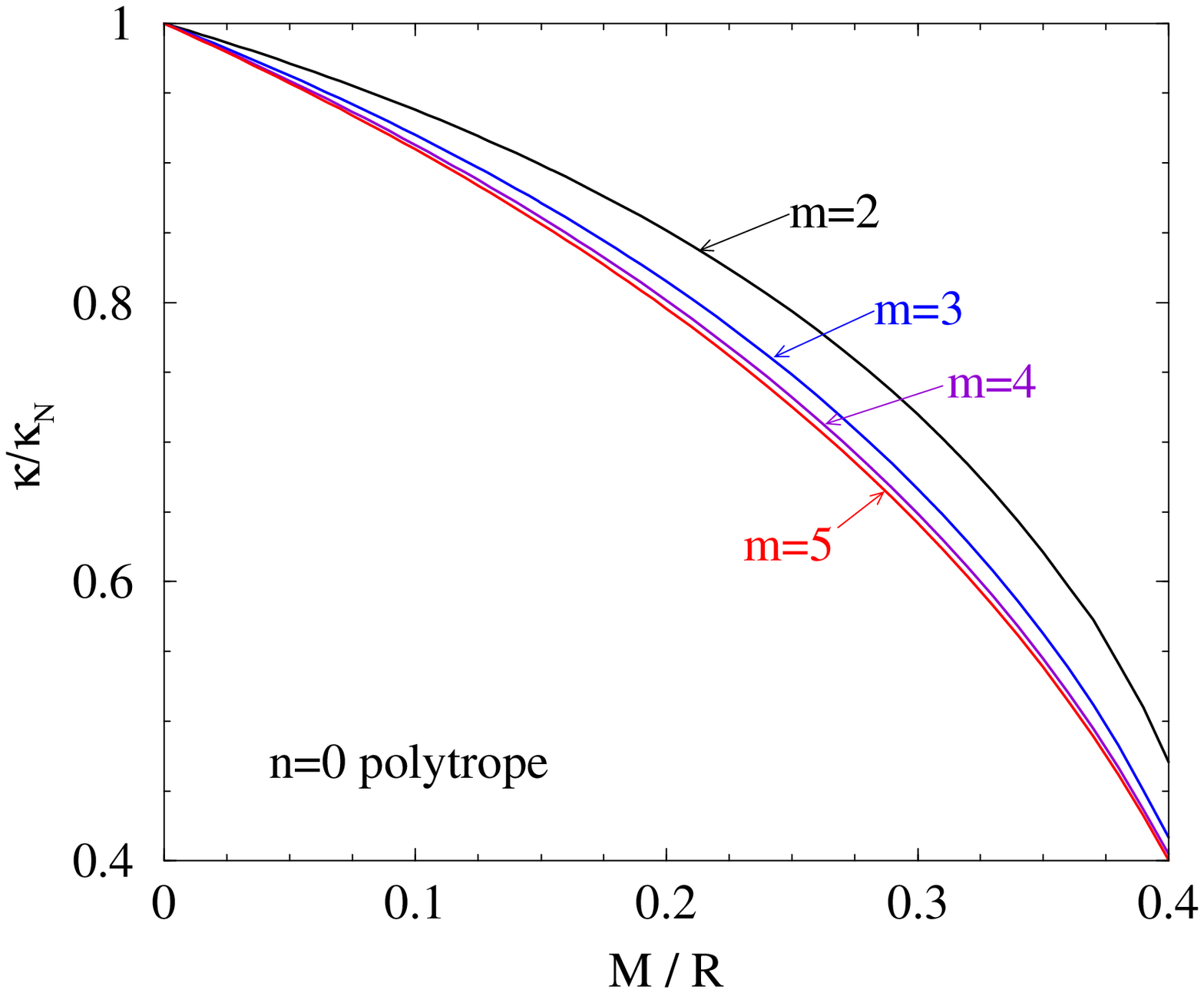}}
\caption{Left panel: Shift in the normalized frequencies,  
$\kappa/\kappa_{\mbox{\tiny N}}$, of the axial-led hybrid modes whose  
Newtonian counterparts are the $l=m=2,3,4$~and~5 r-modes.  These  
frequencies have been computed for a uniform density model  
($n=0$ polytrope) in the small $M/R$ regime.  The frequencies computed  
using our fully relativistic code (solid curves) agree well with the  
post-Newtonian solution (dashed lines) and deviate, as expected, by a  
correction of order $(M/R)^2$. Right panel: The corresponding mode 
frequencies in the strongly relativistic (large $M/R$) regime.} 
\label{f1} 
\end{figure}

To test the accuracy of our numerical code we compared its results with  
our post-Newtonian solution from Paper II in the small $M/R$ regime  
(see Fig.~\ref{f1}).  The solution (II,5.33)-(II,5.40) gives explicitly those  
axial-led inertial modes of a relativistic uniform density star that correspond 
to the $l=m$ Newtonian r-modes.  For these modes the dimensionless comoving  
frequency in the Newtonian star is simply  
\be 
\kappa_{\mbox{\tiny{N}}} = \frac{2}{(m+1)}, 
\ee 
while the post-Newtonian calculation gives Eq. (II,5.33), or, 
\be 
\kappa_{\mbox{\tiny 1pN}} = \frac{2}{(m+1)}\left[1 
-\frac{8(m-1)(2m+11)}{5(2m+1)(2m+5)}\left(\frac{M}{R}\right) 
\right]. 
\label{GR:ex_sol:freq} 
\ee 
Fig.~\ref{f1} shows that the numerically computed eigenvalues agree well  
with the post-Newtonian solution; the differences being of order $(M/R)^2$ 
as expected.  The numerically computed eigenfunctions also agree  well 
with the post-Newtonian eigenfunctions (II,5.34)-(II,5.40). 
This agreement with our analytic solution gives us confidence  
that our code is able to find the relativistic modes.  Thus, 
we may now explore the fully relativistic regime and, in particular, consider 
modes for which we have not worked out a post-Newtonian solution.  
Fig.~\ref{f1} also shows the numerically computed eigenvalues  
for highly relativistic uniform density models.  
We have used the Newtonian frequency $\kappa_{\rm N}$  
to normalize the results. This makes it easy to see that all modes  
clearly approach their Newtonian values as $M/R\rightarrow 0$. 
Furthermore, the results show that the mode-frequency tends to decrease  
as the star becomes more compact.   
It is anticipated that  
general relativity will have this effect \cite{laf1}. The 
gravitational redshift will tend to decrease the fluid oscillation  
frequencies measured by a distant inertial observer. Also, because 
these modes are rotationally restored they will be affected by the 
dragging of inertial frames induced by the star's rotation. 
Specifically, since the Coriolis force is determined by 
the fluid's  
angular velocity relative to that of the local inertial frame 
$\bar\om(r)= \Omega -\omega(r)$ it decreases, 
and the modes oscillate less rapidly, as the dragging of inertial  
frames becomes more pronounced. 
This is, of course, not too surprising given \ref{GR:ex_sol:freq}. 
 
\begin{figure}[h] 
\centerline{\epsfysize=7cm \epsfbox{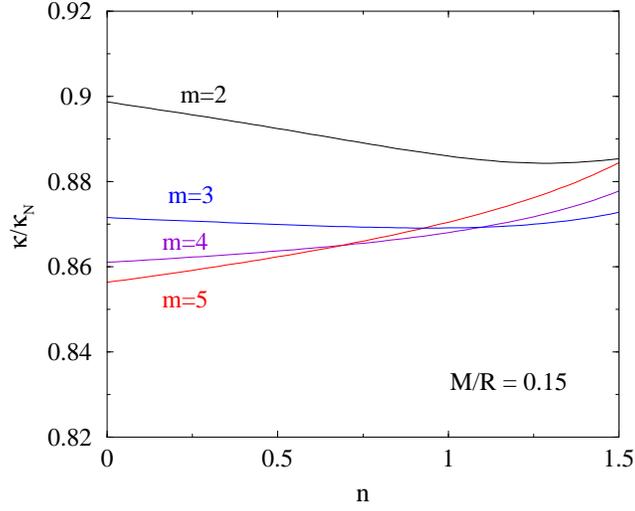} } 
\caption{Shift in the frequencies of the modes shown in  
Fig.~\ref{f1} for a sequence of stars of varying polytropic  
index, $n$. Each model in the sequence is chosen to have the same compactness, 
$M/R=0.15$. } 
\label{f3} 
\end{figure}

 
Next we consider how the relativistic analogues to the Newtonian r-modes 
are affected by changes in the stiffness of the equation of state.  
We can easily do this by determining the modes for a varying  
polytropic index, $n$, at a given compactness.  
Fig.~\ref{f3} provides such results for  $M/R=0.15$ which would be a typical 
value for realistic neutron stars.  
From this data we see that the mode frequencies differ 
from the Newtonian values by $10-15\%$, depending on the 
stiffness of the equation of state.    
In order to confirm that these results are typical, we have carried out  
calculations for a much larger set of modes (inluding also  
polar-led inertial modes). Results for all modes whose  
Newtonian frequencies 
$\kappa_{\mbox{\tiny N}}$ were listed in Tables 5 and 6 of Paper~I,  
show shifts in the frequency that accord well with the results 
in Fig.~\ref{f3}.

 
We now turn to a discussion of the eigenfunctions.  Because the  
Newtonian studies have found the $l=m$ r-modes to be the most  
unstable to gravitational radiation-reaction, we would like to  
understand the nature of these modes in relativistic stars.  
As we have already mentioned, these modes become axial-led hybrids 
in relativistic barotropic models. 
In terms of the spherical harmonic expansion of the fluid  
displacement, Eq.~(\ref{xi_exp}), the statement that the 
Newtonian modes are pure r-modes with $l=m$ means that only  
the coefficient $U_m(r)$ is nonzero (to lowest order in $\Omega$).   
It has the form 
\be 
U_m(r) = \left(\frac{r}{R}\right)^{m+1}. 
\ee 
[We have normalized the mode according to Eq.~(\ref{norm_cond}) 
so that $U_m(r)=1$ at the surface of the star, $r=R$.] 
In a relativistic barotrope, on the other hand,  other  
coefficients in Eq.~(\ref{xi_exp}) will be nonzero to lowest order in  
$\Omega$. In addition we can examine the nonzero functions $h_l(r)$  
($\equiv h_{0,l}$) and $H_{1,l}(r)$ in Eq. (\ref{h_components}), the  
spherical harmonic expansion of the perturbed metric.   
 
\begin{figure}[h] 
\centerline{\epsfysize=7cm \epsfbox{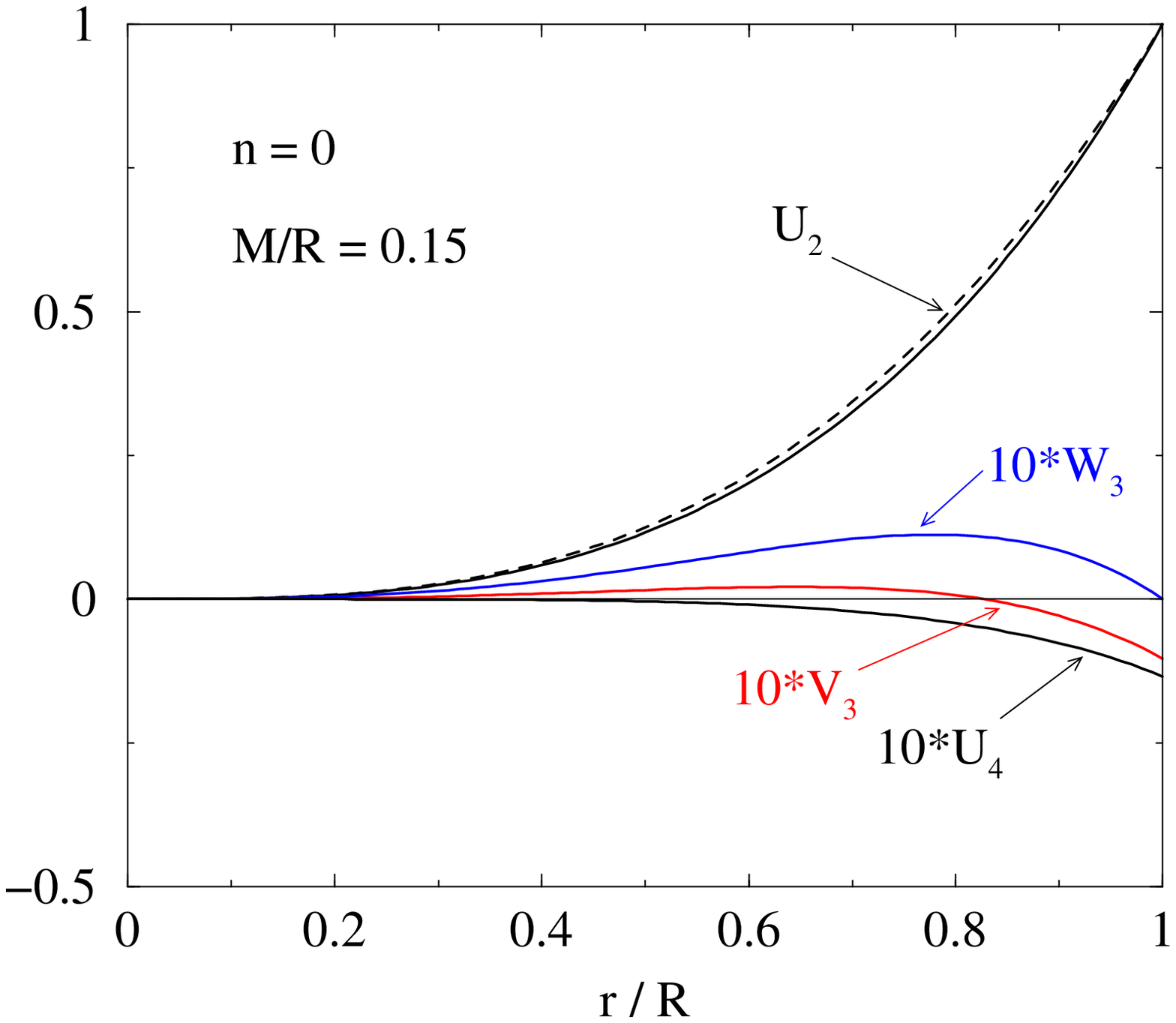} 
\epsfysize=7cm \epsfbox{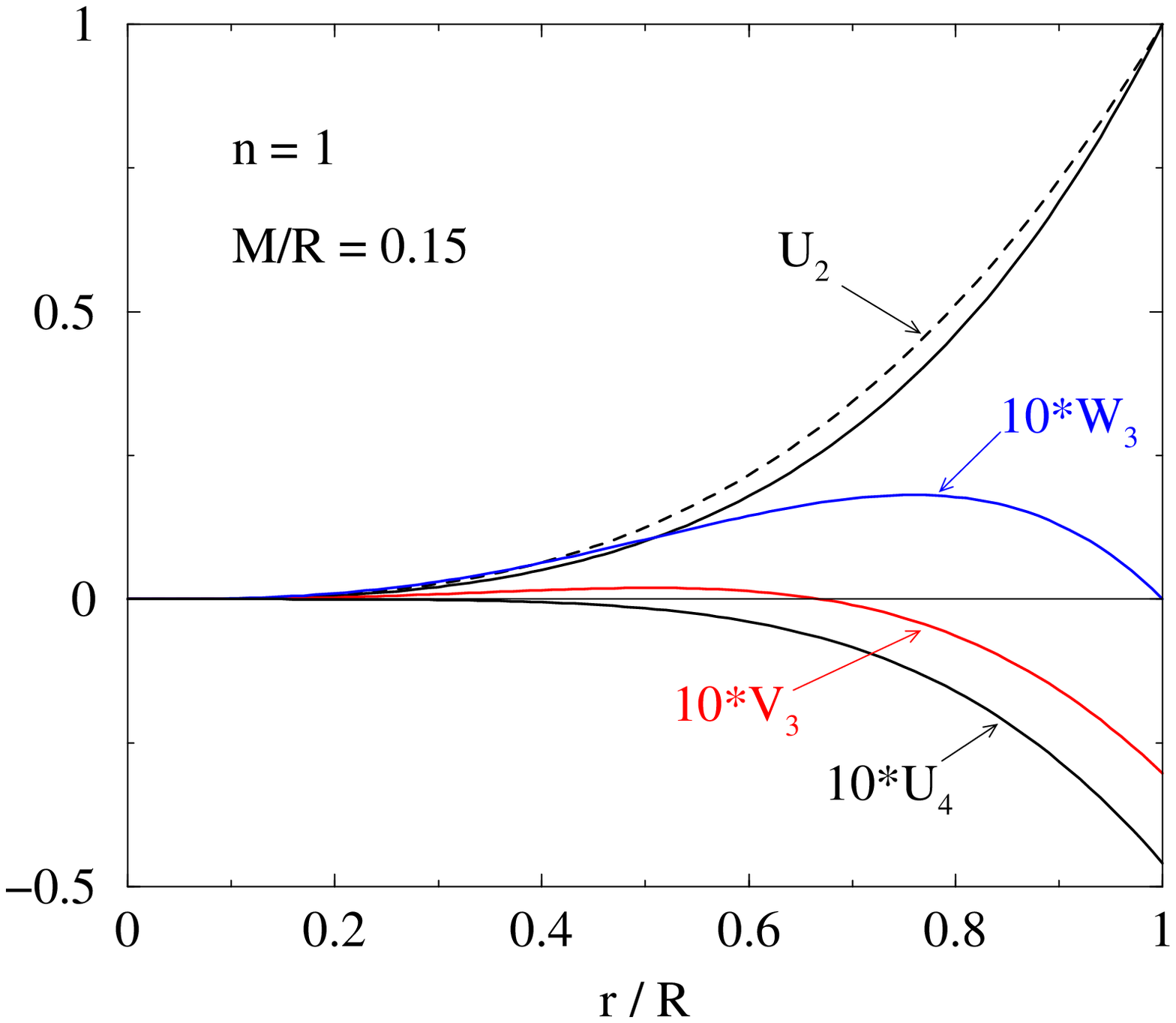}} 
\caption{The relativistic axial-led inertial mode (solid curves) whose  
Newtonian counterpart is the $l=m=2$ r-mode (dashed curve). We show the  
fluid functions $U_l(r)$, $V_l(r)$ and $W_l(r)$ for $l\leq 4$ in a uniform  
density model (left panel) and an $n=1$ polytrope (right panel); each with 
compactness $M/R=0.15$.  The vertical scale is set by the normalization  
of $U_2(R)=1$  [cf. Eq. (\ref{norm_cond})];  
however, the other functions have been scaled up by a factor of 10 to make  
them visible in the figure. The fluid functions with $l>4$ are of order  
$0.05\%$ or smaller and are not shown.}   
\label{f5} 
\end{figure}

\begin{figure}[h] 
\centerline{\epsfysize=7cm \epsfbox{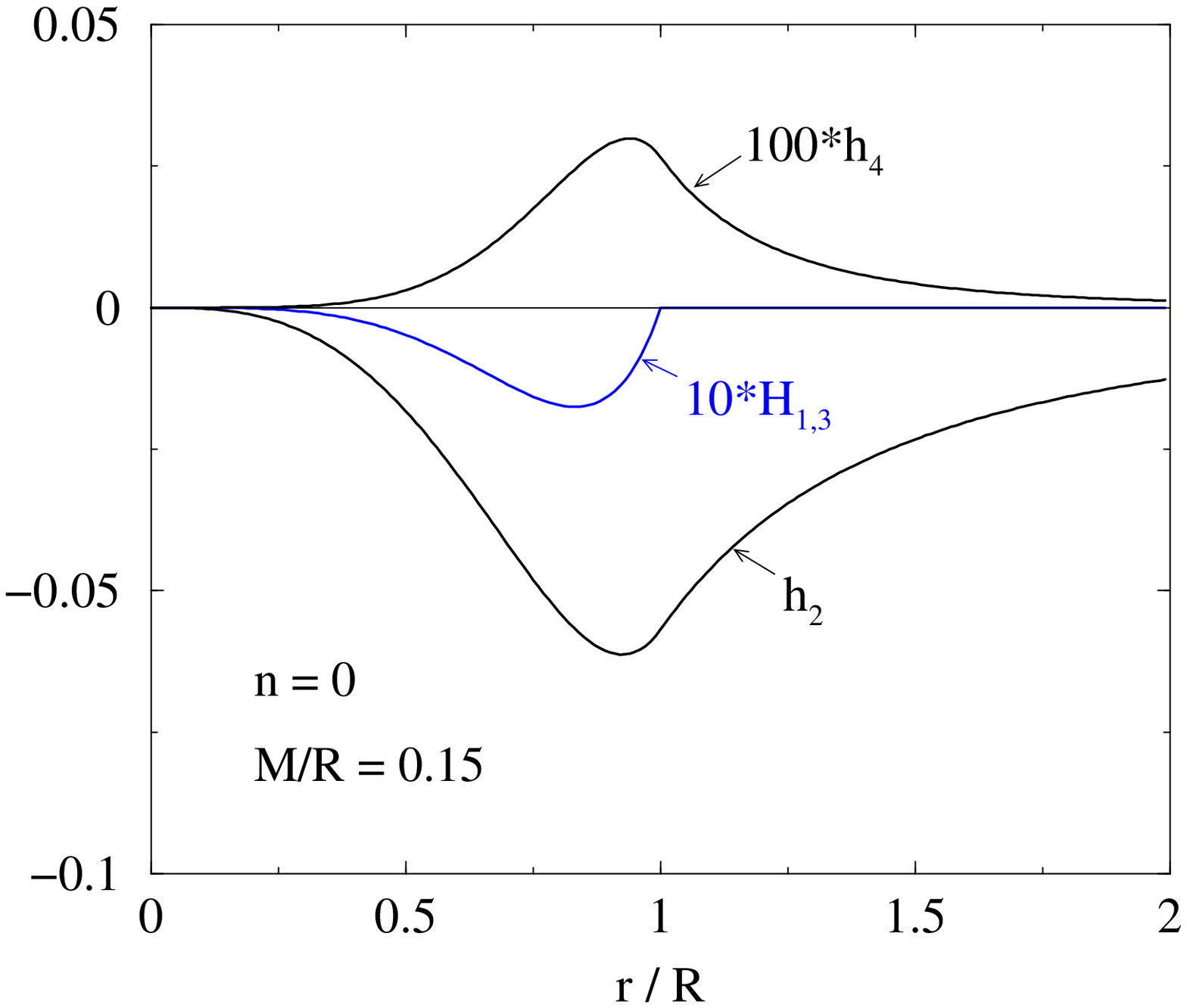} 
\epsfysize=7cm \epsfbox{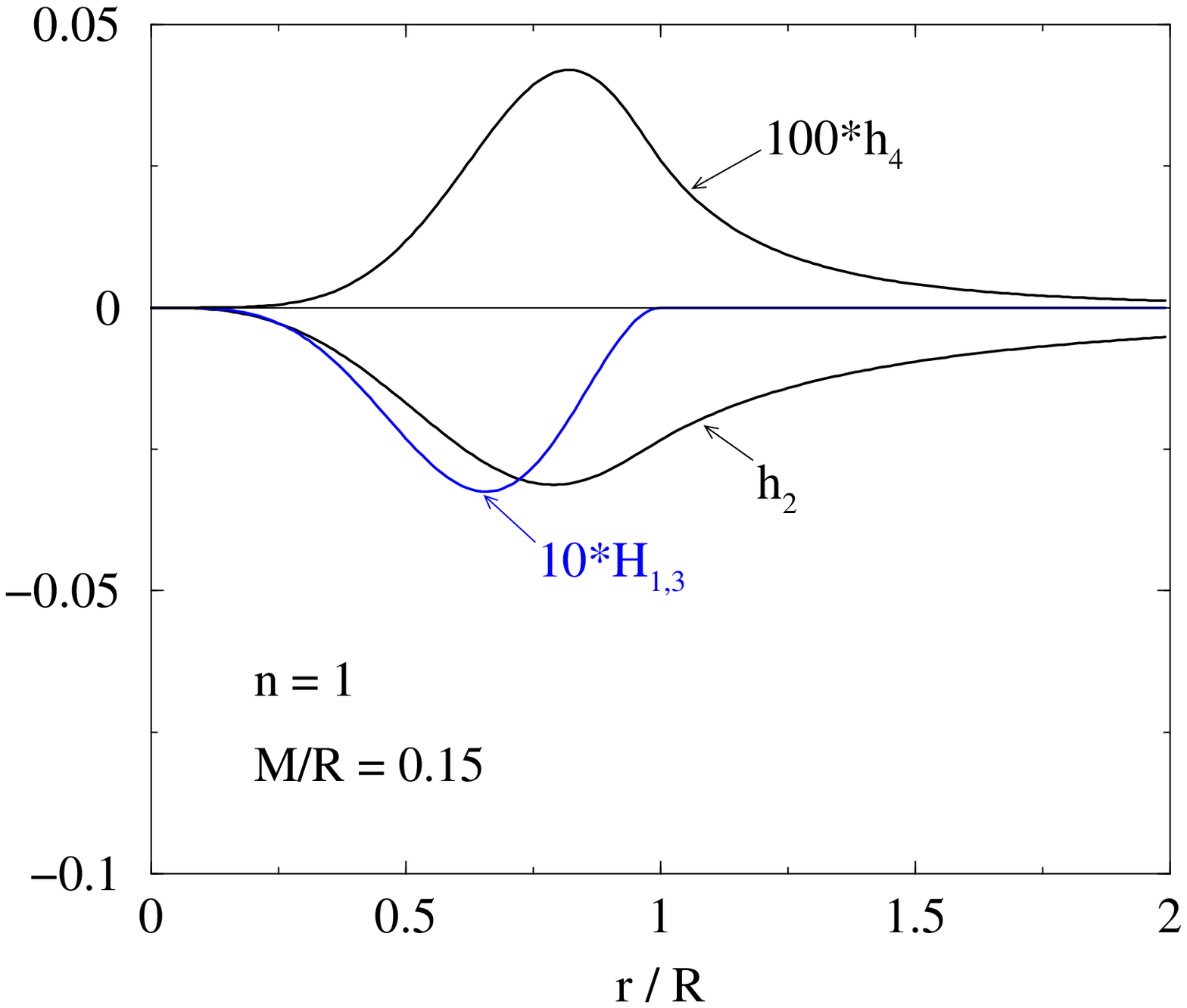}} 
\caption{The metric functions $h_l(r)$ and $H_{1,l}(r)$ for $l\leq 4$ for  
the same mode as in Fig~\ref{f5}. The vertical scale is again set by the  
overall normalization of $U_2(R)=1$, as in Fig~\ref{f5}. 
The metric functions with $l=3$ and 4 have been scaled up to make them  
visible in the figure, while those with $l>4$ are of order $0.005\%$ or  
smaller and are not shown.} 
\label{f6} 
\end{figure}

Figs.~\ref{f5} and \ref{f6} show some of these nonvanishing  
coefficients for the axial-led inertial mode whose Newtonian counterpart  
is the $l=m=2$ r-mode.  The mode is shown for two relativistic models:  
a uniform density star ($n=0$ polytrope) and an $n=1$ polytrope; both  
with compactness $M/R=0.15$.  At this compactness, the mode of the  
$n=0$ model has eigenvalue $\kappa = 0.5991$ and the mode of the $n=1$  
model has eigenvalue $\kappa = 0.5907$.  [In Newtonian gravity the  
eigenvalue is $\kappa_{\mbox{\tiny{N}}}=2/(m+1)=2/3$.] 
Fig.~\ref{f5} shows the fluid displacement functions $U_l(r)$, $W_l(r)$,  
and $V_l(r)$ for $l\leq 4$ as well as the Newtonian r-mode function 
$U_2(r)=(r/R)^3$ (dashed curve).  Meanwhile,  Fig.~\ref{f6}  
shows the metric  
functions $h_l(r)$ and $H_{1,l}(r)$ for $l\leq 4$.  The amplitudes of  
all of these functions are determined by the normalization condition,  
Eq.~(\ref{norm_cond}), and reveal that the relativistic corrections  
to the Newtonian r-mode eigenfunction are only of the order of a few  
percent even for these highly relativistic models.  The fact that  
$h_2(r)$ dominates the perturbed metric reveals directly that this  
mode couples most strongly to current quadrupole radiation, cf. 
estimates based on the Newtonian mode results \cite{lom98,aks99}. 
 
 
 
\begin{figure}[h] 
\centerline{\epsfysize=7cm \epsfbox{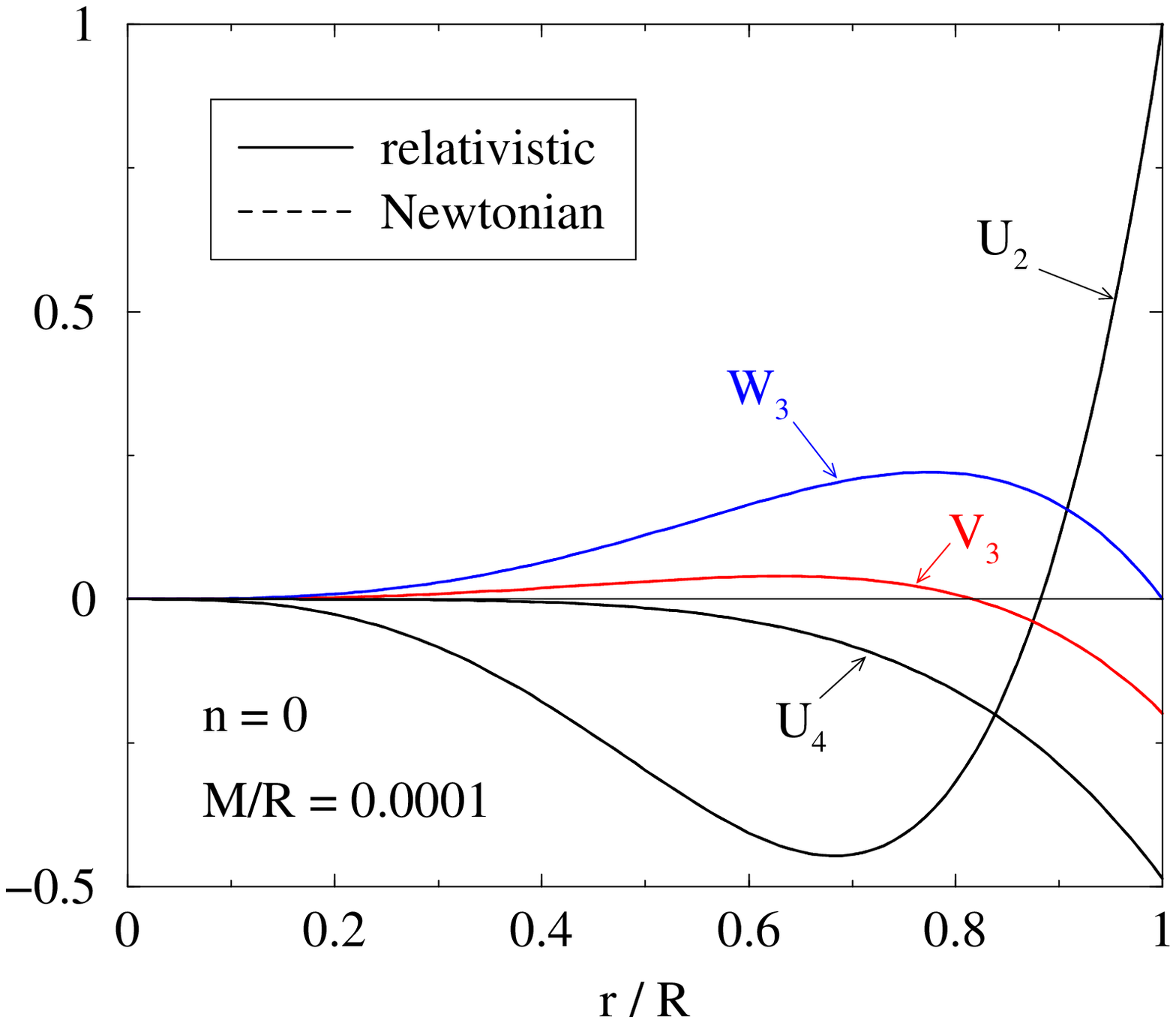} 
\epsfysize=7cm \epsfbox{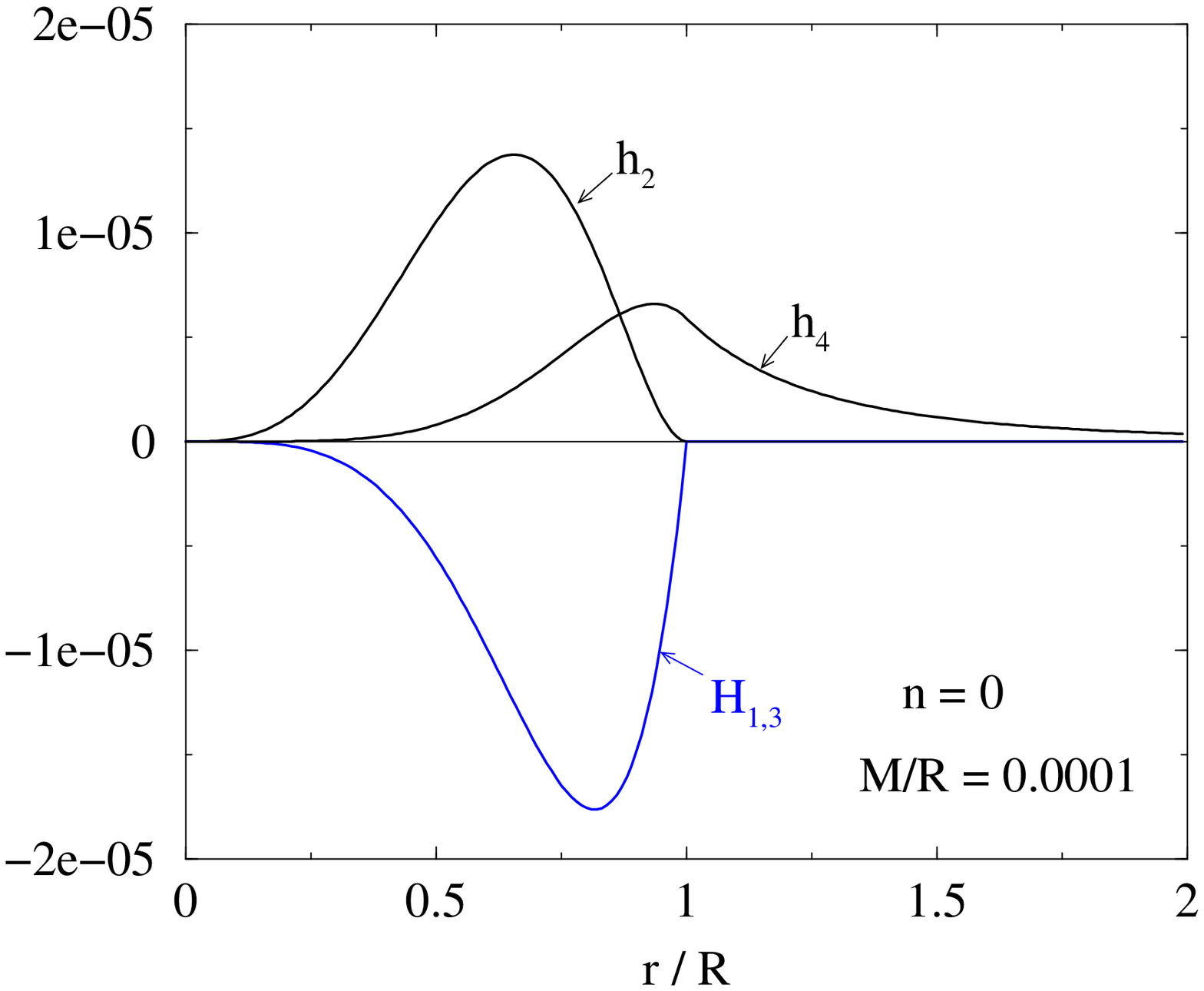}} 
\caption{An axial-led hybrid mode of a relativistic uniform density  
barotrope (solid curves). Also shown (dashed curves) is the Newtonian  
counterpart to this mode: the $m=2$ axial-led hybrid with frequency  
$\kappa_{\mbox{\tiny{N}}}=0.4669$ (see Paper I).  The Newtonian and  
relativistic fluid functions are indistinguishable because the mode is  
shown in the weakly relativistic regime, for a star with compactness
$M/R=10^{-4}$. 
The left panel shows the fluid functions $U_l(r)$, $V_l(r)$ and $W_l(r)$  
while the right panel shows the metric functions $h_l(r)$ and $H_{1,l}(r)$,  
all for $l\leq 4$.  All of the functions are shown to scale; thus the scale 
of the right panel reveals the size of the relativistic corrections at this 
nearly Newtonian compactness.  The functions with $l>4$ are smaller than  
those shown (in both panels) by a factor of order $10^{-5}$ or smaller 
and are not shown.} 
\label{f7} 
\end{figure}

\begin{figure}[h] 
\centerline{\epsfysize=6cm \epsfbox{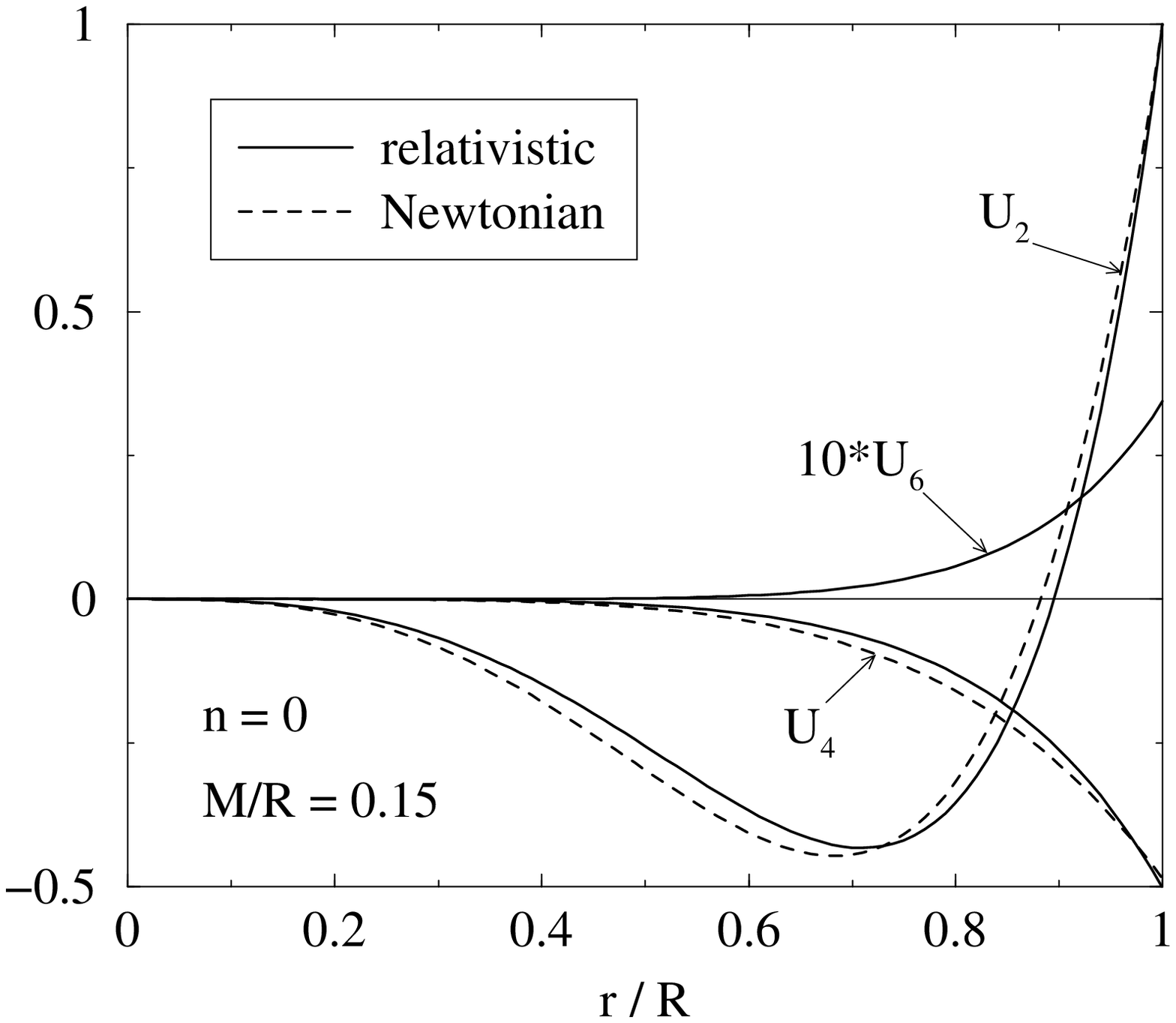} \epsfysize=6cm \epsfbox{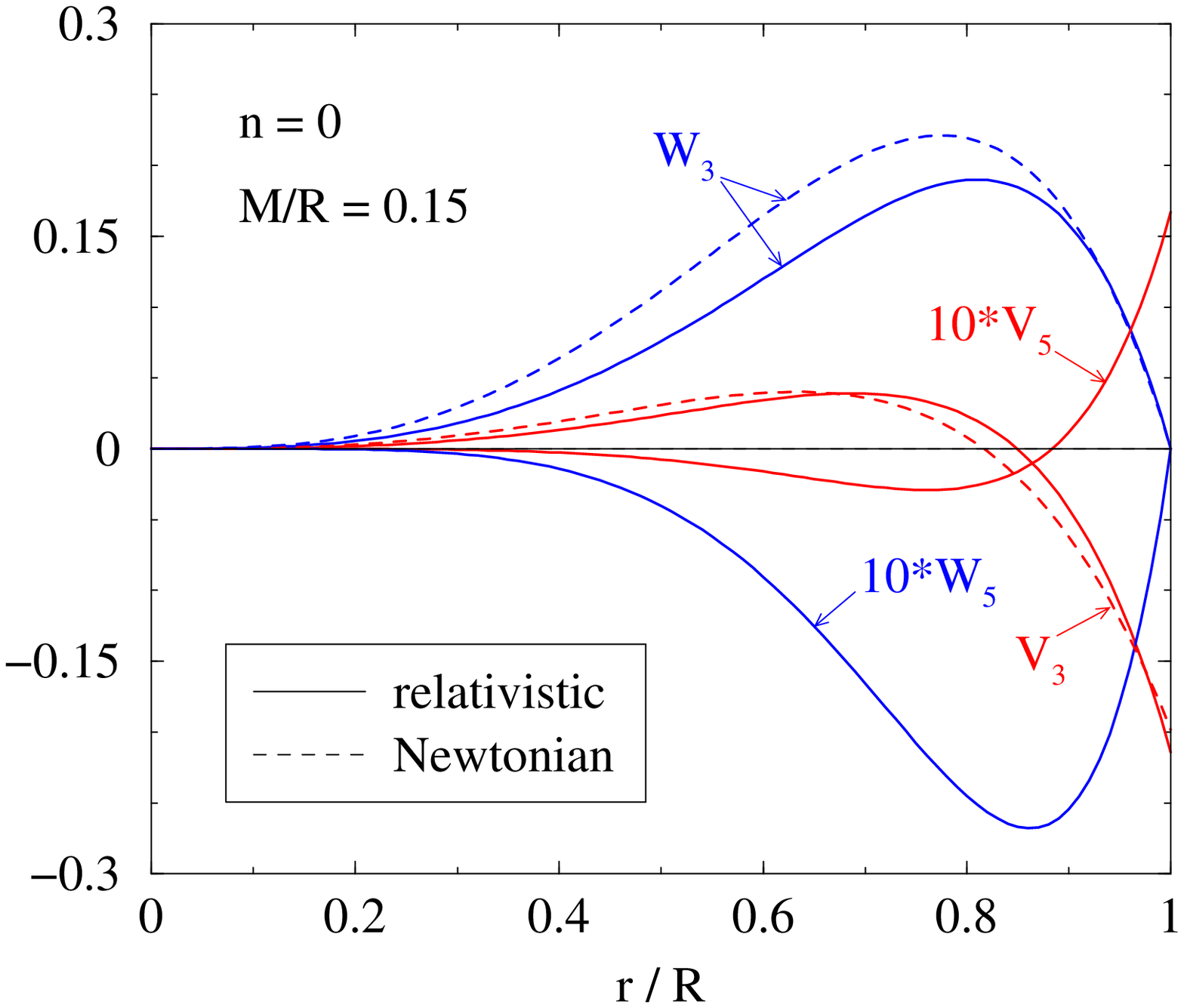}} 

\centerline{\epsfysize=6cm \epsfbox{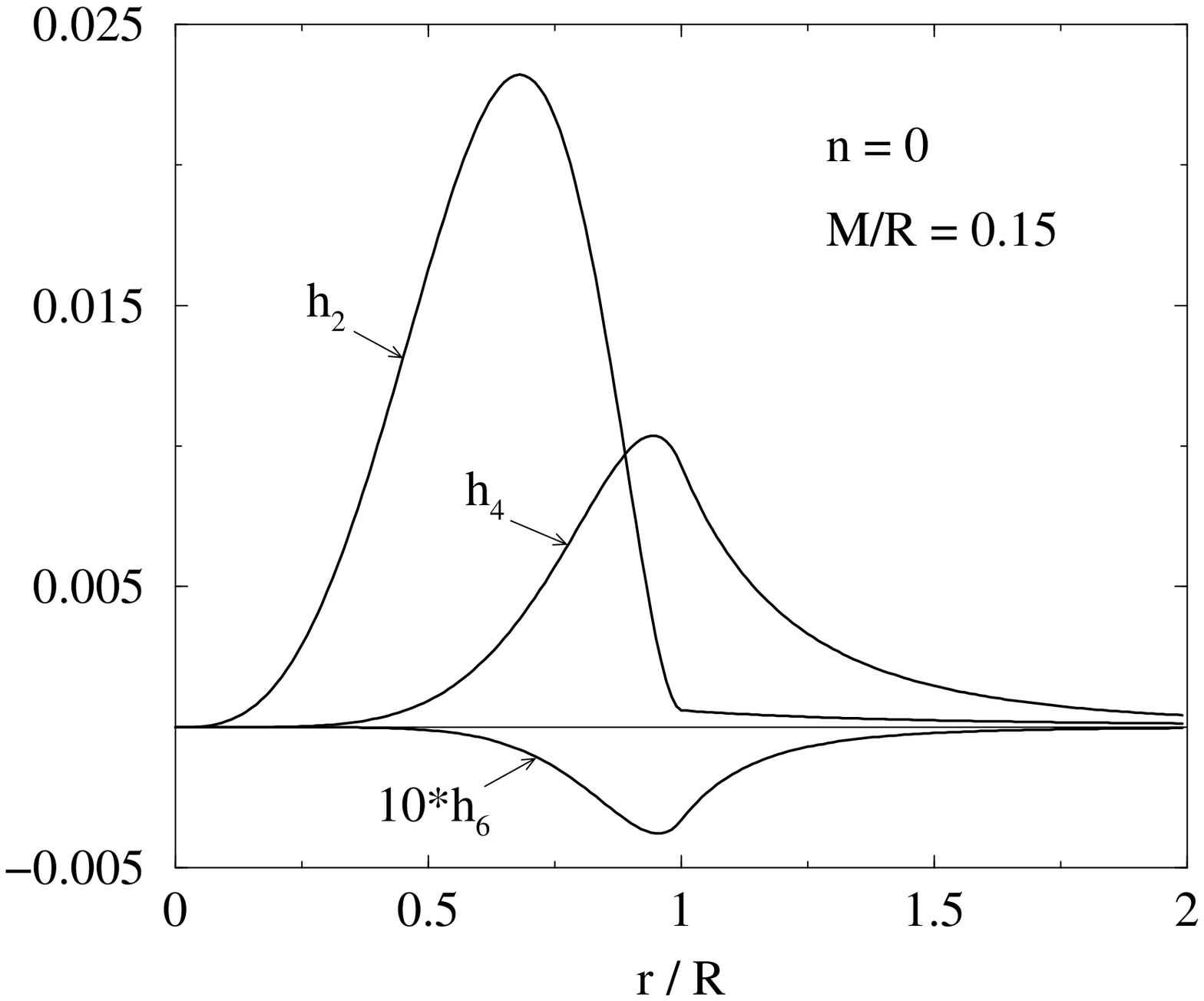} \epsfysize=6cm \epsfbox{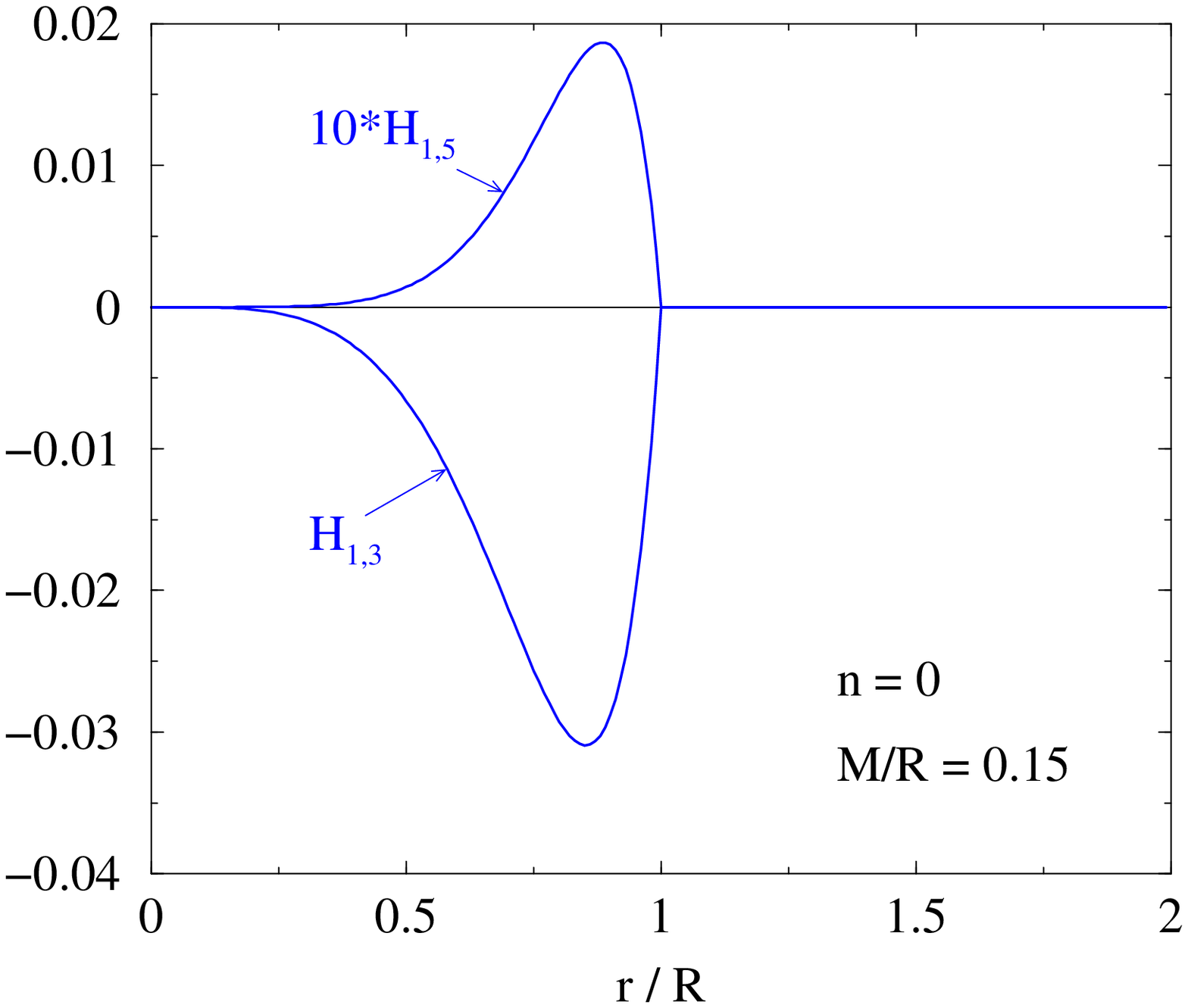}}

\caption{ The same mode as in Fig.~\ref{f7}, but for a strongly  
relativistic uniform density star ($n=0$) with compactness $M/R=0.15$.  
Upper left panel: The  
axial-parity fluid functions $U_l(r)$ for $l\leq 6$.  The Newtonian 
functions (dashed curves) are also shown for comparison.  
Upper right panel: The polar-parity fluid 
functions $W_l(r)$ and $V_l(r)$. 
Lower left panel:  The axial metric  
functions $h_l(r)$ for $l\leq 6$. 
Lower right panel: The  polar metric  
functions $H_{1,l}(r)$. In all cases,  the functions 
with $l>6$ are of order $0.1\%$ or smaller and therefore not shown.  } 
\label{f8} 
\end{figure}

\begin{figure}[h] 
\centerline{\epsfysize=6cm \epsfbox{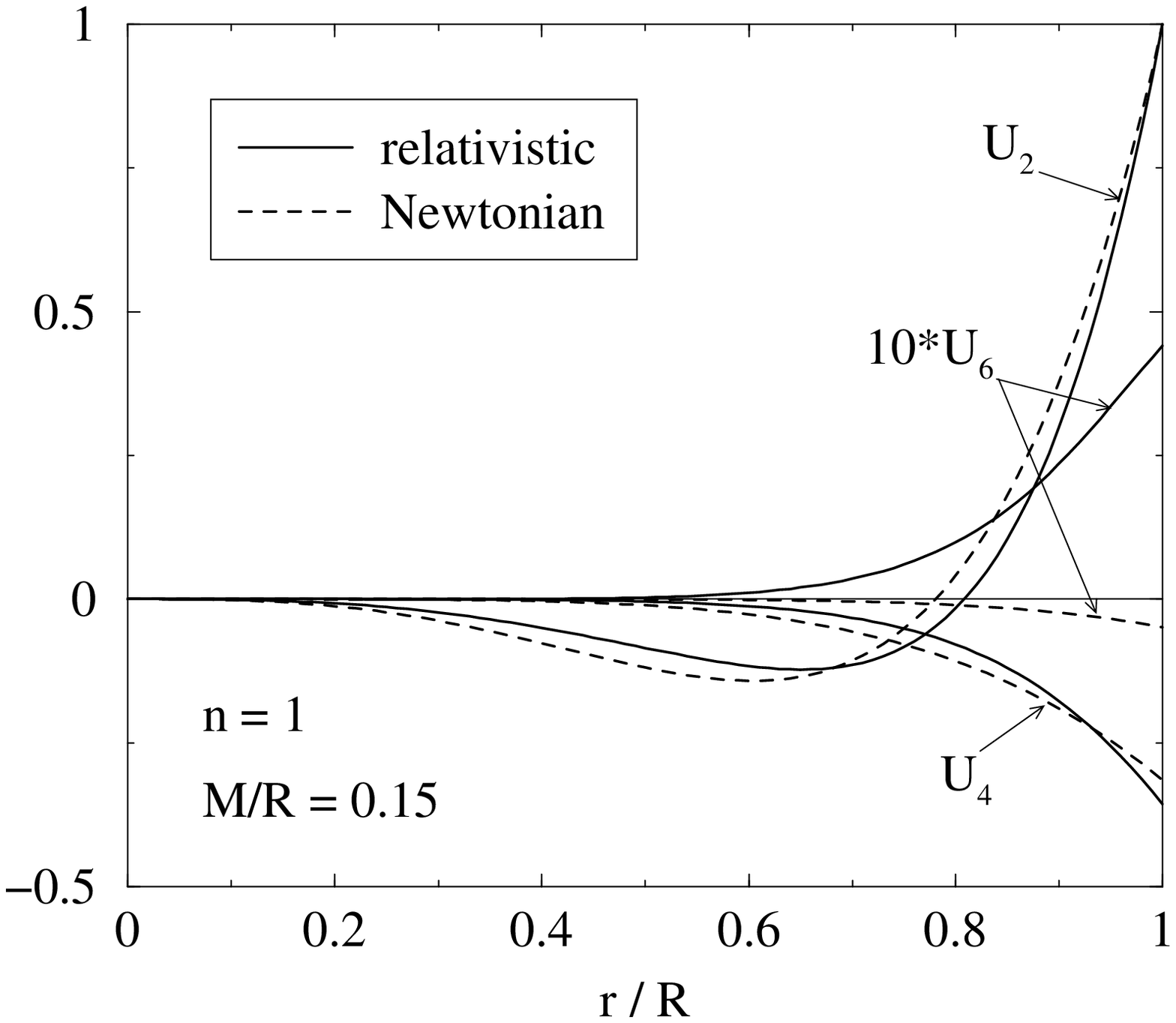} \epsfysize=6cm \epsfbox{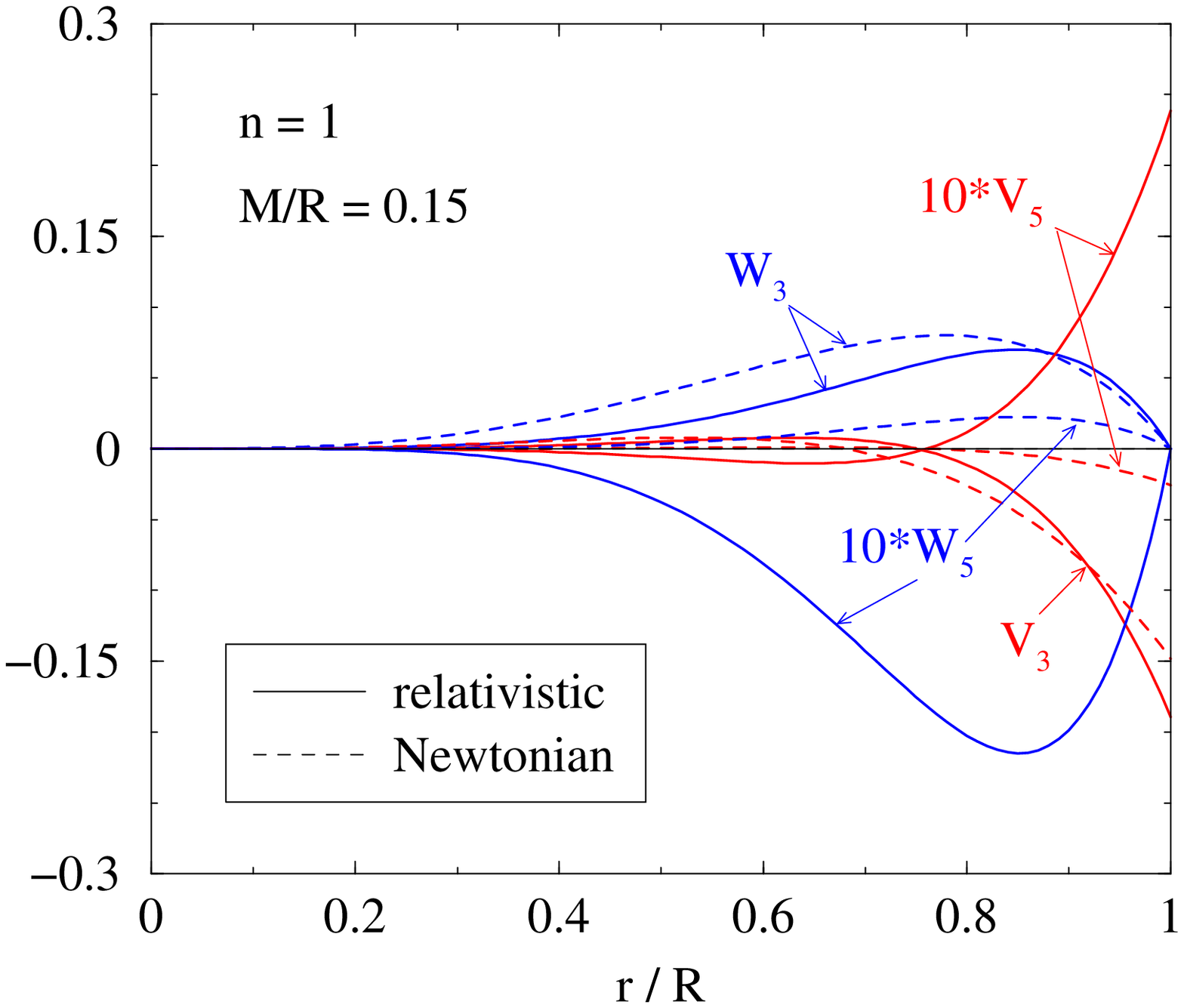}} 

\centerline{\epsfysize=6cm \epsfbox{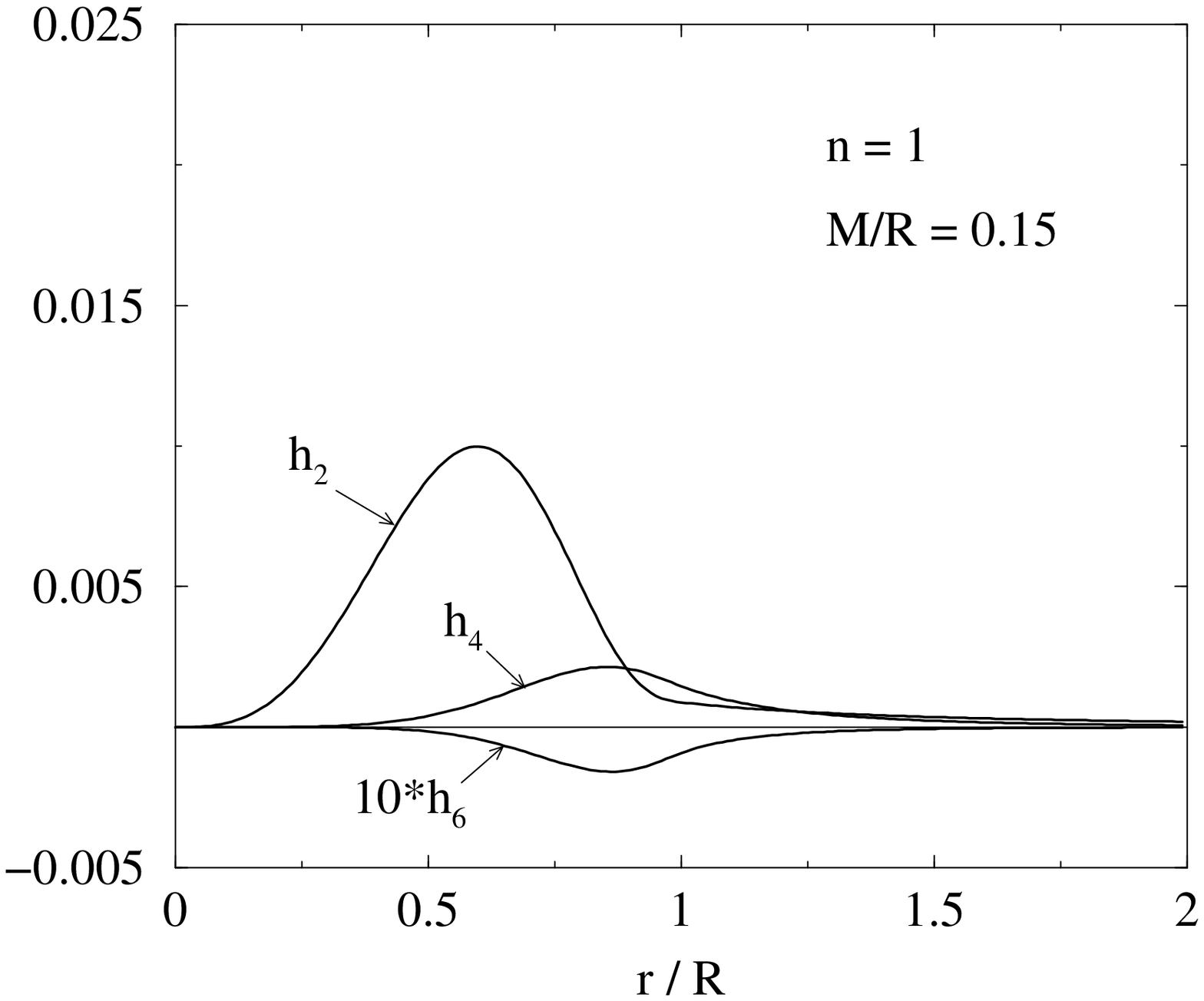} \epsfysize=6cm \epsfbox{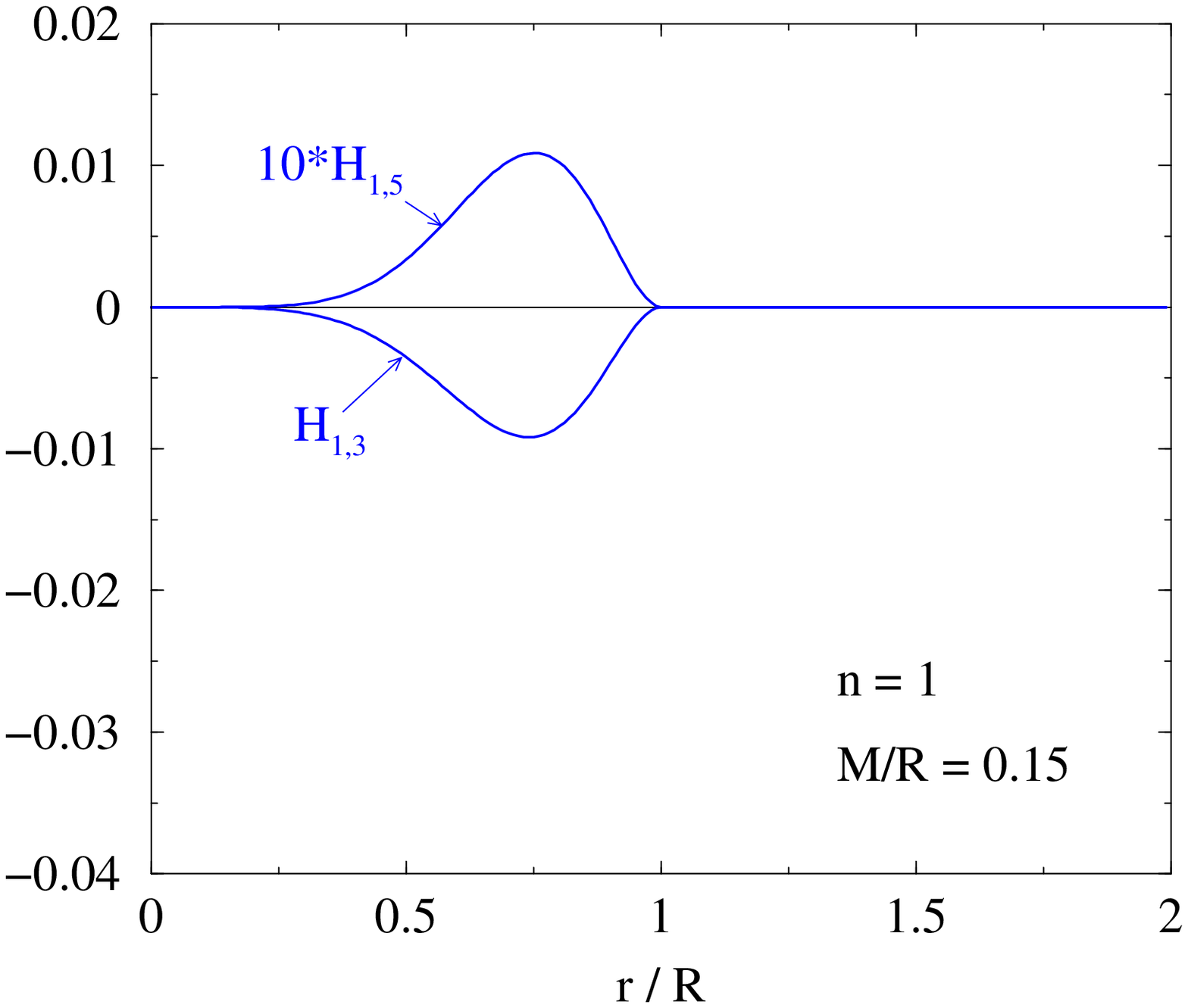}}

\caption{The same as Fig.~\ref{f8} but for a relativistic $n=1$ polytrope.} 
\label{f9} 
\end{figure}

Given the existence of a large number of inertial modes of a rotating star,  
it is interesting to ask whether other such modes may be driven unstable by
gravitational 
radiation. That this is, indeed, the case was shown in Paper~I (see also
\cite{yl}).  
A particular axial-led mode which is interesting since one would 
{\it a priori} 
expect it to couple strongly  
to current quadrupole radiation is shown in the  
series of figures \ref{f7}-\ref{f9}. This mode is the relativistic  
counterpart of one of the Newtonian $m=2$ axial-led hybrids studied  
in Paper I. (In particular, it is the mode whose Newtonian frequency is  
$\kappa_{\mbox{\tiny{N}}}=0.4669$ in the uniform density model and  
$\kappa_{\mbox{\tiny{N}}}=0.5173$ in the $n=1$ polytrope, cf. Paper~I.) 
 
Fig.~\ref{f7} shows the Newtonian mode in a uniform density model  
together with its relativistic counterpart at the nearly Newtonian 
compactness of $M/R=10^{-4}$. The Newtonian and weakly relativistic  
fluid functions $U_l(r)$, $W_l(r)$, and $V_l(r)$ for $l\leq 4$ are  
indistinguishable - the relativistic corrections being of order $10^{-5}$,  
as one would expect.  This scale for the relativistic corrections is also  
indicated by the leading metric functions $h_l(r)$ and $H_{1,l}(r)$ and  
by the fact that the functions with $l>4$ are smaller than those shown in  
both panels by a factor of order $10^{-5}$ or smaller.   
 
As discussed in Paper I, it is possible to find exact, analytic expressions  
for the Newtonian inertial modes of a uniform density star.  The explicit  
forms of the fluid functions shown in Fig.~\ref{f7} are given in Table 3  
of Paper I (but with a different normalization).  The functions with $l>4$  
are identically zero for the Newtonian uniform density model, so the  
eigenfunction shown is indeed an exact solution.  (For the Newtonian $n=1$  
model the fluid eigenfunctions are similar but not identical to those shown  
and have nonvanishing higher $l$ terms of order $0.5\%$ or smaller.) 
 
The fact that $U_2(r)$ is of order unity relative to the other fluid 
functions suggests that this mode ought to lead to significant current  
quadrupole radiation. Since it satisfies the CFS instability criterion the mode
will be driven unstable by gravitational radiation and one might  
even expect it to make as  
important a contribution to the gravitational-wave driven spin-down  
of a hot, rapidly rotating neutron star as the much-discussed $l=m=2$  
r-mode. However, a more detailed calculation of the growth timescale of the
unstable mode  
does not support these expectations. Instead, one finds that 
the current quadrupole radiation from this mode is negligible compared to higher
$l$ multipole moments for the Newtonian models. In Paper I, it was shown 
that the current quadrupole associated with this mode actually vanishes  
identically for the uniform density model and that it is negligibly small for  
the $n=1$ polytrope. The relativistic calculation provides us with a  
way to understand this result since we can now examine the perturbed metric in 
the exterior spacetime for a weakly relativistic model.   
From the results in  Fig.~\ref{f7} we see directly that the metric function
$h_2(r)$  
essentially vanishes outside the star compared to the function $h_4(r)$, 
thus  
implying that the exterior perturbed metric (i.e. the emerging radiation) 
is dominated by the $l=4$ current multipole.   
 
It is interesting to consider whether this result will still be  
true for strongly relativistic stars.  In Figs.~\ref{f8} and \ref{f9} 
we show the same mode in two stellar models with compactness  
$M/R=0.15$.  For this compactness, the mode of the $n=0$ model  
(Fig.~\ref{f8}) has eigenvalue $\kappa=0.3879$ and the  
mode of the $n=1$ model (Fig.~\ref{f9}) has eigenvalue $\kappa=0.4313$. 
That is, the frequencies are 17\% smaller than the Newtonian ones. 
For clarity, we display the coefficients of the axial and  
polar-parity terms in the spherical harmonic expansions (\ref{xi_exp})  
and (\ref{h_components}) in separate panels.   
The upper left panel of each figure shows the axial-parity fluid  
functions $U_l(r)$ for $l\leq 6$, while the upper right panel shows the polar  
parity fluid functions $W_l(r)$, and $V_l(r)$ for $l\leq 5$.  (Their  
Newtonian counterparts are also shown for comparison.)  
The lower left and right panels show, respectively, the axial  
metric functions $h_l(r)$ and the polar metric functions  
$H_{1,l}(r)$ for $l\leq 6$.  These figures indicate that the relativistic  
corrections to this mode are only of the order of $1\%$ even  
for these strongly relativistic models. 
The results also show that the metric function $h_2(r)$ is {\it not} 
completely dominated by $h_4(r)$ in the exterior spacetime. For both stars,  
$h_2$ is of order $0.1\%$ at the surface of the star (where, as  
usual, the normalization is fixed relative to $U_2(R)=1$).  This is  
comparable to $h_4$ for the $n=1$ star and only a factor of 10  
smaller than $h_4$ for the $n=0$ star.  This is an interesting result since it 
suggests that the  
Newtonian prediction that the radiation field is dominated by the  
$l=4$ current multipole is not borne out by the fully relativistic  
calculation. Indeed, in Sect.~\ref{Sect:timescale} we will see that  
the $l=2$ current multipole dominates over the $l=4$ multipole. We 
will also show that the growth timescale of the mode is 
nevertheless much longer than that of the $l=m=2$ r-mode. 
  
 
 
\begin{figure}[h] 
\centerline{\epsfysize=7cm \epsfbox{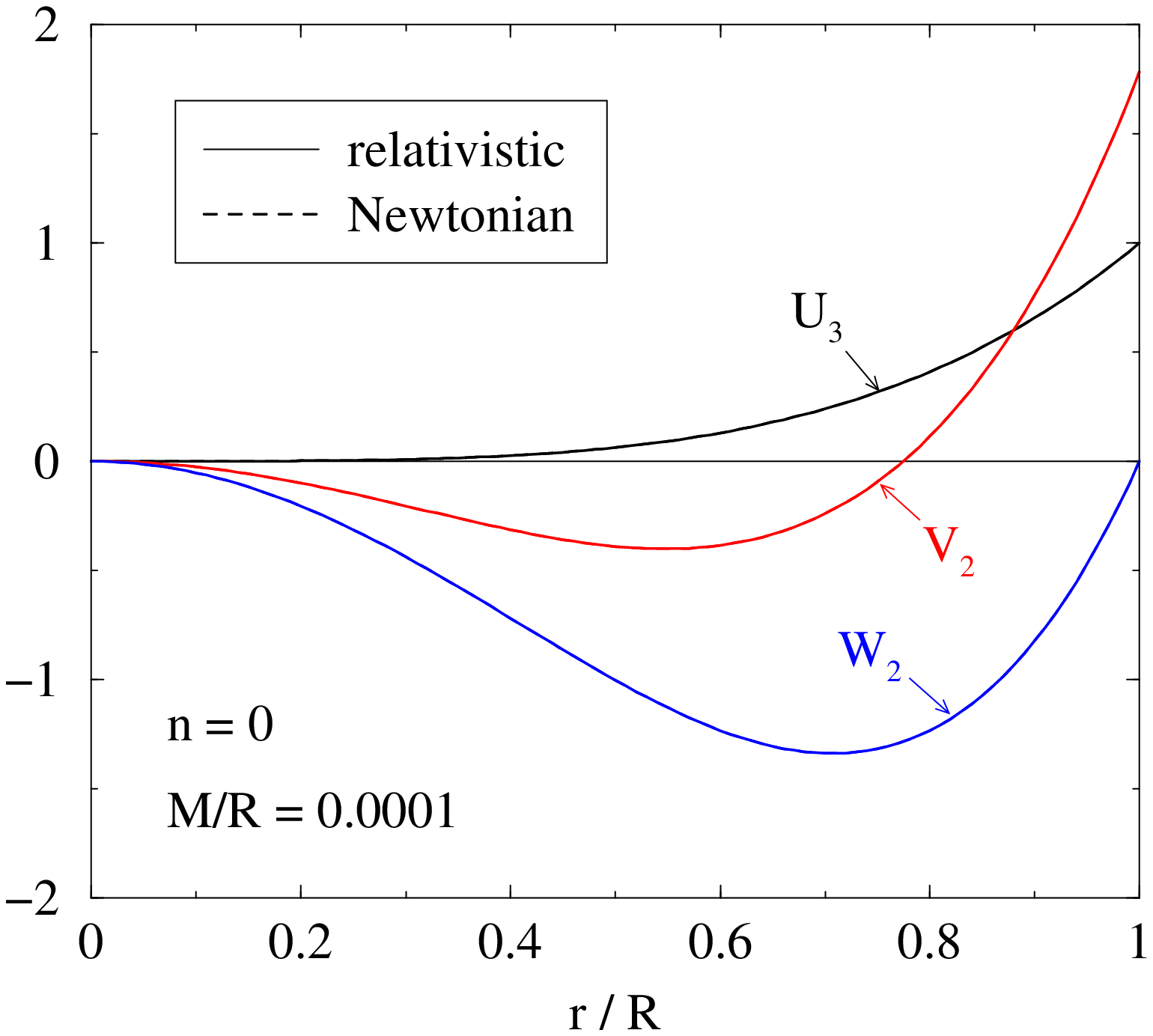} 
\epsfysize=7cm \epsfbox{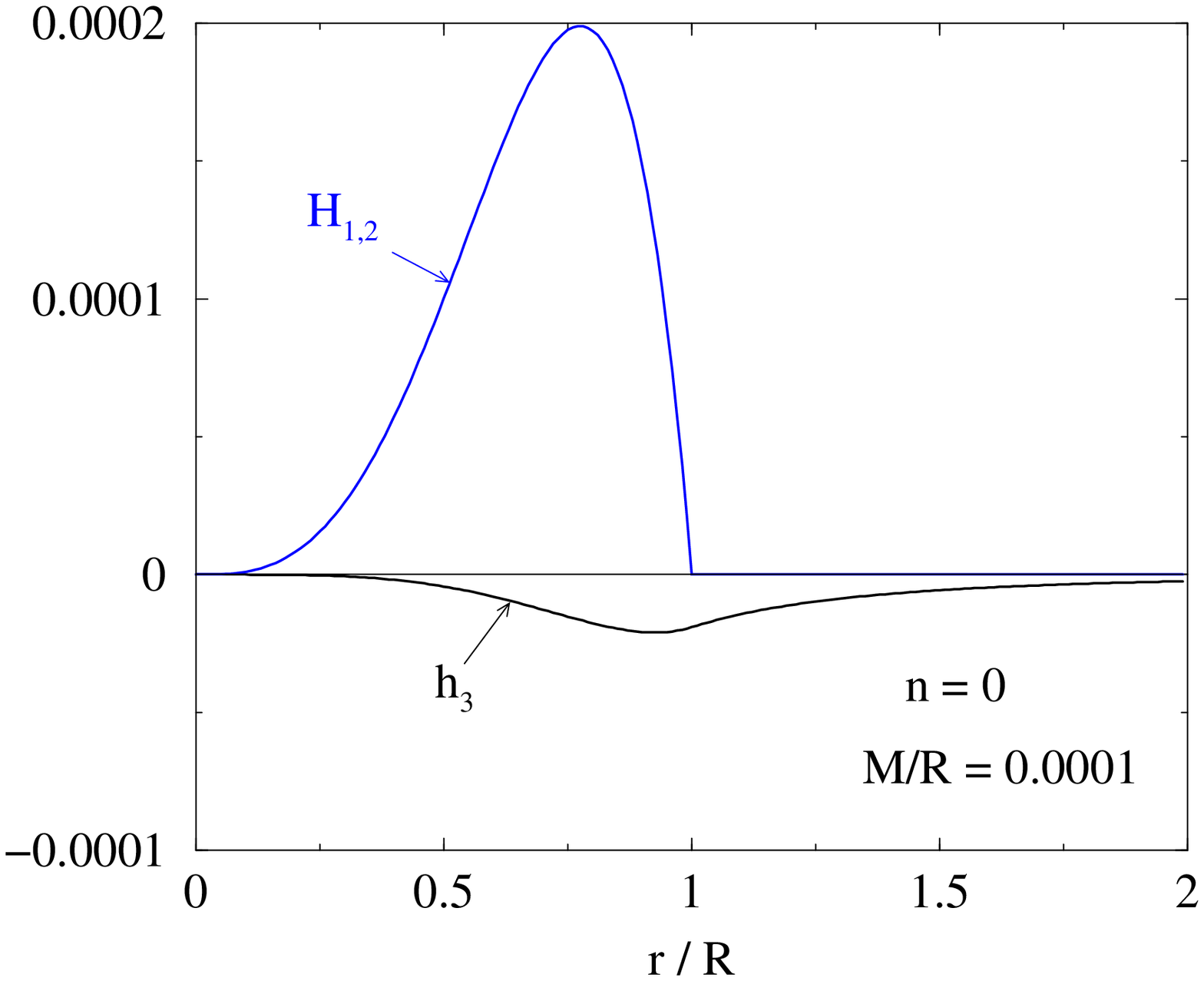}} 
\caption{A polar-led hybrid mode of a relativistic uniform density  
barotrope.  The Newtonian counterpart to this mode (dashed curves) is  
the $m=2$ polar-led hybrid with frequency $\kappa_{\mbox{\tiny{N}}}=1.232$  
(see Paper I). As in Fig.~\ref{f7}, the Newtonian and relativistic fluid  
functions are indistinguishable because the mode is shown in the weakly  
relativistic regime, with compactness $M/R=10^{-4}$.  The left panel shows  
the fluid functions $U_l(r)$, $V_l(r)$ and $W_l(r)$ while the right panel  
shows the metric functions $h_l(r)$ and $H_{1,l}(r)$, all for $l\leq 3$.  
The functions with $l>3$ are smaller than those shown (in both panels) by  
a factor of order $10^{-4}$ or smaller and are not shown.} 
\label{f12} 
\end{figure}

\begin{figure}[h] 
\centerline{\epsfysize=6cm \epsfbox{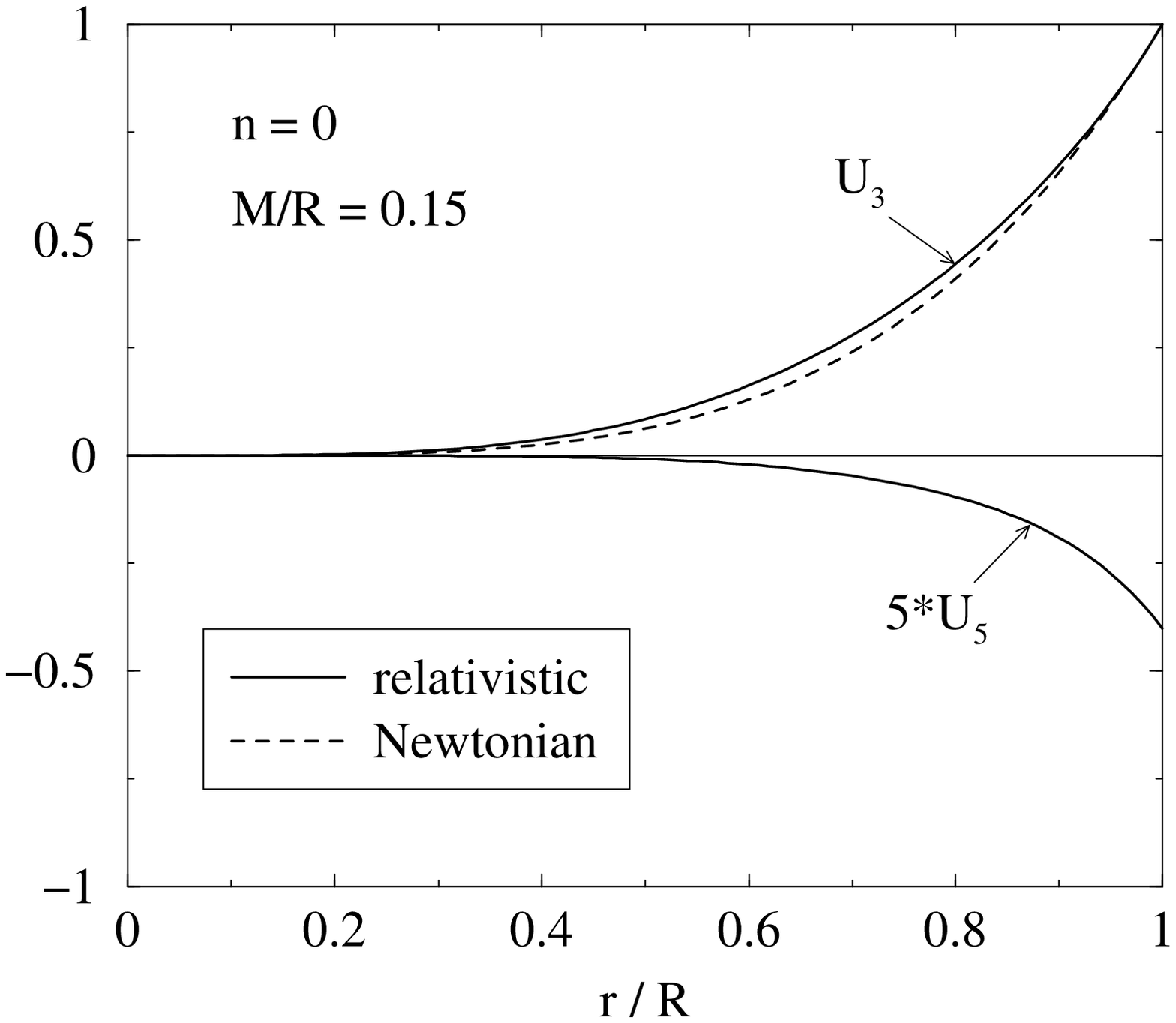} \epsfysize=6cm \epsfbox{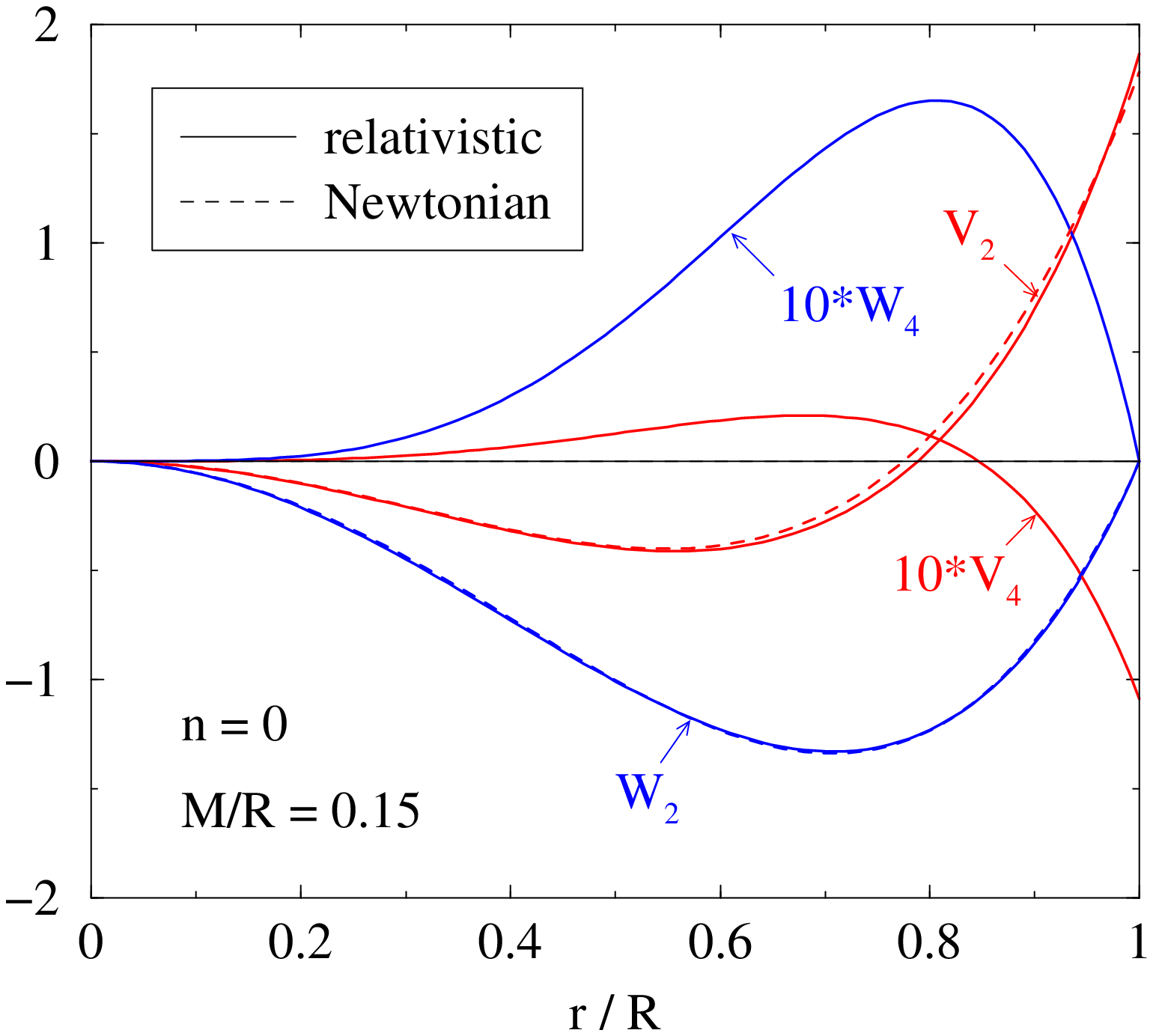}}
 
\centerline{\epsfysize=6cm \epsfbox{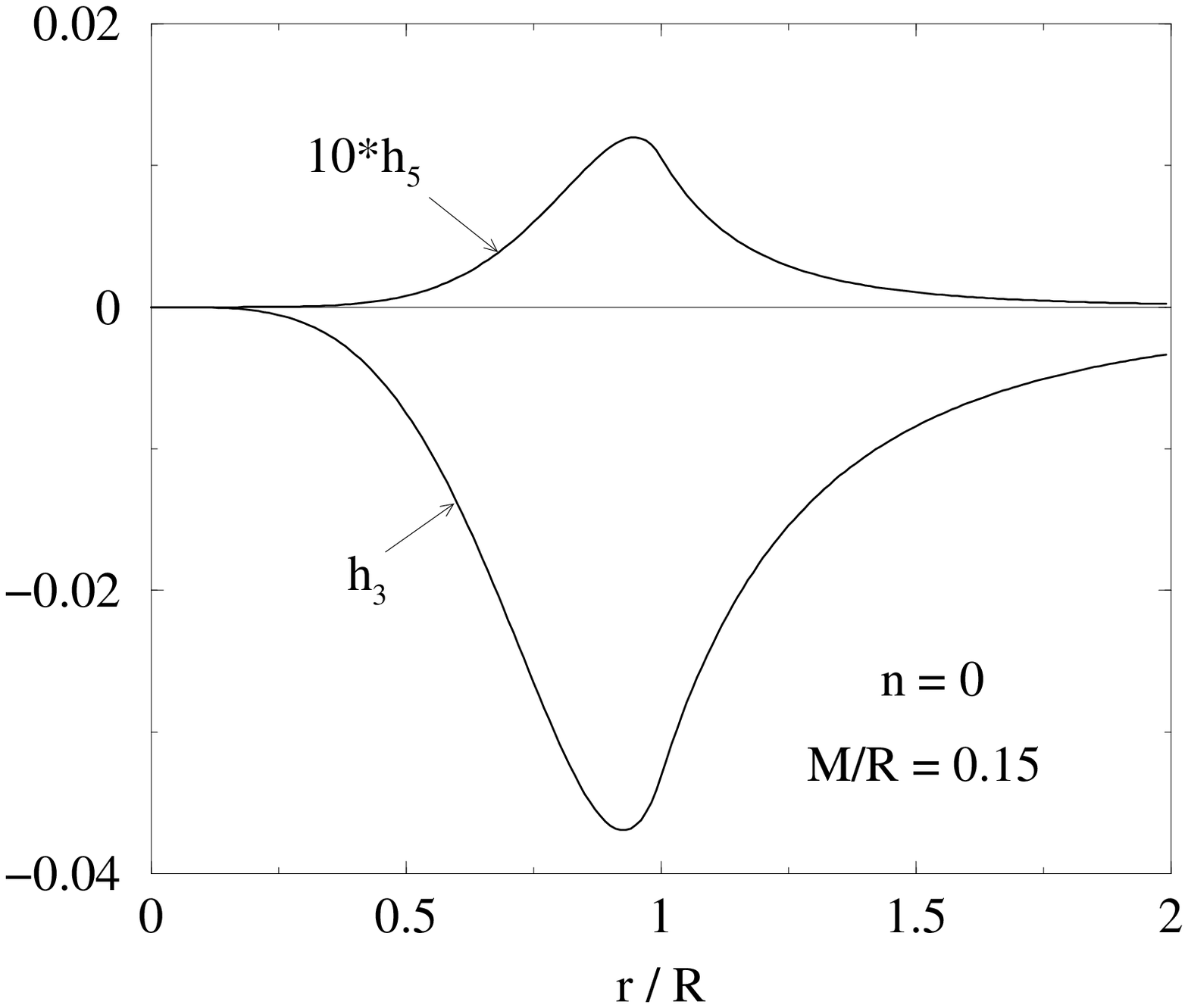} \epsfysize=6cm \epsfbox{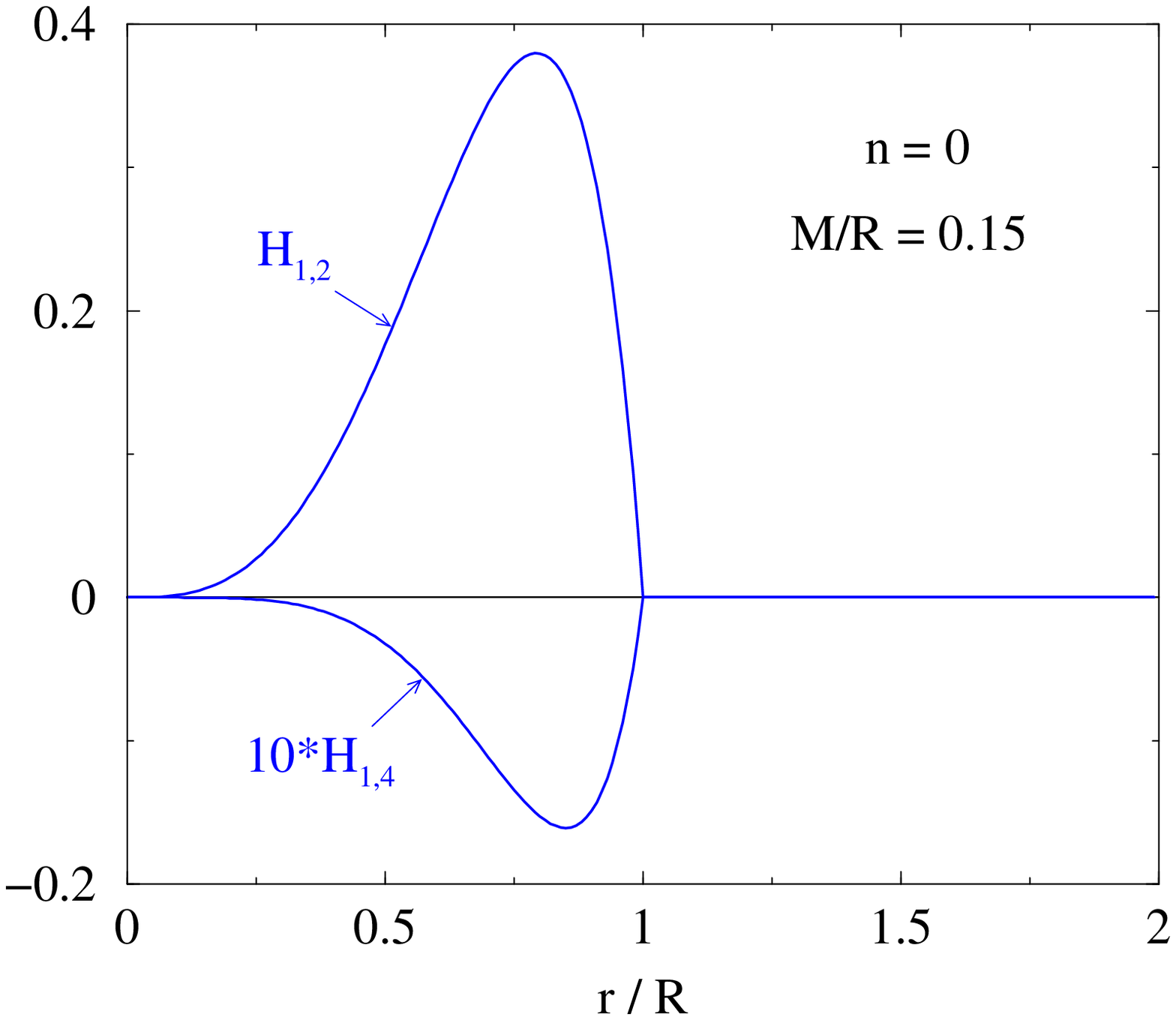}}

\caption{The same mode as in Fig.~\ref{f12}, but for a strongly  
relativistic uniform density star  ($n=0$) with compactness $M/R=0.15$. 
Upper left frame: The  
axial fluid functions $U_l(r)$. The Newtonian 
functions (dashed curves) are also shown for comparison.  
Upper right frame: The polar fluid 
functions $W_l(r)$ and $V_l(r)$.  
Lower left frame: The axial metric  
functions $h_l(r)$. Lower right frame:  
The polar metric functions   $H_{1,l}(r)$. 
 In all cases, the functions 
with $l>5$ are of order $0.5\%$ or smaller and therefore not shown.} 
\label{f13} 
\end{figure}

\begin{figure}[h] 
\centerline{\epsfysize=6cm \epsfbox{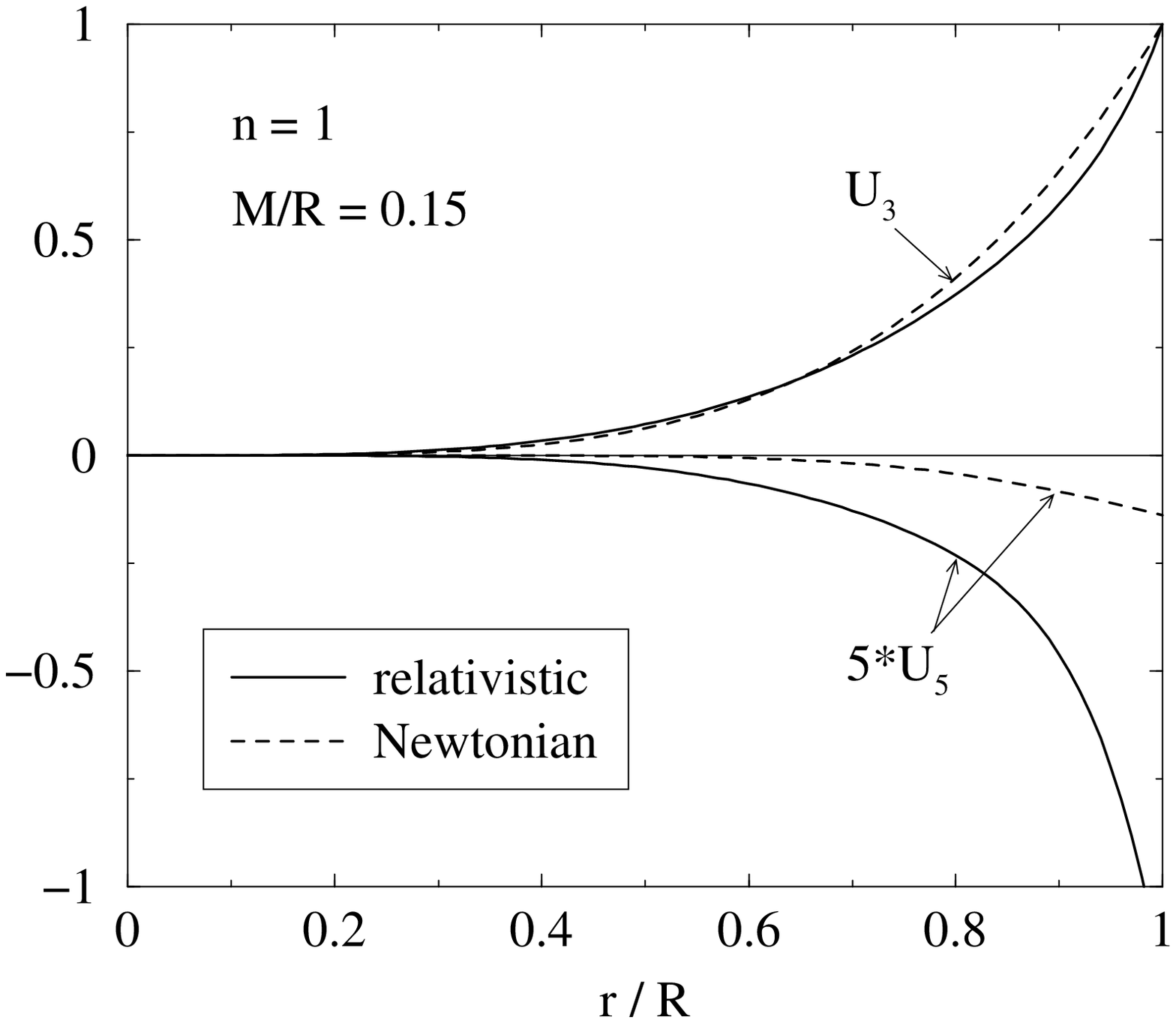} \epsfysize=6cm \epsfbox{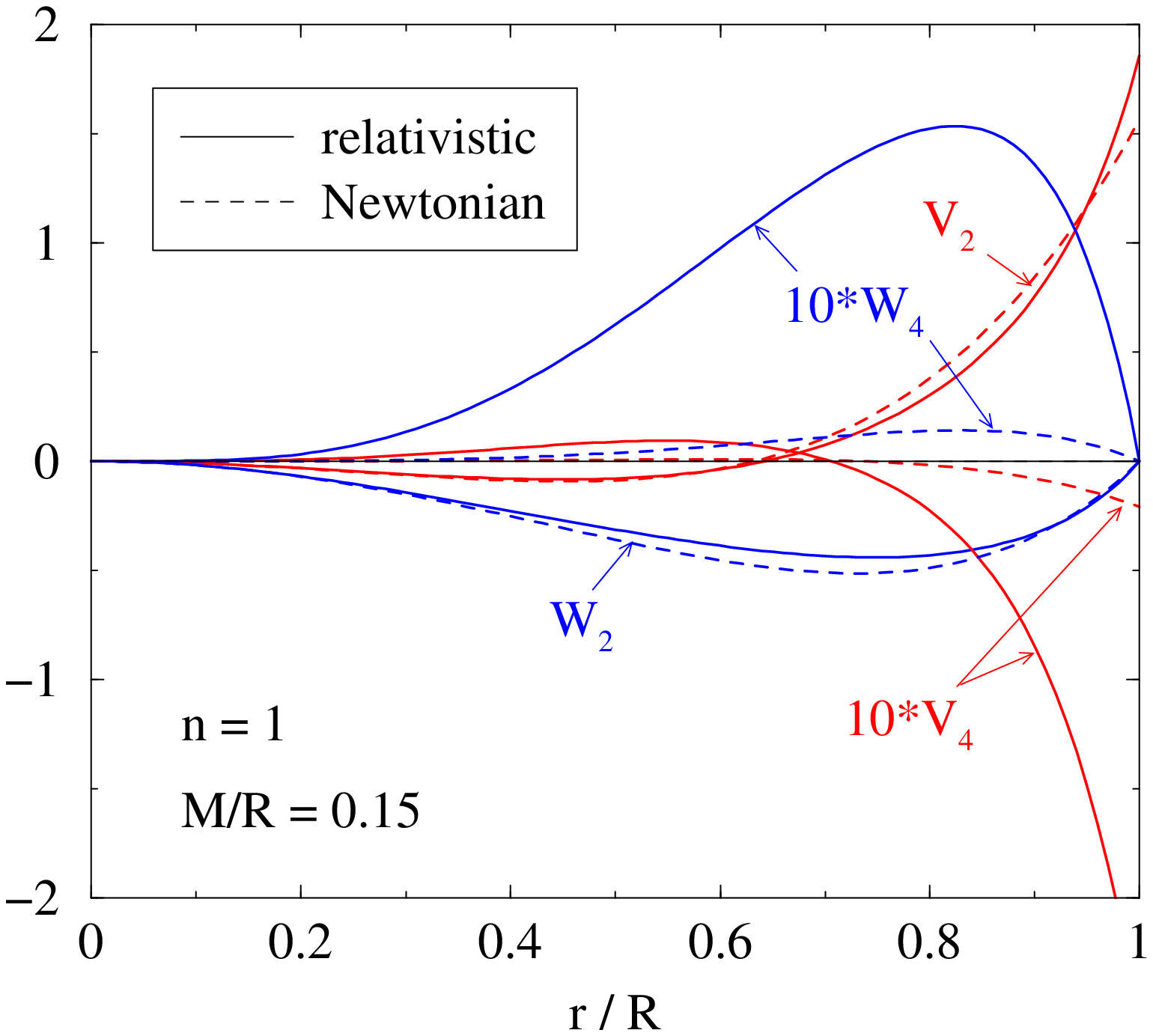}}
 
\centerline{\epsfysize=6cm \epsfbox{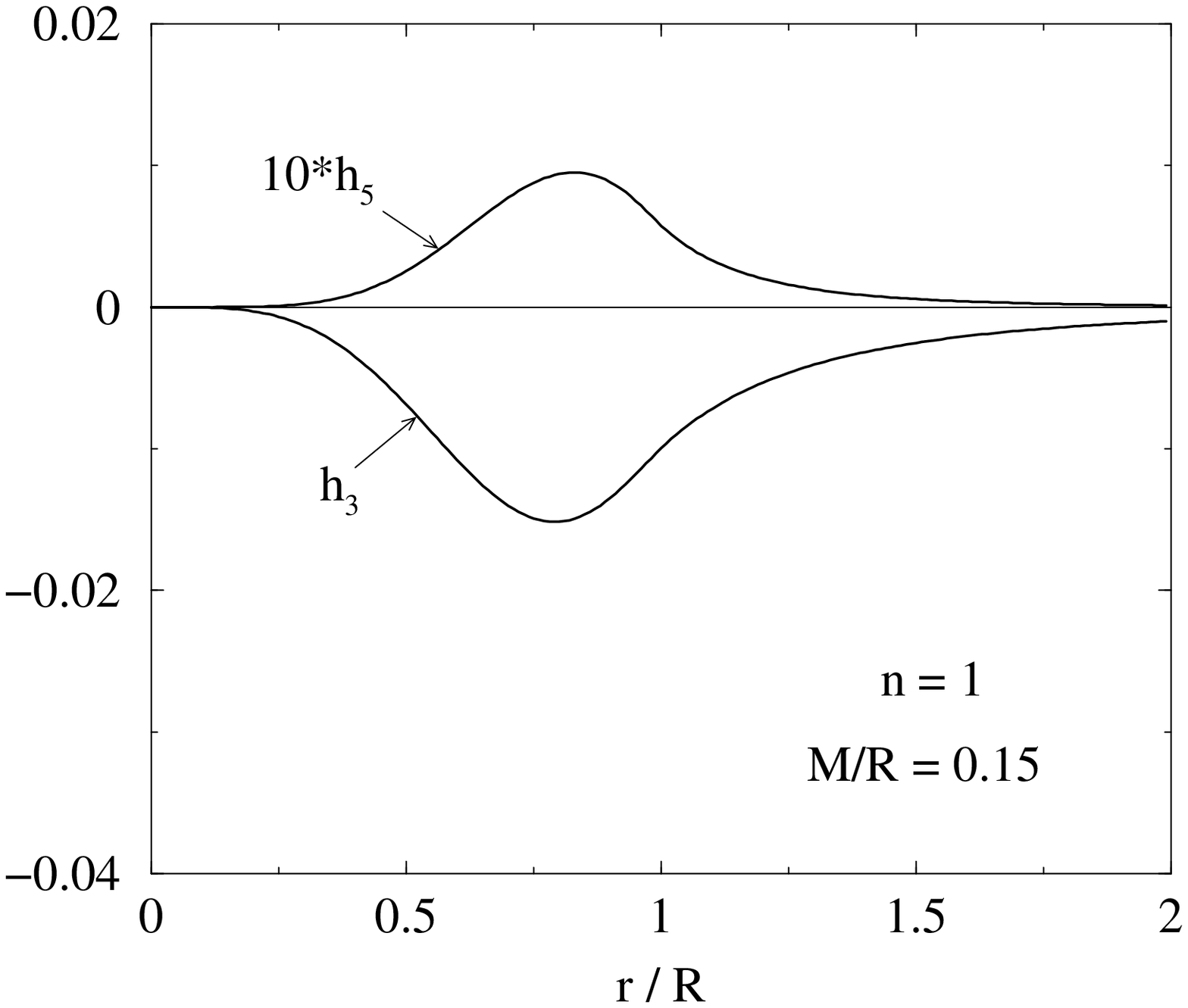} \epsfysize=6cm \epsfbox{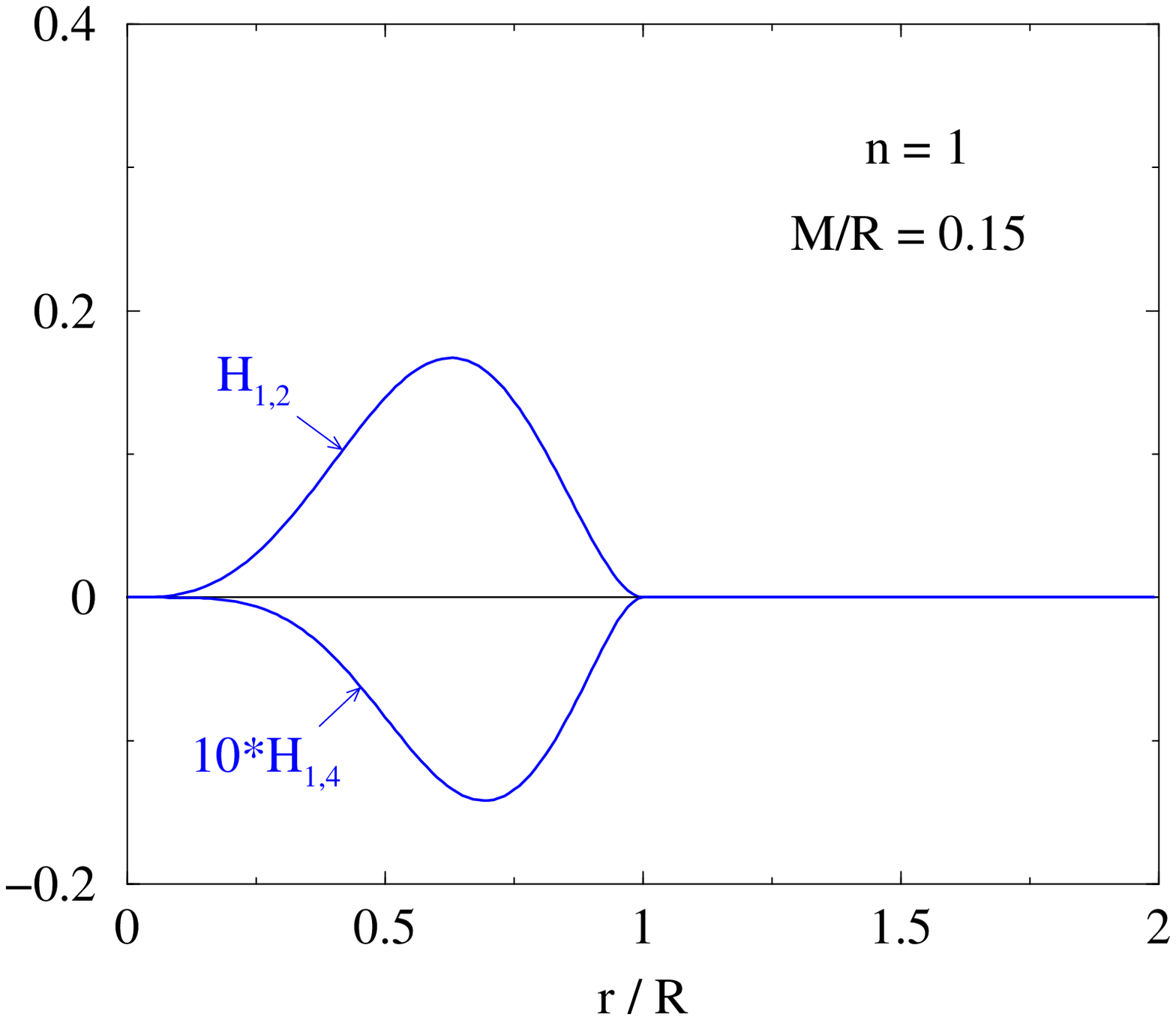}}

\caption{The same as Fig.~\ref{f13} but for a relativistic $n=1$ polytrope.} 
\label{f14} 
\end{figure}

Finally, we present a set of results for the fastest growing 
unstable polar-led inertial mode in Figs.~\ref{f12}-\ref{f14}. 
This mode is the  
relativistic counterpart of one of the Newtonian $m=2$ polar-led hybrids  
studied in Paper I. (In particular, it is the mode whose Newtonian  
frequency is $\kappa_{\mbox{\tiny{N}}}=1.232$ in the uniform density  
model and $\kappa_{\mbox{\tiny{N}}}=1.100$ in the $n=1$ polytrope.) 
Fig.~\ref{f12} shows the Newtonian mode in a uniform density model  
together with its relativistic counterpart at the nearly Newtonian 
compactness of $M/R=10^{-4}$. Again, the Newtonian and weakly  
relativistic fluid functions $U_l(r)$, $W_l(r)$, and $V_l(r)$ for  
$l\leq 3$ are indistinguishable --- the relativistic corrections and  
leading metric functions $h_l(r)$ and $H_{1,l}(r)$ being of order  
$10^{-4}$.  
As with the axial-led mode shown in Fig.~\ref{f7},  
an exact expression  
for the mode eigenfunction in the Newtonian uniform density star is given  
in Paper I (Table 4) with the fluid functions vanishing identically for 
$l>3$.  (Again, for the Newtonian $n=1$ model the fluid eigenfunctions are 
similar but not identical to those shown in Fig.~\ref{f12} and have  
nonvanishing higher $l$ terms of order $1\%$ or smaller.) 
 
In Figs.~\ref{f13}-\ref{f14} we show the mode in two stellar models  
with compactness $M/R=0.15$.  At this compactness, the mode of the  
$n=0$ model (Fig.~\ref{f13}) has eigenvalue $\kappa=1.028$ 
and the mode of the $n=1$ model (Fig.~\ref{f14}) has eigenvalue  
$\kappa=0.9068$.  This means that the frequencies are again about 17\%  
smaller than the Newtonian results. For clarity, we display separately the  
coefficients of the axial and polar-parity terms in the spherical  
harmonic expansions (\ref{xi_exp}) and (\ref{h_components}).   
These figures show that the  
relativistic corrections to this mode are, again, only of the order  
of $1\%$ even for these strongly relativistic models. 
We can also see that the metric function $h_3(r)$ dominates  
the perturbed metric exterior to the star, which implies that this  
mode couples most strongly to current octupole radiation. 
 

 
 

 
\section{Radiation Reaction Timescales} 
\label{times_sect} 
 
Almost all previous estimates of the strength of the r-mode  
instability are based on Newtonian and post-Newtonian calculations. 
One of the main goals of this work is to determine whether general  
relativity will have a significant effect on the stability of the modes, 
and, if so, to determine whether it will make them more or less unstable. 
This is obviously a crucial issue. In particular since it is known  
that relativistic effects change the instability point  
for the f-modes considerably \cite{sf}. This fact underlines why  
radiation driven instabilities need to be studied  
in the full framework of general relativity. As far as unstable inertial  
modes are concerned, there has so far been two studies of the  
associated growth times. Both concern the r-modes of non-barotropic 
relativistic stars. Yoshida and Futamase \cite{yf01} implemented the  
near-zone boundary conditions discussed by  Lindblom, Mendell and Ipser
\cite{nearzone},  
while Ruoff and Kokkotas \cite{rk02} solved the complex eigenvalue  
problem~\footnote{Correcting an earlier calculation by Andersson \cite{na98}  
that was inconsistent in that it included some terms of order 
$\Omega^2$, while neglecting others.} posed by the slow-rotation equations 
when a condition of outgoing radiation at infinity is imposed.   
These two studies provide useful insights into the growth of  
the unstable r-modes in general relativity, but the problem is still 
far from well understood.  
In particular, there are no estimates of  
the radiation reaction timescales for general inertial modes 
of a fully relativistic star in the current literature.   
This is an unfortunate gap since it seems reasonable to expect that 
barotropic models (for which all relativistic inertial modes  
are hybrids) will be relevant for rapidly spinning stars.    
In addition, it was shown in Paper I that, many 
of the hybrid inertial modes are  unstable due to the emission of gravitational
radiation.  
Their growth timescales were estimated using a post-Newtonian calculation 
and found to be considerably longer than that associated with the  
purely axial r-mode. 
It is clearly relevant to ask whether these results remain  
accurate when relativistic effects are included in the analysis.  
In fact, we have already discussed why this will not be the case 
for the particular axial-led mode illustrated in Figs.~\ref{f7}-\ref{f9}.  
 
In this section, we present the first calculation of the gravitational 
radiation-reaction timescales for general inertial modes of fully  
relativistic barotropic stars. Our method for determining  
these timescales is significantly different from those used 
in \cite{yf01,rk02}. It is close to the post-Newtonian  
approach in spirit, but based on a fully  
relativistic analysis.  
When the energy radiated per cycle is small compared to the energy  
of the mode, $E$, the imaginary part of the mode frequency is accurately  
approximated by the expression, 
\be 
\frac{1}{\tau} = -\frac{1}{2E}\frac{dE}{dt}. 
\label{tau_expr} 
\ee 
The rate of energy change is determined by a competition  
between dissipative effects such as viscosity (which tend to damp  
these modes) and gravitational radiation reaction (which drives the  
unstable modes).  To estimate the contribution to $\tau$ from  
gravitational radiation we must calculate both the mode energy and  
the rate at which gravitational waves carry energy away from the star.

\subsection{Calculation of the mode energy} 
 
We compute the mode energy using the Lagrangian perturbation formalism 
\cite{fi92,f78,fs75}. The appropriate quantity to calculate is the  
canonical energy, $E_c$. For canonical displacements $E_c$ agrees  
with the physical second order change in the energy associated with  
the mode to lowest order in perturbation theory \cite{fs78a}.   
For a mode with behavior $e^{i(\sigma t+m\varphi)}$ the canonical  
energy and angular momentum are related by $E_c = -(\sigma/m)J_c$.  
Thus, instead of computing the canonical energy directly we perform  
the (much simpler) calculation of $J_c$ using the following expression  
\cite{fi92,f78,fs75}, 
\be 
J_c = - \frac12 \Re \int n_\alpha \left\{ 
U^{\alpha\beta\gamma\delta}\pounds_\varphi\xi^*_\beta\nabla_\gamma\xi_\delta 
+V^{\gamma\delta\alpha\beta}\pounds_\varphi\xi^*_\beta h_{\gamma\delta} 
- \frac{1}{32\pi}\ep^{\alpha\gamma\mu\nu}\ep^{\beta\delta\rho}_{\ \ \ \,\nu} 
\pounds_\varphi h^*_{\gamma\delta}\nabla_\beta h_{\mu\rho}  
\right\} dV 
\label{J_cdef} 
\ee 
where $*$ denotes complex conjugation, $n_\alpha = e^\nu\nabla_\alpha t$ is  
the past-directed timelike normal to a $t=\mbox{constant}$ hypersurface, 
$dV=\sqrt{{}^3g}d^3x$ is the volume element on that surface and  
\bea 
U^{\alpha\beta\gamma\delta} &=& 
(\ep+p)u^\alpha u^\gamma q^{\beta\delta} 
+p(g^{\alpha\beta}g^{\gamma\delta}-g^{\alpha\delta}g^{\beta\gamma}) 
-\Gamma p q^{\alpha\beta}q^{\gamma\delta} 
\nn\\ 
&&\nn\\ 
2V^{\alpha\beta\gamma\delta} &=& 
(\ep+p)(u^\alpha u^\gamma q^{\beta\delta}+u^\beta u^\gamma q^{\alpha\delta} 
- u^\alpha u^\beta q^{\gamma\delta}) 
-\Gamma p q^{\alpha\beta}q^{\gamma\delta}. 
\nn 
\eea 
 
This expression for $J_c$ and the corresponding expression for $E_c$ below 
are the canonical angular momentum and energy for the physical perturbation, 
for the real (or imaginary) part of the complex perturbation $[h_{\alpha\beta},
\xi^\alpha]$. Care must be taken in comparing their values to those of  
other papers \cite{fi92,f78,lom98} that define $J_c$ (or $E_c$)  
for a complex perturbation to be the sum of its values for the real and  
imaginary parts.     
 
For our slowly rotating star, we need to compute the mode energy only to  
leading order in $\Omega$.  Taking into account the ordering  
(\ref{ordering}), a laborious but straightforward calculation gives,  
\be 
E_c = -\frac{\sigma}{m} J_c = \frac14\sigma\int(\ep+p)\left\{ 
\kappa\Omega e^{-\nu}\xi^\alpha\xi^*_\alpha + \half i u^\alpha 
\left(\xi^\beta h^*_{\alpha\beta}-\xi^{*\beta}h_{\alpha\beta}\right) 
\right\}dV. 
\ee 
[Note that it is necessary to use the fact that $\xi^\alpha$ is a  
canonical displacement to reduce expression (\ref{J_cdef}) to this form.] 
 
We now substitute for $\xi^\alpha$ and $h_{\alpha\beta}$ their 
spherical harmonic expansions (\ref{xi_exp}) and (\ref{h_components}) 
and perform the angular integration.  This leaves us with an expression 
for the canonical energy involving the variables $U_l(r)$, $V_l(r)$, 
$W_l(r)$ and $h_l(r)$ associated with the hybrid mode eigenfunctions, 
\be 
E_c = \frac{\sigma R \bar\ep}{4\kappa\Omega} I,  
\label{E_c1} 
\ee 
with $\bar\ep$ the average energy density and 
\be 
I \equiv \sum_{l=m}^\infty \int_0^1 e^{\lambda-\nu} 
\left(\frac{\ep+p}{\bar\ep}\right) 
\left\{ 
\left[e^{2\lambda} - \frac{4r(\nu'+\lambda')}{l(l+1)}\right] W^2_l 
+l(l+1)V^2_l+l(l+1)U_l(U_l+h_l) 
\right\} d\left(\frac{r}{R}\right). 
\label{E_c2} 
\ee 
Given one of our numerical solutions for a particular eigenmode it  
is straightforward to perform this radial integral numerically and  
compute the mode energy. It is also straightforward to show that (\ref{E_c1})  
reduces to Eq.~(64) of Paper~I in the Newtonian limit, once we account for  
the difference between the inertial and the rotating frames and for the fact 
that (I,64) involves the sum of the real and imaginary parts of a complex  
perturbation, giving it an extra factor of two relative to (\ref{E_c1}).

\subsection{Calculation of the radiated power} 
\label{sect:dEdt} 
 
We now compute the rate at which energy is emitted in gravitational 
waves, simplifying the analysis by using a gauge-invariant expression  
(derived in Appendix~\ref{appendix2}) for the asymptotic power radiated by an
axial mode. 
We otherwise closely follow a calculation of Ipser~\cite{ip71} (our bibliography
notes  
two minor corrections to his equations.)   
 
Recall that the perturbation equations we derived in 
Sect.~\ref{sect:pert} are relevant only in the near zone.  Because the  
mode frequency is proportional to the star's angular velocity, however, 
the slow rotation approximation allows one to extend the near zone  
arbitrarily far beyond the star by making the angular velocity sufficiently 
small (see Table~\ref{tab1}). 
Hence, to find the more general equations valid in the wave 
zone as well as in the near zone, we need only concern ourselves with  
metric perturbations of the exterior vacuum spacetime.  Furthermore, 
the equations relevant here are those governing the perturbations of a 
{\it spherical} background.  The leading rotational terms are of order 
$\om r$, and outside the star the metric function $\om(r)$ has the form, 
 
\be 
\om(r) = \frac{2J}{r^3}, 
\ee  
with $J$ the angular momentum of the star~\cite{h67}. Such terms can be  
neglected in the slow rotation approximation because 
\be 
\om r \lesssim \left(\frac{M}{r}\right)\left(\frac{R}{r}\right)\Omega R << 1. 
\ee 
Hence, the relevant components of the perturbed Einstein equation 
($\delta G_r^{\ \varphi}=0$ and $\delta G_\theta^{\ \varphi}=0$), are 
\bea 
0&=& 
i\sigma \left[h_{0,l}'-\frac{2}{r}h_{0,l}\right] 
+\left[\sigma^2  
- \frac{(l^2+l-2)}{r^2}\left(1-\frac{2M}{r}\right)\right]h_{1,l} 
\label{h1from h0} \\ 
&& \nn \\ 
0&=& 
\left(1-\frac{2M}{r}\right)^2 h_{1,l}'  
+ \left(1-\frac{2M}{r}\right)\frac{2M}{r} h_{1,l} 
- i\sigma h_{0,l},  
\label{h0fromh1} 
\eea 
where we have restored the ``0'' subscript on the metric function  
$h_{0,l} (\equiv h_l)$ to distinguish it from $h_{1,l}$. 
 
These equations may be combined to give wave equations for either  
$h_{0,l}$ or $h_{1,l}$.  It is not difficult to show that the  
resulting equation for $h_{0,l}$ reduces in the near zone to  
Eq.~(\ref{h_l''_ext}), which we used to impose boundary conditions  
on our interior solution.  The wave equation for $h_{1,l}$ is, 
\bea 
0 &=& \left(1-\frac{2M}{r}\right)^2 h_{1,l}''  
- \frac{2}{r}\left(1-\frac{2M}{r}\right)\left(1-\frac{5M}{r}\right)h_{1,l}' 
\nn \\ 
&& \nn \\ 
&& + \left[\sigma^2 - \frac{(l^2+l-2)}{r^2}\left(1-\frac{2M}{r}\right)  
        - \frac{4M}{r^3}\left(2-\frac{5M}{r}\right)\right]h_{1,l}. 
\label{h1full} 
\eea 
 
Following Ipser~\cite{ip71}, we make use of the fact that the near zone  
extends well into the nonrelativistic region (see Table~\ref{tab1}):  
Eq.~(\ref{h1full}) is easily solved in the nonrelativistic limit 
($M/r\rightarrow 0$) and the solution will be valid both in the wave  
zone and in the outer part of the near zone.  Thus, we will be able to  
impose outgoing-wave boundary conditions at $r\rightarrow\infty$, and  
still match to our near-zone solution, Eq.~(\ref{h_ext}).   
In the nonrelativistic zone, Eq.~(\ref{h1full}) is simply 
\be 
h_{1,l}'' - \frac{2}{r} h_{1,l}'  
+ \left[\sigma^2- \frac{(l^2+l-2)}{r^2}\right]h_{1,l} = 0 
\ee 
with solution 
\be 
h_{1,l} = \frac{i \sigma^{l+2} R^l}{(l-1)(2l-1)!!} \, r^2 \,  
\left[ A n_l (\sigma r) + B j_l (\sigma r) \right] 
\label{h1soln} 
\ee 
where $A$ and $B$ are constants, $n_l$ and $j_l$ are spherical Bessel  
functions and the overall normalization has been chosen for later  
convenience.  This solution for $h_{1,l}$ gives rise to a solution  
for $h_{0,l}$ using Eq.~(\ref{h0fromh1}), which, in the nonrelativistic 
zone becomes simply, 
\be 
i\sigma h_{0,l} = h_{1,l}'. 
\ee 
 
In the outer part of the near zone ($\sigma r << 1$) we have 
\be 
h_{0,l} \simeq A \left(\frac{R}{r}\right)^l 
\left\{1+ 
\frac{B}{A}\frac{(l+2)(\sigma r)^{2l+1}}{(l-1)(2l+1)[(2l-1)!!]^2} 
\right\} 
\left[ 1 + O(\sigma^2 r^2) \right]. 
\ee 
 
In the wave zone ($\sigma r >> 1$) we have 
\be 
h_{0,l} \simeq \frac{(\sigma R)^{l+1}}{(l-1)(2l-1)!!} \left(\frac{r}{R}\right) 
\left[ A \sin(\sigma r - l\pi/2) + B \cos(\sigma r - l\pi/2)\right]. 
\label{waveh0} 
\ee 
 
On this wave zone solution we now impose the boundary condition that  
the radiation be purely outgoing.  With time dependence $e^{i\sigma t}$  
this requires that $B=iA$. Our solution then becomes, 
\bea 
\mbox{wave zone:}\hspace{4em} 
h_{0,l} &\simeq& \frac{i A (\sigma R)^{l+1}}{(l-1)(2l-1)!!}  
\left(\frac{r}{R}\right) e^{-i(\sigma r - l\pi/2)} 
\\ 
&&\nn\\ 
\mbox{near zone:}\hspace{4em} 
h_{0,l} &\simeq& A \left(\frac{R}{r}\right)^l 
\left\{1+\frac{i(l+2)(\sigma r)^{2l+1}}{(l-1)(2l+1)[(2l-1)!!]^2} 
\right\} 
\left[ 1 + O(\sigma^2 r^2) \right]. 
\label{nearh0} 
\eea 
The normalization constant $A$ is fixed by matching Eq.~(\ref{nearh0}) 
in the outer (nonrelativistic) part of the near zone to the solution   
(\ref{h_ext}), which is valid throughout the near zone.  This gives  
$A = {\hat h}_{l,0}$ and justifies our neglect in Sect.~\ref{sect:pert}  
of the singular solution to Eq.~(\ref{h_l''_ext}) by the fact that the  
singular solution is indeed of order $(\sigma r)^{2l+1}$ relative to  
the regular solution. The constant ${\hat h}_{l,0}$ is, in turn, 
fixed by the matching to the interior solution, Eq.~(\ref{cont_cond}), 
whose normalization ultimately is set by Eq. (\ref{norm_cond}). 
 
We now have a solution for our axial-parity metric perturbation valid  
in the entire domain $r\in[0,\infty)$.  To compute the rate of energy  
radiation we use the gauge-invariant expression  
 
\be  
\left\langle\frac{dE}{dt}\right\rangle = - \frac{1}{32\pi }(l-1)l(l+1)(l+2)  
		\lim_{r\rightarrow\infty}\left|\frac{k_0}{r}\right|^2, 
\label{dedt0}\ee  
where  
\be 
	k_0 := h_0 - \frac12\partial_t h_2. 
\ee 
In Regge-Wheeler gauge we have $h_2=0$, $k_0 = h_0$, and hence we obtain 
\be 
\frac{dE}{dt} = -  
\frac{l(l+1)(l+2)\left(\sigma R\right)^{2l+2}|\hat h_{l,0}|^2} 
{32\pi(l-1)[(2l-1)!!]^2 R^2}. 
\label{dEdt} 
\ee

\subsection{The gravitational radiation reaction timescale} 
\label{Sect:timescale} 
 
We now combine our expressions (\ref{E_c1}) for the mode energy and  
(\ref{dEdt}) for the power radiated to find the gravitational radiation  
reaction timescale~(\ref{tau_expr}).  Restoring factors of $G$ and $c$, we 
write the timescale as 
\be 
\frac{1}{\tau_{\mbox{\tiny GR}}} = \sum_{l\geq 2} \frac{1}{\tilde\tau_l}  
\left(\frac{\Omega^2}{\pi G\bar\ep}\right)^{l+1} 
\label{GRtime1} 
\ee 
where  
\be 
\frac{1}{\tilde\tau_l} =  
\frac{c}{R} \left(\frac{3GM}{4c^2R}\right)^l 
\frac{l(l+1)(l+2)\kappa(\kappa-m)^{2l+1}|\hat h_{l,0}|^2} 
{16(l-1)[(2l-1)!!]^2 \ I} 
\label{GRtimescale} 
\ee 
with the integral $I$ defined by Eq.~(\ref{E_c2}).  
 
Having obtained this expression, we are in a position where we can evaluate the
growth  
timescales associated with the unstable inertial modes of relativistic stars.  
As a quick check on our calculation, we plug into Eqs.~(\ref{E_c2}) and  
(\ref{GRtimescale}) our post-Newtonian solution (II,5.33)-(II,5.40) for  
the corrections to the $l=m$ Newtonian r-modes of a uniform density star.  
The resulting timescale associated with the $l=m=2$ multipole agrees 
with previously published results \cite{ks99,lf,ak}: 
\be 
\tau_{\mbox{\tiny GR}} = 1.56 \mbox{s }  
\left(\frac{1.4 M_\odot}{M}\right)^4 
\left(\frac{R}{12.53 \mbox{km}}\right)^5 
\left(\frac{\pi G\bar\ep}{\Omega^2}\right)^3. 
\ee 
(See, for example, Table 8 of Paper I.) 
 
\begin{figure}[h] 
\centerline{\epsfysize=7cm \epsfbox{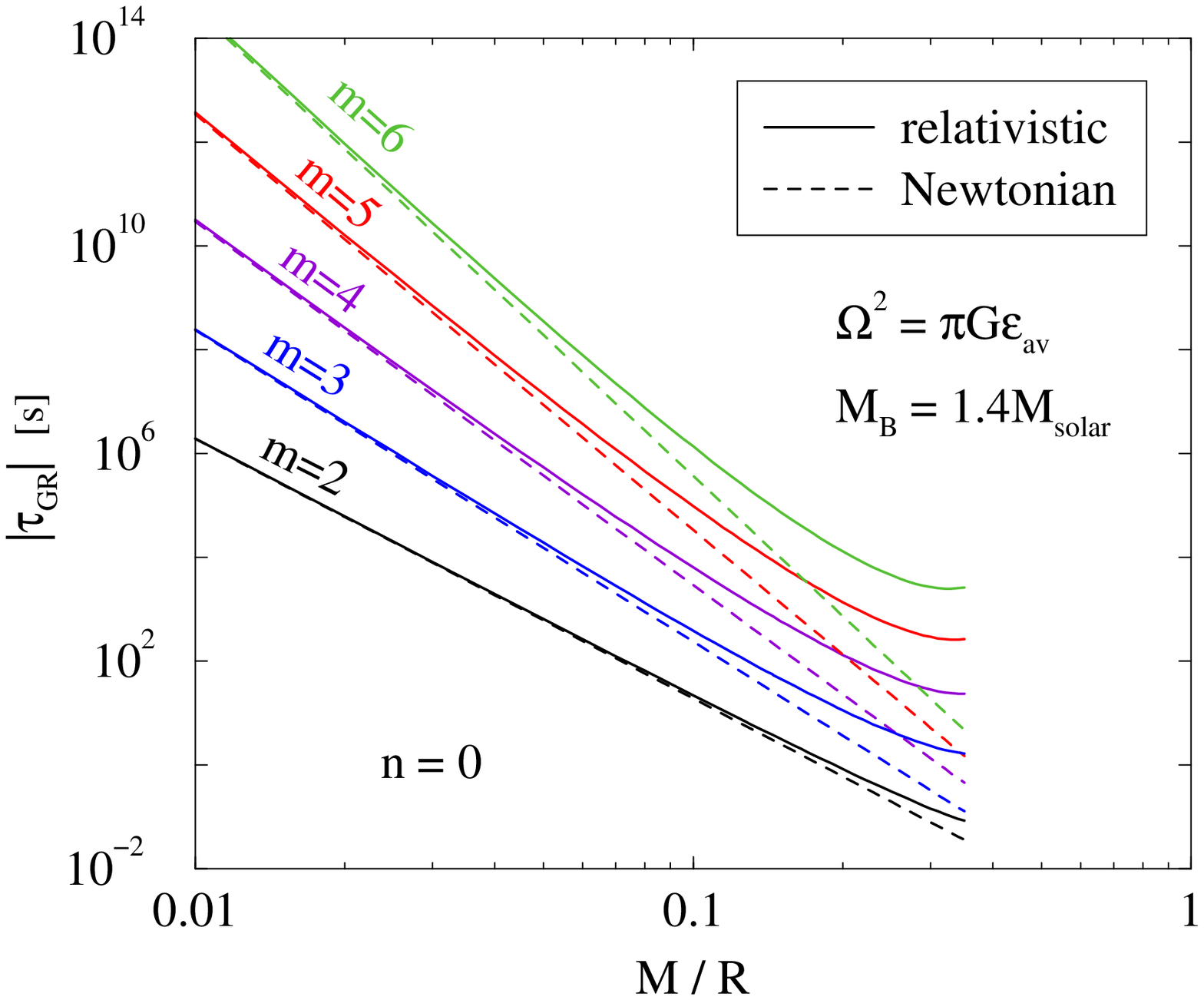} 
\epsfysize=7cm \epsfbox{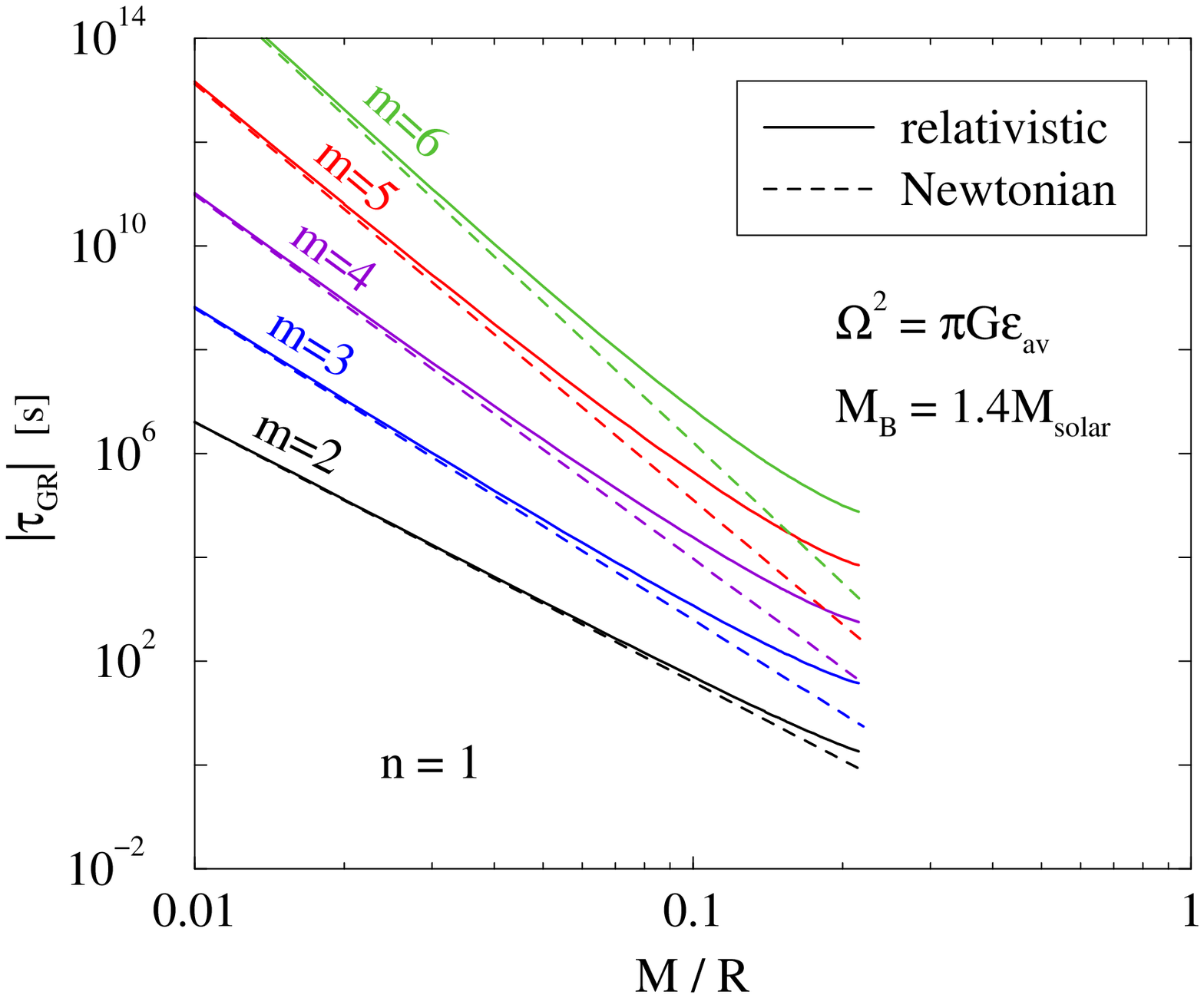}} 
\caption{Gravitational radiation reaction timescales for the fastest growing 
$l=m$ Newtonian r-modes (dashed lines) and their relativistic axial-hybrid  
counterparts (solid curves). The timescales are shown as a function of  
compactness for uniform density stars (left panel) and $n=1$ polytropes  
(right panel) of fixed baryon mass, $M_B=1.4M_\odot$.} 
\label{f17} 
\end{figure}

\begin{figure}[h] 
\centerline{\epsfysize=7cm \epsfbox{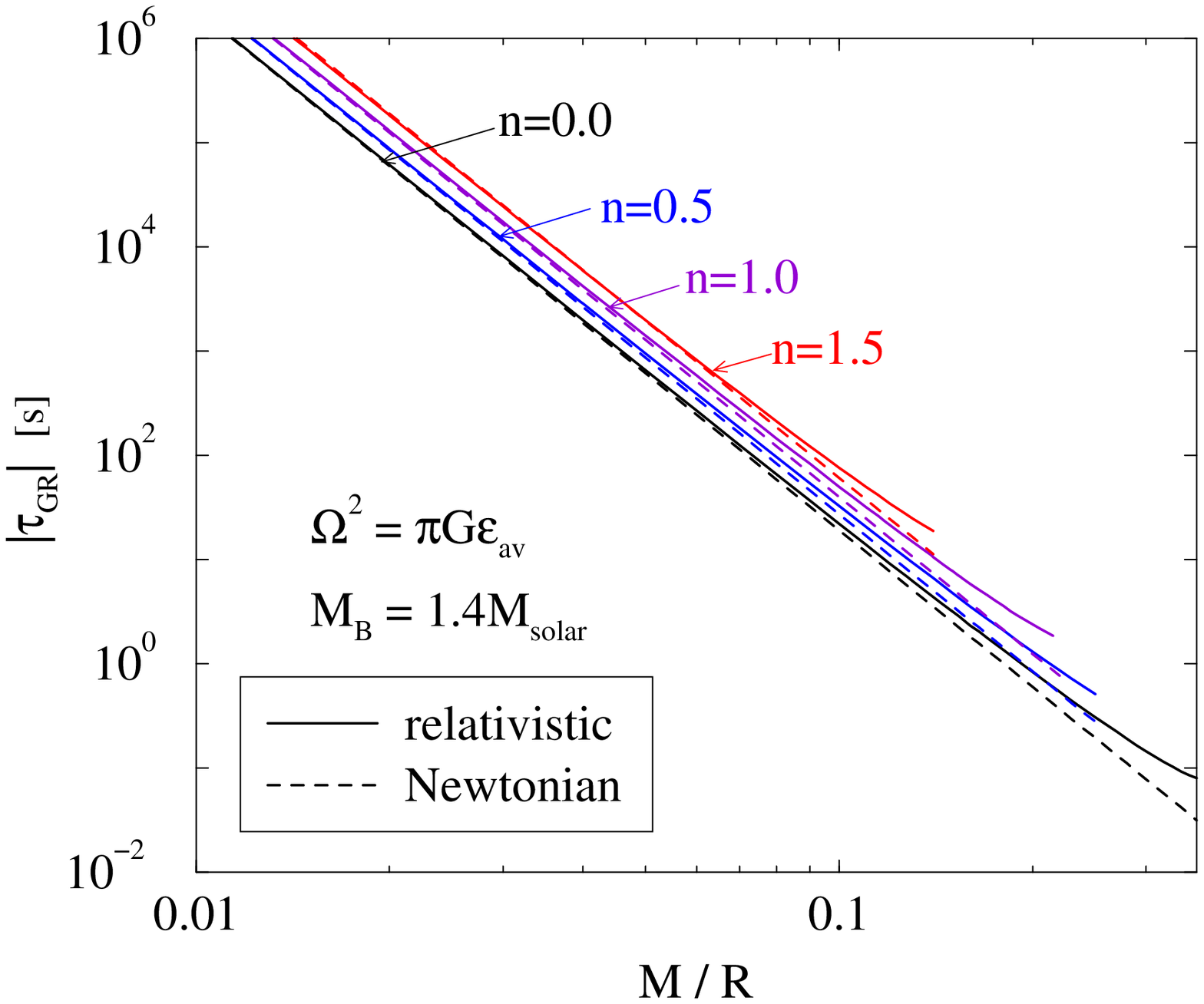} 
\epsfysize=7cm \epsfbox{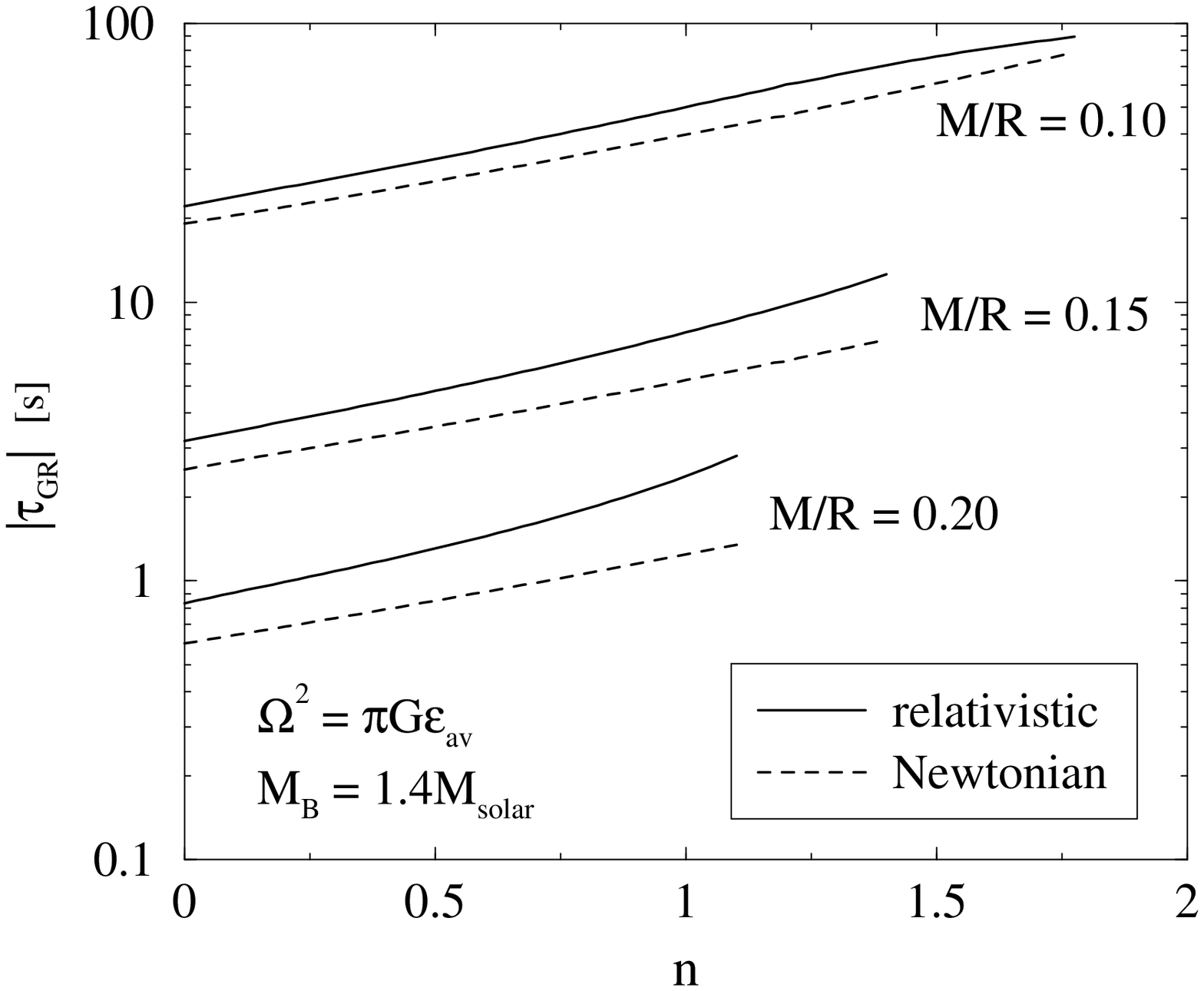}} 
\caption{Gravitational radiation reaction timescales for the $l=m=2$ Newtonian 
r-mode (dashed lines) and its relativistic counterpart (solid curves).  The  
timescales are plotted versus compactness for fixed polytropic index (left  
panel) and versus polytropic index for fixed compactness (right panel).   
All of the stellar models have the same baryon mass, $M_B=1.4M_\odot$.} 
\label{f18} 
\end{figure}

As a further check on the calculation we verify that, for weakly relativistic  
models, the timescale~(\ref{GRtime1}) exhibits the same scaling with  
the stellar parameters 
as the r-mode growth timescale estimated  
from the Newtonian calculations. The Newtonian timescales for the $l=m$  
r-modes exhibit the following scaling with the neutron star mass and  
radius  \cite{ks99,ak}: 
\be 
\tau_{\mbox{\tiny GR}} = T_l \, 
\frac{GM}{c^3} 
\left(\frac{GM}{c^2 R}\right)^{-(l+3)} 
\left(\frac{\pi G\bar\ep}{\Omega^2}\right)^{(l+1)} 
\ee 
where $T_l$ is a constant that depends only on $l$ and the star's EOS.   
with 
To test our formula for the growth timescale we keep the baryon mass, $M_B$,  
fixed at $1.4 M_\odot$ and set $\Omega^2 = \pi G\bar\ep$.  One would then  
expect a log-log plot to clearly reveal that the timescale  
depends on the star's compactness as  $(M/R)^{-(l+3)}$ for low $M/R$.  
That this is, indeed, the case can be seen from the data in  
Fig.~\ref{f17}, which compares the Newtonian and  
relativistic growth timescales of the modes whose Newtonian analogues are  
the first five $l=m$ r-modes.    
Fig.~\ref{f18} illustrates our results in a different way by 
showing the dependence on the polytropic index, $n$, of the timescale for  
the $l=m=2$ r-mode and its relativistic counterpart. 
 
These two figures suggest that for highly relativistic stars, the relativistic 
calculation tends to give a slightly {\it longer} growth timescale than  
that of a Newtonian star with the same EOS, baryon mass and compactness.   
This suggests that general relativity tends to {\it stabilize} the modes  
making the r-mode instability slightly weaker than previously expected 
(in accordance with the results for nonbarotropic stars \cite{yf01,rk02}). 
That this should be the case is natural: All inertial modes have relatively  
low frequencies since $\sigma \sim \Omega$. This means that the associated 
gravitational waves will suffer significant backscattering by the  
spacetime curvature as they escape to infinity. As the star becomes increasingly
compact more of the curvature potential,  which can be approximated by 
\be 
V \approx \left( 1 - {2M \over r} \right){ (l-1)(l+2) \over r^2} 
\label{vpot}\ee 
cf.~Eq.~(\ref{h1full}), is unveiled. Hence, one would expect low-frequency modes
of  
oscillation to radiate less efficiently as the star becomes more compact. 
 
We should point out that it is somewhat misleading to do a direct comparison 
of Newtonian and relativistic models, because there is no one-to-one 
correspondence between the two.  Although a weakly relativistic polytrope  
agrees well in all of its physical characteristics (mass, radius, compactness  
etc.) with the Newtonian model having the same polytropic index and central  
density, this is not true for strongly relativistic models.  
Newtonian and relativistic stars with the same polytropic index can be  
constructed so as to agree with respect to two of their physical  
characteristics, but in general they will differ on the others and this 
will affect the growth timescales of unstable modes. 
 
We would nevertheless like to quantify how our fully relativistic 
radiation timescales differ from the Newtonian ones.  
To account for the lack of one-to-one correspondence between Newtonian and 
relativistic models, we have constructed a number of different relativistic 
stars that agree in two of their physical characteristics with a given  
Newtonian polytrope of the same index, $n$.  The fiducial Newtonian models  
are chosen in such a way that a $1.4M_\odot$ star has a radius of
$12.53\mbox{km}$.  
This facilitates comparison with results from the literature, which made use  
of stars with these parameters \cite{lf,yl,lom98,aks99,owenetal}.   
It is natural to choose one of the  
physical characteristics on which the relativistic and Newtonian stars agree  
to be the baryon mass, $M_B$, (i.e. the rest mass) of the star.   
(For relativistic stars the baryon mass is slightly higher than the total  
gravitational mass, whereas these quantities are the same for Newtonian  
stars.)  For the other physical characteristic on which the stars agree we 
choose such quantities as the radius, central pressure, central energy  
density and so on.  We list the characteristics of these various models in  
Table~\ref{tab2}. In the table we compare a number of  
relativistic uniform density ($n=0$)  
stars with our fiducial Newtonian uniform density star, labelled N0.   
Similiarly, we compare a set of relativistic $n=1$ polytropes  
with our fiducial Newtonian $n=1$ polytrope, labelled N1.  All of these 
stars are constructed so as to have a baryon mass of $1.4M_\odot$ and also 
to agree with the fiducial Newtonian model on one of their other physical 
characteristics.  The models are listed in order of increasing compactness 
and the final column of the table indicates the second physical characteristic 
on which the stars agree.   
 
 
\begin{table}[!] 
\begin{tabular}{cccccccc} 
\hline\hline 
Model & $M/R$ & $M$ [$M_\odot$] & $R$ [km] & $\rho_c$
[$10^{15}$~g/$\mbox{cm}^3$]  
& $\ep_c$ [$10^{35}$~erg/$\mbox{cm}^3$] & $p_c$ [$10^{34}$~dyne/$\mbox{cm}^2$] &
same	 
\\ 
\hline 
N0 & 0.1650  & 1.400  & 12.53  & $0.3380$ & $3.038$  & $2.506$  &   - 
\\ 
\hline 
1 & 0.1440  & 1.269  & 13.02  & $0.2734$ & $2.457$  & $2.506$  &   $p_c$	 
\\ 
2 & 0.1490  & 1.264  & 12.53  & $0.3053$ & $2.744$  & $2.940$  &    $R$	 
\\ 
3 & 0.1538  & 1.260  & 12.10  & $0.3380$  & $3.038$  & $3.408$  &  $\rho_c$,
$\ep_c$ 
\\ 
4 & 0.1650  & 1.248  & 11.17  & $0.4251$   & $3.820$  & $4.763$  &   $M/R$	 
\\ 
\hline 
N1  & 0.1650  & 1.400 & 12.53 & $1.1120$ & $9.994$  & $8.245$ &   -	 
\\ 
\hline 
5          & 0.1390  & 1.306 & 13.87 & $0.7819$ & $7.852$  & $8.245$ &   $p_c$ 
\\ 
6          & 0.1493  & 1.300 & 12.85 & $0.9813$ & $9.994$  & $11.746$ &  
$\ep_c$ 
\\ 
7          & 0.1529  & 1.297 & 12.53 & $1.0601$ & $10.855$ & $13.272$ &   $R$ 
\\ 
8          & 0.1552  & 1.296 & 12.33 & $1.1120$ & $11.426$ & $14.319$ &  
$\rho_c$ 
\\ 
9          & 0.1650  & 1.291 & 11.55 & $1.3552$ & $14.150$ & $19.702$ &   $M/R$

\\  
\hline\hline 
\end{tabular} 
\caption{Comparison between i) a fiducial Newtonian uniform density star  
(labelled N0) and four relativistic uniform density models (labelled 
1 to 4), and ii)  a fiducial Newtonian $n=1$ model (N1) with  
five relativistic models with the same poytropic index (5-9). 
 All of the relativistic models are constructed so as to have a  
baryon mass of $1.4M_\odot$ and also to agree with the fiducial Newtonian  
model on one of their other physical characteristics (indicated in the final 
column).  Gravitational radiation reaction timescales for unstable modes of  
these stellar models are presented in Tables~\ref{tab4} and \ref{tab5}.} 
\label{tab2} 
\end{table} 

The effect of this ambiguity in comparing the relativistic and Newtonian 
models is indicated in Table~\ref{tab4}, which presents the gravitational 
radiation reaction timescales for the fastest growing $l=m$ Newtonian r-modes 
and their relativistic hybrid counterparts.  (Some of these Newtonian  
timescales have been computed in previous work
\cite{lom98,aks99,owenetal,lf,yl}.)  
We see that, depending on which physical characteristics one chooses to equate,
the  
timescales for the analogous mode of the Newtonian and relativistic models  
can differ by as much as an order of magnitude, with the relativistic mode  
generally having the longer growth time.  The fastest relativistic growth  
times are obtained by equating the compactness, $M/R$, of the relativistic  
and Newtonian models.  In this case, the relativistic growth times are  
typically weaker than the corresponding Newtonian growth times by only a  
factor of a few.  
 
In connection with the results listed in Table~\ref{tab4} it is relevant  
to make two observations. First of all, the tabulated data suggest that 
for each $l$ there is a variation of about a factor of two in $\tilde{\tau}_l$ 
between the various relativistic models (for the same equation of state).  
It is relevant to point out that had we instead tabulated the  
combination $\tilde\tau_l M^{l+2}/R^{l+3}$ (i.e. accounted for the  
expected scaling with mass and radius) then the variation between  
the models would have been much smaller. For $l=2$ we would have found  
a variation of about 6\% between the uniform density models, while  
the result for the polytropes vary by about 11\%. This indicates that  
the variation in the growth timescale between  
models 1-4 and models 5-9 is mainly due to the differences 
in mass and radius. The second feature worth noticing from the  
data in Table~\ref{tab4} is that, while the relativistic timescales 
differ from the Newtonian ones by only a factor of 2-3 for the quadrupole mode, 
the difference increases with $l$. For example, for $l=6$ the difference 
is at least an order of magnitude. This result can likely be explained 
in terms of backscattering from the curvature potential in the exterior 
spacetime. From (\ref{vpot}) we see that the ``height'' of the potential 
increases as $(l-1)(l+2)$. Thus we would expect the difference between  
our fully relativistic results and the Newtonian ones to increase with $l$   
roughly as  $(l-1)(l+2)/4$ (after rescaling with the result 
for the quadrupole mode). Our numerical results are in reasonable agreement with
this  
expectation.  
 
\begin{table}[!] 
\begin{tabular}{cc|ccccc} 
\hline\hline 
Model	& $M/R$	  & $l=m=2$ & $l=m=3$ & $l=m=4$ & $l=m=5$ & $l=m=6$ 
\\ 
\hline 
N0 & 0.1650  & $-1.56\times 10^{0}$ & $-1.17\times 10^{1}$  
& $-8.79\times 10^{1}$ & $-6.19\times 10^{2}$	& $-4.11\times 10^{3}$   
\\ 
\hline 
1 & 0.1440  & $-3.84\times 10^{0}$ & $-5.60\times 10^{1}$  
& $-7.43\times 10^{2}$ & $-8.87\times 10^{3}$	& $-9.76\times 10^{4}$   
\\ 
2 & 0.1490  & $-3.27\times 10^{0}$ & $-4.70\times 10^{1}$  
& $-6.14\times 10^{2}$ & $-7.19\times 10^{3}$	& $-7.76\times 10^{4}$   
\\ 
3 & 0.1538  & $-2.82\times 10^{0}$ & $-4.01\times 10^{1}$  
& $-5.16\times 10^{2}$ & $-5.94\times 10^{3}$	& $-6.31\times 10^{4}$   
\\ 
4 & 0.1650  & $-2.03\times 10^{0}$ & $-2.82\times 10^{1}$  
& $-3.52\times 10^{2}$ & $-3.92\times 10^{3}$	& $-4.01\times 10^{4}$   
\\ 
\hline 
N1 & 0.1650  & $-3.26\times 10^{0}$ & $-3.11\times 10^{1}$  
& $-2.84\times 10^{2}$ & $-2.37\times 10^{3}$	& $-1.81\times 10^{4}$   
\\ 
\hline 
5 & 0.1390  & $-1.09\times 10^{1}$ & $-2.23\times 10^{2}$  
& $-3.80\times 10^{3}$ & $-5.57\times 10^{4}$	& $-7.32\times 10^{5}$ 
\\ 
6 & 0.1493  & $-7.95\times 10^{0}$ & $-1.59\times 10^{2}$  
& $-2.63\times 10^{3}$ & $-3.72\times 10^{4}$	& $-4.70\times 10^{5}$ 
\\ 
7 & 0.1529  & $-7.15\times 10^{0}$ & $-1.42\times 10^{2}$  
& $-2.33\times 10^{3}$ & $-3.25\times 10^{4}$	& $-4.06\times 10^{5}$ 
\\ 
8 & 0.1552  & $-6.70\times 10^{0}$ & $-1.33\times 10^{2}$  
& $-2.17\times 10^{3}$ & $-3.01\times 10^{4}$	& $-3.71\times 10^{5}$ 
\\ 
9 & 0.1650  & $-5.15\times 10^{0}$ & $-1.01\times 10^{2}$  
& $-1.61\times 10^{3}$ & $-2.18\times 10^{4}$	& $-2.61\times 10^{5}$ 
\\  
\hline\hline 
\end{tabular} 
\caption{Gravitational radiation reaction timescales, $\tilde\tau_l$, in  
seconds for unstable ($\tilde\tau_l<0$) modes of the stellar models listed  
in Table~\ref{tab2}. The growth timescales listed here are  
those of the fastest growing $l=m$ Newtonian r-modes and their relativistic  
counterparts.  In general, the relativistic models (1-9) produce longer  
timescales than those of their Newtonian analogues (N0 and N1), suggesting  
that general relativity tends to stabilize the modes slightly.} 
\label{tab4} 
\end{table} 
  
Finally, we want to confirm the expectation that the analogue 
of the $l=m=2$ Newtonian r-mode remains the fastest growing 
unstable mode also when the growth times are estimated in  
full general relativity. That this is the case can be seen  
from  Table~\ref{tab5} where we list the growth timescales for a number of
unstable 
inertial modes.  We compare the timescales of modes of the Newtonian models 
(N0 and N1) with those of the corresponding modes of relativistic models  
4 and 9.  (These are the models with the same compactness as their  
Newtonian counterparts, and which lead to the fastest growth times, cf.
Table~\ref{tab4}).   
For each mode  
considered, we list (in four consecutive rows) the data for the four  
different stellar models --- all of which have baryon mass $M_B=1.4M_\odot$  
and compactness $M/R = 0.165$.  We list the frequency of the mode in each  
star as well as the growth timescales associated with the various current  
multipole moments of the mode.  We also list (enclosed in parentheses) the  
timescales associated with some of the inertial mode {\it mass} multipoles of  
the Newtonian $n=1$ polytrope.  These mass multipoles are higher order in  
$\Omega$ than can be computed within our slow rotation formalism. However,  
they have been calculated by Yoshida and Lee using a self-consistent third order
 
Newtonian formalism (see Table 4 of Ref. \cite{yl}) and we list them here  
simply for ease of comparison with our new results. The point is to compare  
the new relativistic timescales with the previously published Newtonian  
timescales (see, in particular, Tables 7-9 of Paper I, Table 4 of \cite{yl}  
and Table 1 of \cite{owenetal}). 
 
\begin{table}[!] 
\begin{tabular}{cccccccc} 
\hline\hline 
$m$  &Parity	&Model	&$\kappa$	&$\tilde\tau_2$	&$\tilde\tau_3$	 
&$\tilde\tau_4$	&$\tilde\tau_5$	 
\\ 
\hline 
1    &   a	&N0	&0.6120		&$\cdots$	&$-9.79\times 10^6$ 
&$\cdots$	&$-\infty  $	 
\\ 
     &   	&4	&0.5008		&$\cdots$	&$-5.21\times 10^6$ 
&$\cdots$	&$-4.19\times 10^{16}$	 
\\ 
     &   	&N1	&0.6906		&$(-2.46\times 10^5)$	 
&$-1.25\times 10^8$	&$\cdots$	&$-1.22\times 10^{20}$	 
\\ 
     &   	&9	&0.5630		&$\cdots$	&$-4.65\times 10^7$ 
&$\cdots$	&$-2.06\times 10^{17}$	 
\\ 
     &		&	&		&		&		 
&		&		 
\\ 
2    &   a	&N0	&0.6667		&$-1.56\times 10^0$	&$\cdots$ 
&$-\infty $	&$\cdots$	 
\\ 
     &   	&4	&0.5903		&$-2.03\times 10^0$	&$\cdots$ 
&$-2.09\times 10^9$	&$\cdots$	 
\\ 
     &   	&N1	&0.6667		&$-3.26\times 10^0$	 
&$(-3.49\times 10^2)$	&$-\infty $     &$\cdots$	 
\\ 
     &   	&9	&0.5796		&$-5.15\times 10^0$	&$\cdots$ 
&$-8.51\times 10^8$	&$\cdots$	 
\\ 
     &		&	&		&		&		 
&		&		 
\\ 
     &   p	&N0	&1.2319		&$\cdots$	&$-4.77\times 10^4$ 
&$\cdots$	&$-\infty  $	 
\\ 
     &   	&4	&1.0039		&$\cdots$	&$-2.41\times 10^4$ 
&$\cdots$	&$-6.07\times 10^{12}$	 
\\ 
     &   	&N1	&1.1000		&$(-1.71\times 10^3)$	 
&$-3.37\times 10^4$	&$\cdots$	&$-3.13\times 10^{14}$	 
\\ 
     &   	&9	&0.8780		&$\cdots$	&$-3.41\times 10^4$ 
&$\cdots$	&$-1.56\times 10^{12}$	 
\\ 
     &		&	&		&		&		 
&		&		 
\\ 
     &   a	&N0	&0.4669		&$-\infty $	&$\cdots$	 
&$-3.88\times 10^5$	&$\cdots$	 
\\ 
     &   	&4	&0.3786		&$-1.33\times 10^4$	&$\cdots$ 
&$-1.16\times 10^6$	&$\cdots$	 
\\ 
     &   	&N1	&0.5173		&$<-10^{18}$     
&$(-8.39\times 10^4)$	&$-1.85\times 10^6$	&$\cdots$	 
\\ 
     &   	&9	&0.4206		&$-8.29\times 10^2$	&$\cdots$ 
&$-7.47\times 10^6$	&$\cdots$	 
\\ 
     &		&	&		&		&		 
&		&		 
\\ 
3    &   a	&N0	&0.5000		&$\cdots$	&$-1.17\times 10^1$ 
&$\cdots$	&$-\infty $	 
\\ 
     &   	&4	&0.4278		&$\cdots$	&$-2.82\times 10^1$ 
&$\cdots$	&$-1.72\times 10^9$	 
\\ 
     &   	&N1	&0.5000		&$\cdots$	&$-3.11\times 10^1$ 
&$(-1.88\times 10^3)$	&$-\infty $      
\\ 
     &   	&9	&0.4259		&$\cdots$	&$-1.01\times 10^2$ 
&$\cdots$	&$-1.04\times 10^9$	 
\\ 
     &		&	&		&		&		 
&		&		 
\\ 
     &   p	&N0	&1.0532		&$\cdots$	&$\cdots$	 
&$-2.00\times 10^4$	&$\cdots$	 
\\ 
     &   	&4	&0.8438		&$\cdots$	&$\cdots$	 
&$-3.93\times 10^4$	&$\cdots$	 
\\ 
     &   	&N1	&0.9049		&$\cdots$	 
&$(-8.62\times 10^3)$	&$-2.71\times 10^4$	&$\cdots$	 
\\ 
     &   	&9	&0.7213		&$\cdots$	&$\cdots$	 
&$-9.69\times 10^4$	&$\cdots$	 
\\ 
     &		&	&		&		&		 
&		&		 
\\ 
     &   a	&N0	&0.3779		&$\cdots$	&$-\infty $	 
&$\cdots$	&$-7.67\times 10^5$	 
\\ 
     &   	&4	&0.3057		&$\cdots$	&$-7.20\times 10^4$ 
&$\cdots$	&$-4.31\times 10^6$	 
\\ 
     &   	&N1	&0.4126		&$\cdots$	&$<-10^{10}$     
&$(-5.30\times 10^5)$	&$-3.97\times 10^6$	 
\\ 
     &   	&9	&0.3369		&$\cdots$	&$-1.03\times 10^4$ 
&$\cdots$	&$-3.36\times 10^7$	 
\\  
\hline\hline 
\end{tabular} 
\caption{Gravitational radiation reaction timescales in seconds for unstable 
($\tilde\tau_l<0$) rotational modes of Newtonian and relativistic stellar  
models.  For each mode, we compare data from four different stellar models:  
a Newtonian uniform density star (model N0), a relativistic uniform density  
star (model 4) a Newtonian $n=1$ polytrope (model N1) and a relativistic $n=1$ 
polytrope (model 9).  All of these models have baryon mass $M_B=1.4M_\odot$  
and compactness $M/R = 0.165$. (Hence the Newtonian models agree with the 
canonical model typically used in the literature with mass $1.4M_\odot$ and  
radius 12.53 km.)  We list the azimuthal index, $m$, the dimensionless  
comoving frequency, $\kappa$, and the parity of the mode (i.e., whether it  
is an axial-led or a polar-led hybrid) as well as the current multipole  
radiation timescales computed to lowest order in our slow-rotation formalism. 
For convenience, we also show (in parentheses) the {\it mass} multipole  
radiation timescales computed by Yoshida and Lee~\cite{yl} for modes of  
the Newtonian $n=1$ polytrope.  (These mass multipoles, which we are unable  
to compute within our slow-rotation formalism, were computed using a  
self-consistent third order calculation.)} 
\label{tab5} 
\end{table} 

As with the timescales listed in Table~\ref{tab4}, most of the relativistic 
growth times presented in Table~\ref{tab5} are basically unchanged compared 
with their Newtonian analogs.  There are, however, cases where there 
is a dramatic difference  between the 
Newtonian and the relativistic results. The best example of this 
is provided by the fourth mode listed in Table~\ref{tab5}. This is the axial-led
inertial mode  
presented in Figs.~\ref{f7}-\ref{f9} in Sect.~\ref{Sect:freqsandfuncs},  
whose current quadrupole moment vanishes (or nearly vanishes) in the Newtonian 
models.   
In Paper I, it was argued based on the Newtonian slow-rotation calculation  
that the growth of this mode is dominated by its $l=4$ current multipole. 
However, by including rotational corrections to higher order in $\Omega$,  
Yoshida and Lee~\cite{yl} were able to compute the $l=3$ {\it mass} multipole  
and found that it drives the mode on an even shorter timescale.  Now, with the 
inclusion of relativistic corrections to the mode, we see that 
the growth of the mode is, in fact, dominated by the $l=2$ current multipole ---
as one 
would have expected in the first place.  This mode is significantly more  
unstable in general relativity than in the Newtonian calculations. However, 
it's growth time is nevertheless still much longer than that of the mode whose
Newtonian 
analogue is the $l=m=2$ r-mode. 
 
To conclude: The data presented in Tables~\ref{tab4} and \ref{tab5} suggest that
the 
relativistic corrections to the timescales are not large enough to alter the  
standard picture of the gravitational-wave driven instability of sufficiently  
hot and rapidly rotating neutron stars.  The fastest growing mode (by at  
least an order of magnitude) is the axial-led inertial mode corresponding  
to the $l=m=2$ Newtonian r-mode. And although the actual growth timescale of  
this mode is uncertain due to the uncertainty in the neutron star EOS, it is  
unlikely to be significantly shorter than has been estimated here.

 

 
\section{Concluding remarks} 
 
In this paper we have studied the inertial modes of rotating relativistic stars.
Numerical results were presented for the mode-eigenfrequencies 
and the associated eigenfunctions of barotropic models.  
These results were shown to be in good agreement with results in the literature
(in particular in the post-Newtonian limit), and provide a  
significant improvement on previous studies as far as the 
strongly relativistic regime is concerned.   
 
We also analyzed the rate at which these modes radiate gravitationally.  
In particular, we studied the growth timescale of various modes that are
unstable  
due to the emission of gravitational waves \cite{fs78a,fs78b}.  
Our calculation was based on  
two ingredients: The energy associated with the mode oscillation was  
determined as the canonical energy  defined by Friedman \cite{f78}, 
while the gravitational-wave luminosity followed from an analysis  
parallel to that of Ipser \cite{ip71}. 
By combining these two quantities we arrived at the required 
damping/growth timescale due to gravitational-wave emission. 
Our approach to the problem is novel, and differs from  
the methods previously used to estimate 
the corresponding timescales for the modes of non-barotropic stars  
\cite{rk02,yf01}. Nevertheless, the final results show the same qualitative
behaviour.  
Most notably, the post-Newtonian estimates for the growth  
rate of an unstable inertial mode are found to be surprisingly accurate 
even for strongly elativistic models. Still, the results show a  
deviation from the post-Newtonian estimates as the star reaches  
compactness similar to that expected of a neutron star 
$M/R \sim 0.1$. Then the efficiency of radiation reaction tends to  
decrease.  
This result is likely due to the fact that the low-frequency  
waves from an inertial mode experience enhanced backscattering 
by the spacetime curvature as the star becomes increasingly  
compact (recall that the spacetime of a spherical star has a curvature 
potential barrier with a peak in the region $R/M\sim 3$). This means that  
general relativistic effects tend to stabilize the inertial modes.  
This is in contrast to the results for the instability 
associated with the acoustic f-modes. As was shown by Stergioulas 
and Friedman \cite{sf}, the f-modes are significantly 
destabilized by relativistic effects.  
 
Despite significant progress in the last few years, 
it is still not clear to what extent the gravitational-wave 
driven instabilities in rotating compact stars are of astrophysical 
relevance. In particular, we do not yet have a clear answer to the question  
of whether the unstable modes may lead to detectable graviational 
waves or whether they limit the spin of nascent neutron stars or of 
old neutron stars spun up by accretion. However, it is important to realize that
serious  theoretical 
challenges need to be overcome if we want to make further progress 
in this area of research. For the unstable r-modes,  
key questions include the role of complex (not well understood) 
interior physics, e.g. the strength of hyperon bulk viscosity \cite{pbj,lo02}, 
and the saturation amplitude set by nonlinear coupling \cite{arras}.

\appendix 
\section{The spectral method used to solve the eigenvalue problem} 
\label{appendix1} 
 
To solve  the set of equations  (\ref{GR:sph_H1})-(\ref{om_r_th}) for the   
 inertial modes of a barotropic relativistic star,  
we use a variant of the method developed  
in Paper I for the analogous Newtonian hybrid/inertial  
mode problem \cite{lf}.  
We express our equilibrium and perturbation variables as a sum over a  
set of basis functions and substitute these series into our system of  
differential equations.  This results in a system of algebraic equations  
for the expansion coefficients, which may then be solved using standard linear 
algebra techniques.  In Paper~I, we made use of power series expansions 
and were able to accurately compute the Newtonian hybrid mode eigenvalues  
and eigenfunctions.  However, for the relativistic problem we have found  
it necessary to use a basis of orthogonal functions and we work instead 
with Chebychev polynomials (or, more specifically, ``Type I'' Chebychev  
polynomials \cite{arfken}). In other words, we use a spectral method 
to approach the problem.  
 
Since the Chebychev polynomials are naturally defined on the domain 
$[-1,1]$, we define a new coordinate 
\be 
y=2\left({r\over R}\right)-1 
\ee 
which maps the interior of the star, $0\leq r \leq R$, to the required domain. 
The Chebychev polynomial of degree $i$ is then given by 
\be 
T_i(y) = \cos(i\arccos y). 
\label{chebdef} 
\ee 
 
We begin by expressing our equilibrium variables in terms of this Chebychev 
basis.  We construct an equilibrium star using standard integration recipes  
and then use Chebychev approximation \cite{numrec} to find a Chebychev series  
that accurately fits each of our equilibrium variables.  In other words, we  
take all of the background variables appearing in Eqs.  
(\ref{GR:sph_V})-(\ref{om_r_th}), such as $\ep(r)$, $p(r)$, $\bar\om(r)$  
etc., and represent each of them by a Chebychev series of the form, 
\be 
B(r) = \sum_{i=0}^\infty b_i \ T_i(y) - \half b_0, 
\label{gen_back_series} 
\ee 
where  the coefficents, $b_i$, are determined from the Chebychev  
approximation algorithm (see Numerical Recipes \cite{numrec}). 
 
Having constructed Chebychev series for our (known) background variables, 
we write each of our (unknown) perturbation variables in terms of an  
expansion in Chebychev polynomials of the form, 
\be 
F_l(r) =  \left(\frac{r}{R}\right)^{l+q} 
\Biggl[\sum_{i=0}^\infty f_{l,i} \ T_i(y) - \half f_{l,0}\Biggr]. 
\label{gen_pert_series} 
\ee 
The factor $(r/R)^{l+q}$ provides for the condition of regularity at 
the origin, which requires the perturbation variables to vanish as an 
appropriate power of $r$ as $r\rightarrow 0$. The axial parity variables 
$U_l(r)$ and $h_l(r)$ have $q=1$, while the polar parity variables 
$W_l(r)$ and $V_l(r)$ have $q=0$. 
 
If we substitute the various Chebychev series represented by 
(\ref{gen_back_series}) and (\ref{gen_pert_series}) into our perturbation 
equations, each term in these equations will take the form of a product of  
two Chebychev series. That is, the generic term in our perturbation  
equations will have the form $B(r)F_l(r)$ with $B(r)$ a known background  
function and $F_l(r)$ an unknown perturbation variable.   
We would like to be able to write such a product as a new expansion in  
Chebychev polynomials. This can be accomplished using the  
identity $2 T_i T_j = T_{i+j}+T_{|i-j|}$, which follows from  
Eq. (\ref{chebdef}) and standard cosine identities.  After some 
careful rearrangement of terms we find for the product of  
the series (\ref{gen_back_series}) and (\ref{gen_pert_series}), 
\be 
B(r) F_l(r) = \half\left(\frac{r}{R}\right)^{l+q} 
\Biggl[\sum_{i=0}^\infty \pi_{l,i} \ T_i(y) - \half \pi_{l,0}\Biggr] 
\label{cheb_prod} 
\ee 
where 
\be 
\pi_{l,i} = \sum_{j=0}^\infty 
\biggl[ b_{i+j} + \Theta(j-1) b_{|i-j|} \biggr] f_{l,j}  
\ee 
with 
\be 
\Theta(k) = \left\{ \ba{l} 
0 \ \ \ \ \ \ \ \ \ \ \ \mbox{for $k<0$} \\ 
1 \ \ \ \ \ \ \ \ \ \ \ \mbox{for $k\geq 0$} 
\ea 
\right. . 
\ee 
 
We also need an expression for the derivatives of our perturbation 
variables in terms of the Chebychev expansions (\ref{gen_pert_series}). 
If we define the Chebychev expansion for the derivative of $F_l(r)$ 
as follows: 
\be 
R\frac{d}{dr} \left[\left(\frac{r}{R}\right)^{-(l+q)}F_l\right] \equiv  
\sum_{i=0}^\infty {\tilde f}_{l,i} \ T_i(y)  
- \half {\tilde f}_{l,0} 
\label{cheb_deriv_series} 
\ee 
and then make use of standard identities involving Chebychev  
polynomials \cite{arfken} it is not too difficult to show that the  
coefficients $\tilde f_{l,i}$ of this series are related to the  
coefficients $f_{l,i}$ of (\ref{gen_pert_series}) by 
\be 
\tilde f_{l,i} - \tilde f_{l,i+2} = 4(i+1)f_{l,i+1}. 
\label{cheb_deriv} 
\ee 
 
Our method of solution is as follows: We expand all of the quantities 
appearing in Eqs. (\ref{GR:sph_V})-(\ref{om_r_th}) in Chebychev series, 
and substitute these expansions into the equations and into the boundary  
and matching conditions.  We then use the formulas (\ref{cheb_prod})  
and (\ref{cheb_deriv}) to express the resulting equations as a linear 
algebraic system of the form 
\be 
Ax = 0 
\label{linalg1} 
\ee 
where $A$ is a known matrix that depends nonlinearly on the parameter  
$\kappa$ and $x$ is a vector whose components are the unknown coefficients  
in the Chebychev series for the variables $h_l$, $U_l$, $V_l$, $W_l$ and  
their derivatives. 
 
To satisfy Eq. (\ref{linalg1}) we must search for those values of $\kappa$ 
for which the matrix $A$ is singular; that is, we must find the zeroes of  
the determinant of $A(\kappa)$. Since $A$ is infinite  
dimensional, we must truncate our spherical harmonic expansions  
(\ref{xi_exp}) and (\ref{h_components}) at some maximum index  
$l_{\mbox{\tiny max}}$ and also truncate our Chebychev expansions  
(\ref{gen_back_series}), (\ref{gen_pert_series}) and  
(\ref{cheb_deriv_series}) at some maximum index $i_{\mbox{\tiny max}}$. 
The resulting finite matrix is band diagonal.  To find its zeroes 
we use standard linear algebra and root finding routines.  We then 
check for convergence of these eigenvalues as we increase 
$l_{\mbox{\tiny max}}$ and $i_{\mbox{\tiny max}}$. 
 
The eigenfunctions associated with these eigenvalues are determined  
by the perturbation equations only up to normalization.  Given a 
particular eigenvalue, we find its associated eigenfunction by replacing  
one of the equations in the system (\ref{linalg1}) with the normalization 
condition (\ref{norm_cond}).  Since this eliminates one of the rows of the  
singular matrix $A$ in favor of the normalization equation, the result is  
an algebraic system of the form 
\be  
\tilde A x = b,  
\label{linalg2}  
\ee  
where $\tilde A$ is now a non-singular matrix and $b$ is a known  
column vector.  We solve this system for the vector $x$ and  
reconstruct the various Chebychev expansions from this solution  
vector of coefficients. 
 
\section{Gauge-invariant expression for the luminosity}  
\label{appendix2} 
 
We present here a brief derivation of the gauge-invariant expression 
(\ref{dedt0}) for the rate at which energy is radiated to future null 
infinity (${\cal I}^+$) by an axial mode.  One first writes the energy 
radiated in an asymptotically regular gauge and then observes that the 
gauge-dependent quantity in the expression can be replaced by a 
quantity that is gauge invariant.

We begin with standard expressions for $dE/dt$.  Written in terms of the  
leading term $\sigma^0$ in the asymptotic shear of 
outgoing null geodesics, in outgoing null coordinates $(u,r,\theta,\varphi)$ 
of a flat asymptotic metric $\eta_{\alpha\beta}$, the instantaneous power 
radiated is 
\be 
\frac{dE}{dt} = - \frac{1}{4\pi}\int_\infty   
 \left|\partial_t\sigma^0(u,\theta,\varphi)\right|^2  d\Omega. 
\ee 
where $\displaystyle \int_\infty:= \lim_{r\rightarrow\infty}\int_{S_r}$, 
with $S_r$ a sphere of constant $r$ and $u$. In an asymptotically regular  
gauge \cite{ip71,f78}, the components $h_{(\mu)(\nu)}$ of the  
metric perturbation in an orthonormal basis $\{{\bf e}_{(\mu)}\}$  
fall off like $1/r$ or faster in an outgoing null direction, and 
(for a real perturbation), Eq. (B1) takes the form,  
\be 
\frac{dE}{dt} = - \frac{1}{16\pi}\int_\infty  
	\left[ \left(\partial_t h_{(\theta)(\varphi)} \right)^2  
		+ \frac{1}{4} \left(\partial_t h_{(\theta)(\theta)}  
		- \partial_t h_{(\varphi)(\varphi)}\right)^2 \right] 
	d\Omega. 
\label{a1}\ee

Because this expression is identical to the flat-space expression in  
terms of the deviation $h_{\alpha\beta}$ of the metric from its Minkowski value,
the radiation field can be treated as a perturbation of flat space.  In  
particular, the parts of $h_{\alpha\beta}$ belonging to different 
representations of the rotation group decouple, and we can  
restrict consideration to perturbations $h_{\alpha\beta}$ belonging to  
an $(l,m)$ representation with axial symmetry.  The gauge  
invariance of the vector $k_\alpha$ in (\ref{dedt0}) arises from this  
decoupling.   
 
  In Regge-Wheeler notation~\cite{rw57}, a (complex) axial $(l,m)$  
perturbation has the form  
\bea 
 h_{tA}  = - h_0\, \Phi^m_{l\,A},		\qquad  
  h_{rA} = - h_1 \Phi^m_{l\,A}, 	\qquad 
 h_{AB}  = \ h_2\ \chi^m_{l\,AB} 
\eea 
with $A,B$ indices on the sphere.  Here, as in Eq. (\ref{h_components}), the
axial vector harmonics have the form  
$\Phi^m_{lA} = \epsilon_A^B\partial_B Y_l^m$, 
with $\epsilon_\theta^\varphi = \frac1{\sin\theta}$,  
$\epsilon_\varphi^\theta = -\sin\theta$, while the axial tensor harmonics  
have components  
\be 
 \chi_{\theta\theta} = \frac{-1}{\sin^2\theta} \chi_{\varphi\varphi}=  
 	\frac1{\sin\theta}(\partial_\theta - \cot\theta) \partial_\varphi Y^m_l 
 				\qquad 
 \chi_{\theta\varphi} = -\frac12\sin\theta(\partial^2_\theta 
 -\cot\theta\partial_\varphi -\frac1{\sin^2\theta}\partial_\varphi^2)Y^m_l. 
\ee  
 
  A gauge transformation associated with a vector field $\zeta^\alpha$  
changes $h_{\alpha\beta}$ by  
$\nabla_\alpha\zeta_\beta+\nabla_\beta\zeta_\alpha$.  The transformation 
preserves the axial  
$(l,m)$ representation to which the physical perturbation belongs   
if and only if $\zeta_\alpha$ is itself an axial vector on the sphere: 
$\zeta_t = \zeta_r = 0, $  
\be  
\zeta_A = \zeta \Phi^m_{l\,A} 
\ee 
Then\\  
\phantom{xxxxxxxxxxxxxxx} 
$\displaystyle 
h_{tA} \rightarrow h_{tA} + \partial_t\zeta\ \Phi^m_{l\,A}, \qquad 
h_{rA} \rightarrow h_{rA} + r^2\partial_r(\frac1{r^2}\zeta)\
\Phi^m_{l\,A},\qquad 
h_{AB} \rightarrow h_{AB} + \zeta \, \chi^m_{l\,AB}, 
$\\ 
and 
\be 
h_0 \rightarrow h_0 - \partial_t\zeta,\ \ \ 	\qquad\qquad 
h_{1} \rightarrow h_1 - r^2\partial_r(\frac1{r^2}\zeta),\qquad\qquad\ \ \ \ \  
h_{2}  \rightarrow h_2 + 2 \zeta,    
\ee 
implying \cite{gs,gm} that the vector $k_\alpha$, whose nonzero  
components are 
\be 
	k_0 = h_0 + \frac12 \partial_t h_2, \qquad  
	k_1 = h_1 + \frac12 r^2 \partial_r (r^{-2} h_2),  
\ee 
is gauge invariant. 
 
When an axial perturbation is written in an asymptotically regular  
gauge, expression (\ref{a1}) is valid. For a complex perturbation,  
  
\bea  
\frac{dE}{dt}(h) &:=& \left\langle\frac{dE}{dt}(\Re h)\right\rangle  
		   +  \left\langle\frac{dE}{dt}(\Im h)\right\rangle \nonumber\\ 
 &=& - \frac{1}{16\pi}\int_\infty  
   \left|\partial_t h_2\right|^2 \left[ \left|\chi_{(\theta)(\varphi)} \right|^2
		+ \frac{1}{4} \left|\chi_{(\theta)(\theta)}  
		- \chi_{(\varphi)(\varphi)}\right|^2 \right] d\Omega\nonumber\\ 
 &=& - \frac{1}{16\pi }(l-1)l(l+1)(l+2) \lim_{r\rightarrow\infty} 
 	\left|\frac{\partial_t h_2}{2r}\right|^2. 
\label{a2}\eea 
 
Finally, noting that, in this gauge, $h_2 = O(r)$, $h_0 = O(1)$, we have  
$\displaystyle\frac{\partial_t h_2}{2r} = \frac{k_0}r +\mbox{O}(r^{-2}). $ Thus

\be 
	\frac{dE}{dt}(h) = - \frac{1}{16\pi }(l-1)l(l+1)(l+2)  
		\lim_{r\rightarrow\infty}\left|\frac{k_0}{r}\right|^2, 
\ee 
for a complex perturbation.  A real axial mode then radiates average power 
\be  
\left\langle\frac{dE}{dt}(\Re h)\right\rangle = \left\langle\frac{dE}{dt}(\Im h)
\right\rangle=  
		- \frac{1}{32\pi }(l-1)l(l+1)(l+2)  
		\lim_{r\rightarrow\infty}\left|\frac{ k_0}{r}\right|^2, 
\label{dedt1}\ee  
with the limit, as above, taken along a radially outgoing null 
direction.  Although, for an unstable mode, the perturbed 
metric and its gauge invariant quantities blow up exponentially at 
{\em spatial infinity}, the power radiated to null infinity is well-defined 
and finite. 
 
 
\acknowledgments 
 
We are grateful to the Institute for Theoretical Physics at the University  
of California - Santa Barbara for generous hospitality during the workshop  
``Spin and Magnetism in Young Neutron Stars'' where part of this research  
was conducted. 
KHL acknowledges with thanks the support provided by a Fortner Research  
Fellowship at the University of Illinois and by the Eberly research funds 
of the Pennsylvania State University. This research was also supported  
in part by NSF grant AST00-96399 at Illinois and NSF grant PHY00-90091  
at Pennsylvania. 
JLF was supported in part by NSF Grant PHY00-71044. 
NA  acknowledges support from the Leverhulme Trust via a Prize  
Fellowship, PPARC grant PPA/G/1998/00606 as well as the EU Programme  
'Improving the Human Research Potential and the Socio-Economic Knowledge  
Base' (Research Training Network Contract HPRN-CT-2000-00137).  
 

 
%
 



\end{document}